\documentclass[11pt,a4paper]{article}
\pdfoutput=1

\usepackage{jheppub}
\usepackage{latexsym}
\usepackage{multirow}
\usepackage{color}
\usepackage[usenames,dvipsnames,table]{xcolor}
\usepackage{graphicx}% Include figure files
\usepackage{epsfig}  % Include figure files
\usepackage{epsf}    % Include figure files
\usepackage{dcolumn}% Align table columns on decimal point
\usepackage{bm}% bold math
\usepackage{dcolumn}% Align table columns on decimal point
\usepackage{textcomp}% Align table columns on decimal point
\usepackage{float}
\usepackage{subcaption}
\usepackage{hypcap}
\usepackage[]{hyperref}
\usepackage{makecell}
\usepackage{color}
\usepackage{pifont}
\usepackage{appendix}
\usepackage{amsmath}
\usepackage{multirow,bigdelim}
\usepackage{multicol}  
\usepackage{lineno}
\usepackage[normalem]{ulem}
\usepackage{diagbox}
\usepackage{pdflscape}
\usepackage{afterpage}
\usepackage{capt-of}
\usepackage{rotating}
\usepackage{lscape}
\usepackage{scalerel}
\usepackage{adjustbox}

\usepackage{bm}
\usepackage{orcidlink}
\usepackage{wasysym}
\usepackage{mathrsfs}
\usepackage{caption}
\usepackage[section]{placeins}
\usepackage{tablefootnote}
\usepackage{fontawesome5}
\usepackage{arydshln}

\newcommand{\mc}[3]{\multicolumn{#1}{#2}{#3}}
\newcommand{\mr}[3]{\multirow{#1}{#2}{#3}}

\hypersetup{
 unicode=false,          % non-Latin characters in Acrobat?s bookmarks
 pdftoolbar=true,        % show Acrobat?s toolbar?
 pdfmenubar=true,        % show Acrobat?s menu?
 pdffitwindow=true,     % window fit to page when opened
 pdfstartview={FitH},    % fits the width of the page to the window
 pdfsubject={Neutrino Oscillations Phenomenology},   % subject of the document
 pdfnewwindow=true,      % links in new window
 pdfcreator={RevTeX},
 colorlinks=true,       % false: boxed links; true: colored links
 linkcolor=red,          % color of internal links
 citecolor=blue,        % color of links to bibliography
 filecolor=black,      % color of file links
 urlcolor=blue,           % color of external links
}
  
\newcommand{\ie}{\textit{i.e.}}
\newcommand{\eg}{{\it e.g.}}
\newcommand{\fig}{fig.}

\newcommand{\Refe}{Ref.}
\newcommand{\Refes}{Refs.}
\newcommand{\nue}{\mbox{$\nu_e$}}

\newcommand{\numu}{\mbox{$\nu_{\mu}$}}

\newcommand{\equ}[1]{eq.~(\ref{equ:#1})}
\newcommand{\figu}[1]{\fig~\ref{fig:#1}}


\preprint{IP/BBSR/2024-01}

\title{A plethora of long-range neutrino interactions probed by DUNE and T2HK~\href{https://github.com/pragyanprasu/LRI-LBL-2024}{\faGithubSquare}}
\author[a,b,c]{\orcidlink{0000-0002-9714-8866}Sanjib Kumar Agarwalla,}
\author[d]{\orcidlink{0000-0001-6923-0865}Mauricio Bustamante,}
\author[a,e]{\orcidlink{0000-0002-8363-7693}Masoom Singh,}
\author[a,b]{\orcidlink{0000-0003-3008-480X}Pragyanprasu Swain}

\affiliation[a]{Institute of Physics, Sachivalaya Marg, Sainik School Post, Bhubaneswar 751005, India}
\affiliation[b]{Homi Bhabha National Institute, Training School Complex, Anushakti Nagar, Mumbai 400094, India}
\affiliation[c]{Department of Physics \& Wisconsin IceCube Particle Astrophysics Center, University of Wisconsin, Madison, WI 53706, U.S.A.}
\affiliation[d]{Niels Bohr International Academy, Niels Bohr Institute, University of Copenhagen, DK-2100 Copenhagen, Denmark}
\affiliation[e]{Department of Physics, Utkal University, Vani Vihar, Bhubaneswar 751004, India}

\emailAdd{sanjib@iopb.res.in}
\emailAdd{mbustamante@nbi.ku.dk} 
\emailAdd{masoom@iopb.res.in}
\emailAdd{pragyanprasu.s@iopb.res.in }

\abstract{Upcoming neutrino experiments will soon search for new neutrino interactions more thoroughly than ever before, boosting the prospects of extending the Standard Model.  In anticipation of this, we forecast the capability of two of the leading long-baseline neutrino oscillation experiments, DUNE and T2HK, to look for new flavor-dependent neutrino interactions with electrons, protons, and neutrons that could affect the transitions between different flavors.  We interpret their sensitivity in the context of long-range neutrino interactions, mediated by a new neutral boson lighter than $10^{-10}$~eV, and sourced by the vast amount of nearby and distant matter in the Earth, Moon, Sun, Milky Way, and beyond.  For the first time, we explore the sensitivity of DUNE and T2HK to a wide variety of $U(1)^\prime$ symmetries, built from combinations of lepton and baryon numbers, each of which induces new interactions that affect oscillations differently.  We find ample sensitivity: in all cases, DUNE and T2HK may constrain the existence of the new interaction even if it is supremely feeble, may discover it, and, in some cases, may identify the symmetry responsible for it.}


\arxivnumber{2404.02775}

\begin{document}

\maketitle

%=================================================
\section{Introduction}
\label{sec:intro}
%=================================================

In the search for physics beyond the Standard Model, neutrino flavor transitions provide versatile and exacting probes.  One large class of proposed models of new neutrino physics posits the existence of new flavor-dependent neutrino interactions.  These are interactions beyond the standard weak ones that, by affecting $\nu_e$, $\nu_\mu$, and $\nu_\tau$ differently, could modify the transitions between them relative to the standard expectation.  Because the new interaction is likely feeble, the modifications are likely difficult to spot.  Yet, if the range of the new interaction is long~\cite{He:1990pn,Foot:1990uf, Foot:1990mn, He:1991qd, Foot:1994vd}, then vast repositories of matter located far from the neutrinos may source a large matter potential that could affect flavor transitions appreciably, even if the new interaction is significantly more feeble than weak interactions.

%=================================================
\begin{figure}[t!]
 \centering
 \includegraphics[width=0.85\textwidth]{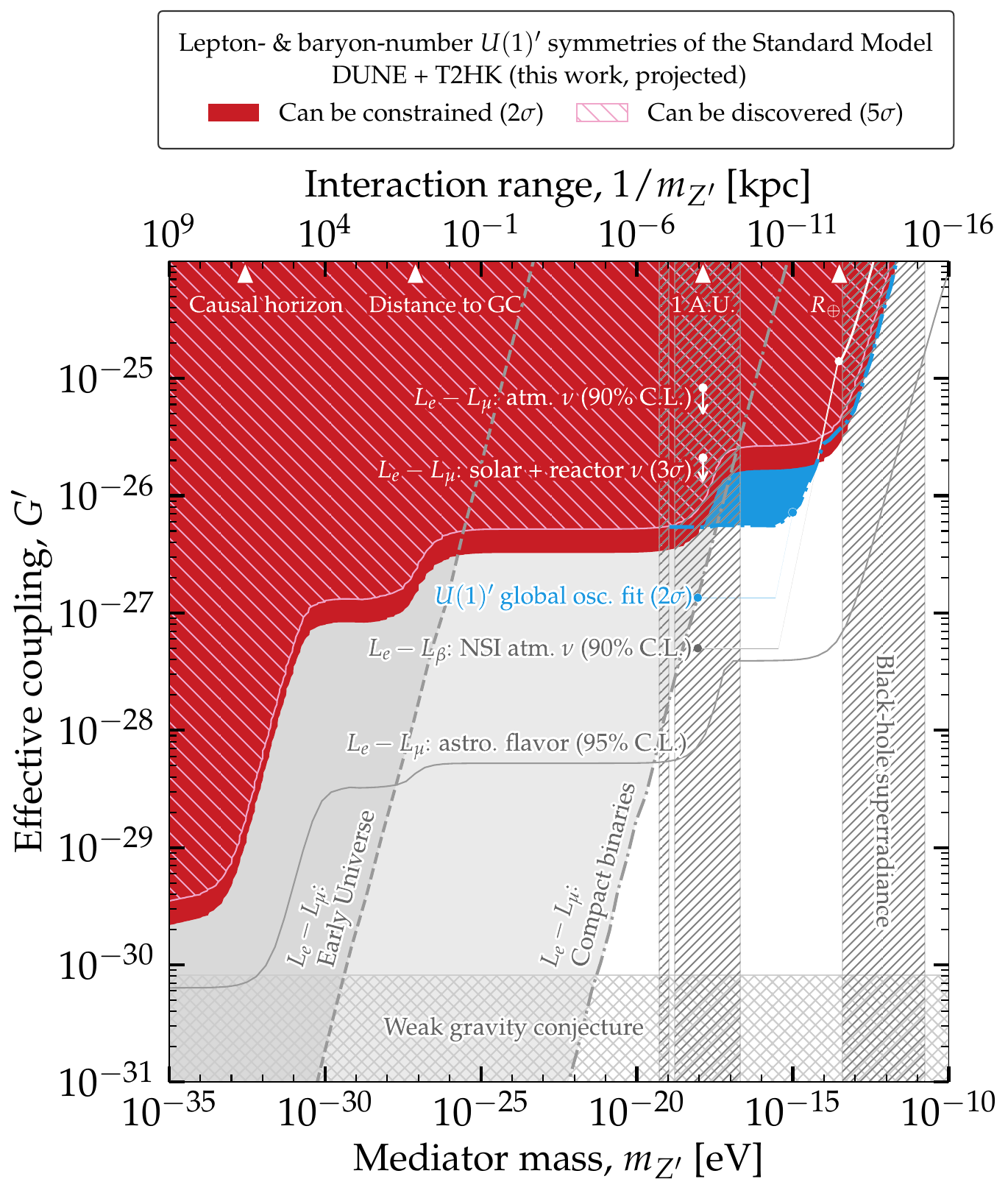}
 \caption{\textbf{\textit{Overview of projected constraints and discovery prospects of long-range neutrino interactions achieved by combining DUNE and T2HK.}}  Results are on the effective coupling of the new gauge boson, $Z^\prime$, that mediates the interaction, across all the candidate $U(1)^\prime$ symmetries that we consider could induce long-range interactions (table~\ref{tab:charges}), and for 10 years of operation of each experiment.  For this figure, we assume that the true neutrino mass ordering is normal. Existing limits are from a recent global oscillation fit~\cite{Coloma:2020gfv}, shown also across all symmetries, and, for specific symmetries, from atmospheric neutrinos~\cite{Joshipura:2003jh}, solar and reactor neutrinos~\cite{Bandyopadhyay:2006uh}, and non-standard interactions~\cite{Super-Kamiokande:2011dam, Ohlsson:2012kf, Gonzalez-Garcia:2013usa}.  The estimated sensitivity from present flavor-composition measurements of high-energy astrophysical neutrinos in IceCube is from~\Refe~\cite{Agarwalla:2023sng}.  Indirect limits~\cite{Wise:2018rnb} are from black-hole superradiance~\cite{Baryakhtar:2017ngi}, the early Universe~\cite{Dror:2020fbh}, compact binaries~\cite{KumarPoddar:2019ceq}, and the weak gravity conjecture~\cite{Arkani-Hamed:2006emk}, assuming a lightest neutrino mass of $0.01$~eV. See sections~\ref{sec:intro}, \ref{sec:constraints_pot}, and  \ref{sec:discovery} for details.  \textit{DUNE and T2HK may constrain long-range interactions more strongly than ever before, or discover them, regardless of which $U(1)^\prime$ symmetry is responsible for inducing them.}}    
 \label{fig:constraints_discovery_dune_t2hk-NMO}
\end{figure}
%=================================================

So far, there is no evidence for such long-range neutrino interactions, but there are stringent constraints on them inferred from observations of atmospheric~\cite{Joshipura:2003jh}, solar~\cite{Grifols:2003gy, Bandyopadhyay:2006uh,Gonzalez-Garcia:2006vic}, accelerator~\cite{Heeck:2010pg, Heeck:2018nzc}, and high-energy astrophysical~\cite{Bustamante:2018mzu, Agarwalla:2023sng} neutrinos, and from their combination~\cite{Davoudiasl:2011sz, Farzan:2016wym, Wise:2018rnb, Dror:2020fbh, Coloma:2020gfv, Alonso-Alvarez:2023tii}. Other constraints do not involve neutrinos, \eg, gravitational fifth-force searches~\cite{Adelberger:2009zz, Salumbides:2013dua}, tests of the equivalence principle~\cite{Schlamminger:2007ht}, black-hole superradiance~\cite{Baryakhtar:2017ngi}, the orbital period of compact binary systems~\cite{KumarPoddar:2019ceq}, and the perihelion precession of planets~\cite{KumarPoddar:2020kdz}; we show some of them in \figu{constraints_discovery_dune_t2hk-NMO}.  Reference~\cite{Singh:2023nek} (also, \Refe~\cite{Wise:2018rnb}) contains a brief review of existing limits, including some shown here in figs.~\ref{fig:constraints_discovery_dune_t2hk-NMO}, \ref{fig:all_symmetry}, and \ref{fig:discovery_all_DUNE+T2HK}.

Long-baseline neutrino oscillation experiments, where the distance between the source and detector is of hundreds of kilometers or more, are particularly well-suited for searching for new neutrino interactions that may affect flavor transitions.  The high precision of their detectors and their well-characterized neutrino beams facilitate identifying subtle deviations from standard expectations~\cite{DUNE:2020fgq,Arguelles:2022tki,NOvA:2024lti}.  In combination with other experiments, they have placed stringent limits on long-range neutrino interactions~\cite{Honda:2007wv,Chatterjee:2015gta,Khatun:2018lzs, Coloma:2020gfv,Mishra:2024riq}.  

In the coming 10--20 years, new long-baseline experiments, larger, using more advanced detection and reconstruction techniques, and more intense neutrino beams, hold an opportunity for important progress.  We prepare to seize it by forecasting the capability of two of the leading long-range neutrino oscillation experiments, the Deep Underground Neutrino Experiment (DUNE)~\cite{DUNE:2021cuw} and Tokai-to-Hyper-Kamiokande (T2HK)~\cite{Hyper-Kamiokande:2016srs, Hyper-Kamiokande:2018ofw}, presently under construction, to constrain, discover, and characterize new flavor-dependent neutrino interactions, and we interpret it in the context of long-range neutrino interactions.

We introduce new neutrino interactions by gauging the accidental global $U(1)^\prime$ symmetries of the Standard Model that involve combinations of lepton numbers, $L_e$, $L_\mu$, and $L_\tau$, and baryon number, $B$; see \Refes~\cite{He:1990pn, Foot:1990uf, Foot:1990mn, He:1991qd, Foot:1994vd} for early works and \Refe~\cite{Langacker:2008yv} for a review.  Gauging one of the several candidate symmetries (more on this later) introduces a new neutral vector gauge boson, $Z^\prime$, whose mass and coupling strength are a priori unknown, and which induces a new Yukawa potential sourced by electrons, neutrons, or protons, depending on the symmetry.  Also depending on the symmetry, the new interaction affects only $\nu_e$, $\nu_\mu$, or $\nu_\tau$, or a combination of them, and modifies flavor transitions differently.  The lighter the mediator, the longer the range of the interaction.  We focus on masses between $10^{-10}$~eV and $10^{-35}$~eV, corresponding to an interaction range between meters and Gpc.  (The complementary case for heavy mediators, studied in the context of contact neutrino interactions, was first studied in \Refes~\cite{Wolfenstein:1977ue,Valle:1987gv,Guzzo:1991hi}; see also \Refe~\cite{Coloma:2020gfv}.)

There are three core ingredients to our analysis; we sketch them below and expand on them later.  First, as in \Refes~\cite{Bustamante:2018mzu, Singh:2023nek, Agarwalla:2023sng}, we use the long-range matter potential sourced by vast repositories of matter in the local and distant Universe: the Earth, Moon, Sun, Milky Way, and the cosmological matter distribution.  Previous calculations of this potential~\cite{Bustamante:2018mzu, Agarwalla:2023sng}, limited to lepton-number symmetries (more on this momentarily) used only the distributions of electrons and neutrons; our new analysis extends that to include also protons.  Second, as in \Refe~\cite{Singh:2023nek}, we base our analysis on detailed simulations of DUNE and T2HK, which grounds our results in realistic detection capabilities.  Third, motivated by \Refe~\cite{Coloma:2020gfv} --- which, unlike us, used present-day oscillation data --- we explore a plethora of candidate $U(1)^\prime$ symmetries that could introduce long-range neutrino interactions, each affecting neutrino oscillations differently (see table~\ref{tab:charges}).  Doing this extends the first forecasts for DUNE and T2HK reported in \Refe~\cite{Singh:2023nek}, which were limited to three candidate symmetries, $L_e-L_\mu$, $L_e-L_\tau$, and $L_\mu-L_\tau$, and allows us to establish whether the sensitivity claimed therein was limited to those three cases, or applies broadly to other symmetries.  \textbf{\textit{While the above ingredients have been accounted for before separately, or in other contexts, the novelty and strength of our analysis lies in combining them.}}

Figure~\ref{fig:constraints_discovery_dune_t2hk-NMO} conveys the essence of our findings; we elaborate on them later.  It summarizes our forecasts on constraining and discovering long-range interactions across the fourteen candidate symmetries that we consider (table~\ref{tab:charges}); the results for individual symmetries are comparable among them and we show them later (figs.~\ref{fig:all_symmetry} and \ref{fig:discovery_all_DUNE+T2HK}).  Figure~\ref{fig:constraints_discovery_dune_t2hk-NMO} shows that DUNE and T2HK may place the strongest constraints on long-range interactions, especially for mediators lighter than $10^{-18}$~eV, and discover them, even if they are subdominant.  This reaffirms the outlook first reported in \Refe~\cite{Singh:2023nek}.  (The sensitivity from flavor measurements of high-energy astrophysical neutrinos~\cite{Agarwalla:2023sng} (see also \Refes~\cite{Bustamante:2018mzu, Ackermann:2022rqc}) could be comparable but, for now, it is subject to large astrophysical uncertainties not captured in \figu{constraints_discovery_dune_t2hk-NMO}.)  

The novel perspective revealed by our results is that \textbf{\textit{DUNE and T2HK may constrain or discover new neutrino interactions with matter --- including long-range ones --- regardless of which symmetry, out of the candidates we consider, induces them.}}  When searching for new interactions, the sensitivity of DUNE and T2HK is not limited to spotting a handful of specific modifications to the flavor transitions, but extends to a broad range of them.  Pivoting on this, we show later that, in some cases, \textbf{\textit{DUNE and T2HK may identify or narrow down which candidate symmetry is responsible for inducing the new interaction; see \figu{confusion-matrix}.}} 

The paper is organized as follows.  Section~\ref{sec:formalism} introduces new neutrino-matter interactions due to various $U(1)^\prime$ symmetries and the long-range matter potential they induce.  Section~\ref{sec:probability_event_rates} illustrates their effect on the neutrino oscillation probability and the event spectra in DUNE and T2HK.  Section~\ref{sec:results} contains our main results: the constraints on the new matter potential, its discovery prospects, their interpretation as being due to long-range interactions, and the separation between different candidate symmetries.  Section~\ref{sec:conclusion} summarizes and concludes.  Appendices \ref{app:U1-charges}--\ref{app:discovery} contain additional details and results.

%=================================================
\section{Long-range neutrino interactions}
\label{sec:formalism}
%=================================================

%=================================================
\subsection{New neutrino-matter interactions from $U(1)^\prime$ symmetries}
\label{sec:formalism_lagrangians}
%=================================================

%=================================================
\begin{figure}[t!]
 \centering
 \includegraphics[width=\textwidth]{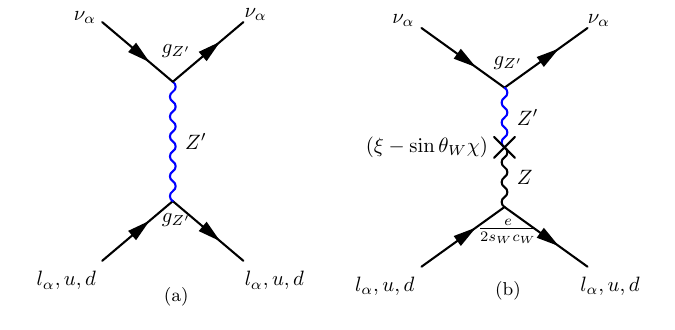}
 \caption{\textbf{\textit{Feynman diagrams of the new neutrino-matter interactions that we consider.}}  The interaction Lagrangian is \equ{full_lagrangian}. Diagram (a) represents the new interaction mediated by a new $Z^\prime$ neutral vector boson, with coupling constant $g_{Z^\prime}$.   Diagram (b) represents the mixing between $Z$ and $Z^\prime$.  In our analysis, we account for the contribution of diagram (a) for all $U(1)^\prime$ symmetries except for $L_\mu-L_\tau$, for which diagram (a) is replaced by diagram (b).  See section~\ref{sec:formalism_lagrangians} for details.}
 \label{fig:feynman_diagram}
\end{figure}
%=================================================

The Standard Model (SM) contains accidental global $U(1)$ symmetries that involve $L_e$, $L_\mu$, $L_\tau$, and $B$. Gauging them individually introduces anomalies. However, certain combinations of them can be gauged anomaly-free, either within the SM particle content or by adding right-handed neutrinos~\cite{Araki:2012ip, Allanach:2018vjg}. Since neutrino oscillations are not affected by flavor-universal gauge symmetries --- say, $B-L$ --- we focus on $U(1)^\prime$ symmetries that are flavor-dependent, namely (table~\ref{tab:charges}), $B-3L_e$, $B-3L_\mu$, $B-3L_\tau$, $B-L_e-2L_\tau$, $B_y+L_\mu+L_\tau$, $B-\frac{3}{2}(L_\mu+L_\tau)$, $L-3L_e$, $L-3L_\mu$, $L-3L_\tau$, $L_e-\frac{1}{2}(L_\mu+L_\tau)$, $L_e+2L_\mu+2L_\tau$, $L_e-L_\mu$, $L_e-L_\tau$, and $L_\mu-L_\tau$.  This is the same list of fourteen candidate symmetries explored in \Refes~\cite{Davoudiasl:2011sz, Araki:2012ip, Coloma:2020gfv, delaVega:2021wpx}.  (Here, $B_y \equiv B_1-yB_2-(3-y)B_3$~\cite{Farzan:2015doa, Coloma:2020gfv}, where $B_1$, $B_2$, and $B_3$ are the baryon numbers of quarks of the first, second, and third generation, respectively, and $y$ is an arbitrary constant that we set to $y=0$ because we consider neutrino interactions with first-generation quarks only.) Their rich phenomenology has been discussed in, \eg,  \Refes~\cite{Ma:1997nq,Lee:2010hf,Davoudiasl:2011sz,Lee:2011uh,Araki:2012ip,Farzan:2015doa,Kownacki:2016pmx,Heeck:2018nzc,Coloma:2020gfv,delaVega:2021wpx,Barman:2021yaz,AtzoriCorona:2022moj,DeRomeri:2024dbv}. We expand on them later. Each one, after being promoted to a local gauge symmetry, generates new flavor-dependent neutrino-matter interactions mediated by a new vector boson, $Z^\prime$.  (In principle, the stringent constraints on new interactions of charged leptons could render the possibility of new neutrino interactions unfeasible, but this limitation can be circumvented by suitable model building; see, \eg, \Refes~\cite{Farzan:2016wym, Joshipura:2019qxz, Almumin:2022rml}.)

Figure~\ref{fig:feynman_diagram} shows the neutrino-matter interactions that we consider between the three active neutrinos, $\nu_e$, $\nu_\mu$, and $\nu_\tau$, and electrons ($e$), and up ($u$) and down ($d$) quarks inside protons and neutrons. Apart from the standard $W^{\pm}$- and $Z$-boson mediated interactions, which we do not show explicitly, for a given $U(1)^\prime$ symmetry, the effective interaction Lagrangian is
\begin{equation}
\label{equ:full_lagrangian}
 \mathcal{L}
 = \mathcal{L}_{Z^\prime}
   +
   \mathcal{L}_{\rm mix} \;.
\end{equation}

The first term on the right-hand side of \equ{full_lagrangian} describes the new neutral-current flavor-dependent neutrino-matter interactions~\cite{He:1990pn, He:1991qd, Heeck:2010pg,Coloma:2022dng}, mediated by $Z^\prime$, whose mass, $m_{Z^\prime}$, and coupling strength, $g_{Z^\prime}$, are a priori unknown, \ie,
\begin{multline}
\label{equ:Lagrangian}
\mathcal{L}_{Z'}
= -g_{Z^\prime}\big( a_u\, \bar{u}\gamma^\alpha u
+ a_d\, \bar{d}\gamma^\alpha d
+ a_e\, \bar{e}\gamma^\alpha e
\\
+ b_e \,\bar{\nu}_e\gamma^\alpha P_L \nu_e
+ b_\mu\,\bar{\nu}_\mu\gamma^\alpha P_L \nu_\mu
+ b_\tau\, \bar{\nu}_\tau \gamma^\alpha P_L \nu_\tau \big)  Z'_\alpha\,,
\end{multline}
where $a_e$, $a_u$, and $a_d$ are the $U(1)^\prime$ charges of the electron, up quark, and down quark, and $b_e$, $b_\mu$, and $b_\tau$ are the charges of $\nu_e$, $\nu_\mu$, and $\nu_\tau$. Appendix~\ref{app:U1-charges} contains the values of the $U(1)^\prime$ charges of the symmetries that we consider.

The above neutrino-matter interactions can be generated upon extending the SM gauge group $SU(3)_{C}\times SU(2)_{L}\times U(1)_{Y}$ by the maximal abelian gauge group $U(1)^\prime=U(1)_{B-L}\times U(1)_{L_{\mu}-L_{\tau}}\times U(1)_{L_\mu-L_e}$ by adding three right-handed neutrinos to the SM particle content and imposing the condition that all new couplings are vector-like and the $U(1)^\prime$ charges of the quarks are flavor-universal. Then, any subset of the $U(1)^\prime$ hypercharge $c_\textsc{bl} (B-L) + c_{\mu\tau} (L_\mu-L_\tau)+ c_{\mu e} (L_\mu-L_e)$ can be gauged in an anomaly-free way~\cite{Araki:2012ip, Heeck:2018nzc, Allanach:2018vjg, Coloma:2020gfv}, with $a_u = a_d = c_\textsc{bl} / 3$, $a_e = b_e = -(c_\textsc{bl} +
c_{\mu e})$, $b_\mu = -c_\textsc{bl} + c_{\mu e} + c_{\mu\tau}$, and
$b_\tau = -(c_\textsc{bl} + c_{\mu\tau})$.

Table~\ref{tab:charges} lists all the $U(1)^\prime$ symmetries that we consider.  We group them according to the texture of the new matter potential that they introduce, $\mathbf{V}_{\rm LRI}$ (section~\ref{sec:yukawa_interaction}); later, we interpret this potential as being due to long-range interactions (LRI).  Different textures affect neutrino oscillations differently; we elaborate on this in section~\ref{sec:prob_var_pot}.  Reference~\cite{Singh:2023nek} explored long-range interactions due to $L_e-L_\mu$, $L_e-L_\tau$, and $L_\mu-L_\tau$ under a similar analysis that we perform here, but when computing their effect on neutrino oscillations (section~\ref{sec:prob_var_pot}) used values of the neutrino mixing parameters from \Refe~\cite{Capozzi:2021fjo}.  We revisit them here using mixing parameters from the NuFIT~5.1 global fit to oscillation data~\cite{Esteban:2020cvm, NuFIT} instead.

%=================================================
\begin{table}[t!]
  \centering
  \footnotesize
  \begin{adjustbox}{width=\linewidth}
  \renewcommand{\arraystretch}{1.4}
  \begin{tabular}{|c|c|c|c|c|}
    \hline
    \mr{3}{*}{\shortstack{Texture 
    \\[0.4em] of $\mathbf{V}_{\rm LRI}$}} &
    \mr{3}{*}{$U(1)^\prime$ symmetry} &  
    \mc{3}{c|}{New matter potential, $\mathbf{V}_{\rm LRI} = \textrm{diag}(V_{{\rm LRI},e}, V_{\rm LRI,\mu}, V_{\rm LRI,\tau})$}
    \\ 
    &
    & 
    \mr{2}{*}{\shortstack{Texture to place limits,
    \\[0.4em] $\mathbf{V}_{\rm LRI} = V_{\rm LRI} \cdot \textrm{diag}(\ldots)$}}
    &
    \mr{2}{*}{\shortstack{General form
    \\[0.4em] of $V_{\rm LRI}$, \equ{pot_total_general}}} 
    &
    \mr{2}{*}{\shortstack{Form of $V_{\rm LRI}$ to convert limits on it 
    \\[0.4em] into limits on $G^\prime$ {\it vs.}~$m_{Z^{\prime}}$, \equ{pot_total_simplified}}}
    \\
    & 
    & 
    &
    &
    \\
    \hline
    %%%%%%
    \mr{6}{*}{$\left(\begin{array}{ccc} \bullet &  &  \\  & 0 & \\ &  & 0 \end{array}\right)$} &
    %%%%%%
    $B-3L_{e}$ &
    $\textrm{diag}(1, 0, 0)$ &
    $9V_{e}-3(V_{p}+V_{n})$ &
    $3(V^{\oplus}_{e}+V^{\leftmoon}_{e}+V^{\rm MW}_{e})+\frac{21}{4}V^{\astrosun}_{e}+\frac{39}{7}V^{\rm cos}_{e}$ 
    \\
    \cline{2-5} 
    %%%%%%
    &
    $L-3L_e$ &
    $\textrm{diag}(1, 0, 0)$ &
    $6 V_e$ &
    $6(V^{\oplus}_{e}+V^{\leftmoon}_{e}+V^{\rm MW}_{e}+V^{\astrosun}_{e}+V^{\rm cos}_{e})$ 
    \\
    \cline{2-5}
    &
    $B-\frac{3}{2}(L_\mu+L_\tau)$ &
    $\textrm{diag}(1, 0, 0)$ &
    $\frac{3}{2}(V_p+V_n)$ &
    $3(V^{\oplus}_{e}+V^{\leftmoon}_{e}+V^{\rm MW}_{e})+\frac{15}{8}V^{\astrosun}_{e}+\frac{12}{7}V^{\rm cos}_{e}$
    \\
    \cline{2-5}
    &
    $L_e-\frac{1}{2}(L_\mu+L_\tau)$ &
    $\textrm{diag}(1, 0, 0)$ &
    $\frac{3}{2}V_e$ &
    $\frac{3}{2}(V^{\oplus}_{e}+V^{\leftmoon}_{e}+V^{\rm MW}_{e}+V^{\astrosun}_{e}+V^{\rm cos}_{e})$ 
    \\
    \cdashline{2-5}
    &
    $L_e+2L_\mu+2L_\tau$ &
    $\textrm{diag}(-1, 0, 0)$ &
    $V_e$ &
    $V^{\oplus}_{e}+V^{\leftmoon}_{e}+V^{\rm MW}_{e}+V^{\astrosun}_{e}+V^{\rm cos}_{e}$ 
    \\
    \cline{2-5}
    &
    $B_{y}+L_\mu+L_\tau$ &
    $\textrm{diag}(-1, 0, 0)$ &
    $V_p+V_n$ &
    $2(V^{\oplus}_{e}+V^{\leftmoon}_{e}+V^{\rm MW}_{e})+\frac{5}{4}V^{\astrosun}_{e}+\frac{8}{7}V^{\rm cos}_{e}$
    \\
    \hline
    %%%%%%
    \mr{4}{*}{$\left(\begin{array}{ccc} 0 &  &  \\  & \bullet & \\ &  & 0 \end{array}\right)$} &
    %%%%%%
    \mr{2}{*}{$B-3L_\mu$} &
    \mr{2}{*}{$\textrm{diag}(0, -1, 0)$} &
    \mr{2}{*}{$3(V_p+V_n)$} &
    \mr{2}{*}{$6(V^{\oplus}_{e}+V^{\leftmoon}_{e}+V^{\rm MW}_{e})+\frac{15}{4}V^{\astrosun}_{e}+\frac{24}{7}V^{\rm cos}_{e}$} 
    \\
    & & & &
    \\
    \cline{2-5}
    %%%%%%
    &
    \mr{2}{*}{$L-3L_\mu$} &
    \mr{2}{*}{$\textrm{diag}(0, -1, 0)$} &
    \mr{2}{*}{$3V_e$} &
    \mr{2}{*}{$3 (V^{\oplus}_{e}+V^{\leftmoon}_{e}+V^{\rm MW}_{e}+V^{\astrosun}_{e}+V^{\rm cos}_{e})$}
    \\ 
    & & & &
    \\
    %%%%%%
    \hline
    \mr{4}{*}{$\left(\begin{array}{ccc} 0 &  &  \\  & 0 & \\ &  & \bullet  \end{array}\right)$} &
    %%%%%%
    \mr{2}{*}{$B-3L_\tau$} &
    \mr{2}{*}{$\textrm{diag}(0, 0, -1)$} &
    \mr{2}{*}{$3 (V_p + V_n)$} &
    \mr{2}{*}{$6(V^{\oplus}_{e}+V^{\leftmoon}_{e}+V^{\rm MW}_{e})+\frac{15}{4}V^{\astrosun}_{e}+\frac{24}{7}V^{\rm cos}_{e}$}
    \\ 
    & & & &
    \\
    \cline{2-5}
    &
    \mr{2}{*}{$L-3L_\tau$} &
    \mr{2}{*}{$\textrm{diag}(0, 0, -1)$} &
    \mr{2}{*}{$3 V_e $} &
    \mr{2}{*}{$3 (V^{\oplus}_{e}+V^{\leftmoon}_{e}+V^{\rm MW}_{e}+V^{\astrosun}_{e}+V^{\rm cos}_{e})$}
    \\ 
    & & & &
    \\
    %%%%%%
    \hline
    \mr{3}{*}{$\left(\begin{array}{ccc} \bullet &  &  \\  & \bullet & \\ &  & 0 \end{array}\right)$} &
    %%%%%%
    \mr{3}{*}{$L_e-L_\mu$} &
    \mr{3}{*}{$\textrm{diag}(1, -1, 0)$} &
    \mr{3}{*}{$V_e$} &
    \mr{3}{*}{$V^{\oplus}_{e}+V^{\leftmoon}_{e}+V^{\rm MW}_{e}+V^{\astrosun}_{e}+V^{\rm cos}_{e}$}
    \\
    & & & &
    \\
    & & & &
    \\
    %%%%%%
    \hline
    \mr{3}{*}{$\left(\begin{array}{ccc} \bullet &  &  \\  & 0 & \\ &  & \bullet \end{array}\right)$} &
    %%%%%%
    \mr{3}{*}{$L_e-L_\tau$} &
    \mr{3}{*}{$\textrm{diag}(1, 0, -1)$} &
    \mr{3}{*}{$V_e$} &
    \mr{3}{*}{$V^{\oplus}_{e}+V^{\leftmoon}_{e}+V^{\rm MW}_{e}+V^{\astrosun}_{e}+V^{\rm cos}_{e}$}
    \\
    & & & &
    \\
    & & & &
    \\
    %%%%%%
    \hline
    \mr{4}{*}{$\left(\begin{array}{ccc} 0 & &  \\  & \bullet & \\ &  & \bullet \end{array}\right)$} &
    %%%%%%
    \mr{2}{*}{$L_\mu-L_\tau$} &
    \mr{2}{*}{$\textrm{diag}(0, 1, -1)$} &
    \mr{2}{*}{$-V_e+V_p+V_n$} &
    \mr{2}{*}{$V^{\oplus}_{e}+V^{\leftmoon}_{e}+V^{\rm MW}_{e}+\frac{1}{4}V^{\astrosun}_{e}+\frac{1}{7}V^{\rm cos}_{e}$}
    \\
    & & & &
    \\
    \cline{2-5}
    &
    %%%%%%
    \mr{2}{*}{$B-L_e-2L_\tau$} &
    \mr{2}{*}{$\textrm{diag}(0, 1, -1)$} &
    \mr{2}{*}{$-V_e+V_p+V_n$} &
    \mr{2}{*}{$V^{\oplus}_{e}+V^{\leftmoon}_{e}+V^{\rm MW}_{e}+\frac{1}{4}V^{\astrosun}_{e}+\frac{1}{7}V^{\rm cos}_{e}$}
    \\
    & & & &
    \\
    \hline
   \end{tabular}
  \end{adjustbox}
  \caption{
  \textbf{\textit{$U(1)^\prime$ gauge symmetries considered in our analysis and the new matter potential they induce.}}  We group symmetries according to the texture of the new matter potential, $\mathbf{V}_{\rm LRI}$, that they induce; equal or similar textures lead to equal or similar sensitivity (sections~\ref{sec:constraints_pot}--\ref{sec:confusion-theory}).  Elements of  $\mathbf{V}_{\rm LRI}$ marked with $\bullet$ represent nonzero entries.  When computing constraints and discovery prospects on the new matter potential, and the distinguishability between competing candidate symmetries, we use for it the form $\mathbf{V}_{\rm LRI} = V_{\rm LRI} \cdot \textrm{diag}(\dots)$, where the texture of the diagonal matrix is indicated in the table for each symmetry, after subtracting terms proportional to the identity.  The general expression for $V_{\rm LRI}$, sourced by electrons, protons, and neutrons, regardless of their source, is from \equ{pot_total_general}.  We use knowledge of the relative abundance of electrons, protons, and neutrons in the Earth ($\oplus$), Moon ($\leftmoon$), Sun ($\odot$), Milky Way (MW), and in the cosmological matter distribution (cos) to convert limits obtained on $V_{\rm LRI}$ into limits on the mass and coupling strength of the $Z^\prime$ mediator, $m_{Z^\prime}$ and $G^\prime$ (sections~\ref{sec:constraints_pot} and \ref{sec:discovery}).  See sections~\ref{sec:intro} and \ref{sec:hamiltonians} for details.
  \label{tab:charges}
  }
\end{table}
%=================================================

The second term on the right-hand side of \equ{full_lagrangian} describes the mixing between the neutral gauge bosons, $Z$ and $Z^\prime$~\cite{Babu:1997st,Heeck:2010pg,Joshipura:2019qxz}, \ie, $\mathcal{L}_{ZZ^\prime} \supset(\xi-\sin\theta_W\chi) Z'_{\mu}Z^{\mu}$, where $\chi$ is the kinetic mixing angle between the two gauge bosons and $\xi$ is the rotation angle between physical states and gauge eigenstates. This induces a four-fermion interaction of neutrinos with matter given by
\begin{equation}
 \label{equ:Vmutau}
 \mathcal{L}_{\rm mix}
 =
 -g_{Z^\prime}\frac{e}{\sin\theta_W \cos\theta_W}(\xi-\sin\theta_W\chi)
 J'_\sigma J_3^\sigma \;,
\end{equation}
where $J^\prime_\sigma = \bar{\nu}_\mu \gamma_\sigma P_L\nu_\mu-\bar{\nu}_\tau \gamma_\rho P_L\nu_\tau$ and $J_3^\rho = -\frac{1}{2}\bar{e}\gamma^\sigma P_L e+\frac{1}{2}\bar{u}\gamma^\rho P_L u-\frac{1}{2}\bar{d}\gamma^\rho P_L d$, $e$ is the unit electric charge, $\theta_W$ is the Weinberg angle, and $P_L$ is the left-handed projection operator. In this case, the contributions from electrons and protons cancel each other out, leaving only neutrons to source the new matter potential.
Because the value of the mixing factor $(\xi-\sin\theta_W\chi)$ is not known (though there are upper limits on it~\cite{Schlamminger:2007ht, Adelberger:2009zz, Heeck:2010pg}), we place bounds instead on the effective coupling $g_{Z^\prime} (\xi-\sin\theta_W\chi)$. In our analysis, we consider the contribution of the mixing potential only for the symmetry $L_\mu-L_\tau$, since in this case there are no muons and taus to source the matter potential via $\mathcal{L}_{Z^\prime}$ that would otherwise be dominant due to the abundance of baryons and electrons. 

%=================================================
\subsection{Long-range interaction potential}
\label{sec:yukawa_interaction}
%=================================================

When the new neutrino interactions stem from purely leptonic symmetries, the matter potential is sourced only by electrons, given the dearth of naturally occurring muons and taus. (In the case of $L_\mu-L_\tau$, neutrons contribute through $Z$--$Z^\prime$ mixing; see above.) When they stem from symmetries that blend baryon and lepton numbers, the potential is sourced by electrons, neutrons, and protons, depending on the specific symmetry.

For a given $U(1)^\prime$ symmetry out of our candidates (table~\ref{tab:charges}), the Yukawa potential, mediated by $Z^\prime$, that is experienced by a neutrino situated a distance $d$ away from an electron ($f = e$), a proton ($f = p$), or a neutron ($f = n$) is
\begin{equation}
 V_{Z^\prime, f}
 =
 G^{\prime 2}
 \frac{1}{4\pi d}
 e^{-m_{Z^\prime}\,d} \;,
 \label{equ:potential_zprime}
\end{equation}
where the interaction range is $1/m_{Z^\prime}$; beyond this distance, the potential is suppressed.  Under the $L_\mu-L_\tau$ symmetry, we ignore the contribution of \equ{potential_zprime}; instead, we consider that a neutrino experiences only a potential due to the mixing between $Z$ and $Z^\prime$ (section~\ref{sec:formalism_lagrangians}), sourced by a neutron, \ie,
\begin{equation}
 V_{ZZ^\prime, n}
 =
 G^{\prime 2}
 \frac{e}{\sin\theta_W\cos\theta_W}
 \frac{1}{4 \pi d}
 e^{-m_{Z^\prime}d} \;.
\label{equ:potential_zzprime}
\end{equation}
In eqs.~(\ref{equ:potential_zprime}) and (\ref{equ:potential_zzprime}), the effective coupling strength is
\begin{equation}
 G^\prime
 =
 \left\{
 \begin{array}{lll}
 g_{Z^\prime} & , & ~{\rm for}~\nu~{\rm interaction~via}~ Z^\prime \\
 \sqrt{g_{Z^\prime} (\xi-\sin \theta_W \chi)} & , & ~{\rm for}~\nu~{\rm interaction~via}~ Z-Z^\prime ~{\rm mixing}\\ 
 \end{array}
 \right.\;.
\label{equ:G_prime}
\end{equation}
Combining eqs.~(\ref{equ:potential_zprime})--(\ref{equ:G_prime}) yields the potential
\begin{equation}
 V_f
 =
 \left\{
 \begin{array}{ll}
  V_{Z^\prime, f} & ,~
  \textrm{for~all~symmetries~but~}L_\mu-L_\tau
  \\
  V_{ZZ^\prime, n} & ,~
  \textrm{for~}L_\mu-L_\tau~\textrm{~and~}f = n \\
  0 & ,~
  \textrm{otherwise}  
 \end{array}
 \right. \;.
\end{equation}

Following \Refe~\cite{Bustamante:2018mzu}, we focus on long-range interactions, with ultra-light mediators with masses $m_{Z^\prime} = 10^{-35}$--$10^{-10}$~eV that result in interaction ranges from a few hundred meters to Gpc; see also \Refe~\cite{Wise:2018rnb}. Based on the methods introduced in \Refe~\cite{Bustamante:2018mzu} (see also \Refe~\cite{Wise:2018rnb}) and developed in \Refes~\cite{Agarwalla:2023sng, Singh:2023nek}, we estimate the total potential sourced by the electrons, protons, and neutrons in nearby and distant celestial objects --- the Earth ($\oplus$), Moon ($\leftmoon$), Sun ($\astrosun$), and the Milky Way (MW) --- and by the cosmological distribution of matter (cos) in the local Universe., \ie,
\begin{equation}
 \label{equ:pot_total}
 V_f
 (m_{Z^\prime}, G^\prime)
 =
 \left.
 \left(
 V_f^\oplus
 +
 V_f^{\leftmoon}
 +
 V_f^{\astrosun}
 +
 V_f^{\rm MW}
 + 
 V_f^{\rm cos}
 \right)
 \right\rvert_{m_{Z^\prime}, G^\prime}
 \;.
\end{equation}
Hence, the potential experienced by $\nu_\alpha$ ($\alpha = e, \mu, \tau$) is
\begin{equation}
 \label{equ:pot_total_general}
 V_{{\rm LRI}, \alpha}
 (m_{Z^\prime}, G^\prime)
 =
 b_\alpha
 \sum_{f = e, p, n}
 \kappa_f
 V_f
 (m_{Z^\prime}, G^\prime)
 \;,
\end{equation}
where $b_\alpha$ is the $U(1)^\prime$ charge of the neutrino (table~\ref{tab:charges2}).  For all symmetries but $L_\mu - L_\tau$, $\kappa_f \equiv a_f$ is the $U(1)^\prime$ charge of an electron, $a_e$, a proton, $a_p = 2a_u + a_d$, or a neutron, $a_n = 2a_d + a_u$ (table~\ref{tab:charges2}).  For $L_\mu - L_\tau$, $\kappa_f = y_f$ is instead their weak hypercharge, and only neutrons contribute, with $y_n = 2y_d + y_u$.
The value of $m_{Z^\prime}$ determines the relative sizes of the contributions of the different celestial objects to the total potential.  We defer to \Refes~\cite{Bustamante:2018mzu, Singh:2023nek} for details on the calculation of these  contributions; below, we sketch it.

We make the assumption that the matter responsible for generating this potential is electrically neutral, \ie, it contains equal abundance of electrons and protons ($N_e = N_p$), and isoscalar, \ie, it contains equal abundance of protons and neutrons ($N_p = N_n$), except for the Sun~\cite{Heeck:2010pg} and the cosmological matter distribution~\cite{Hogg:1999ad, Steigman:2007xt, Planck:2015fie}, as follows.  We treat the Sun ($N_{e,\astrosun} = N_{p,\astrosun} \sim 10^{57}$, $N_{n,\astrosun} = N_{e,\astrosun}/4$) and the Moon ($N_{e,\leftmoon} = N_{p,\leftmoon} = N_{n,\leftmoon} \sim 5 \cdot 10^{49}$) as point sources of electrons, protons, and neutrons, and the Earth ($N_{e, \oplus} = N_{p, \oplus} = N_{n, \oplus} \sim 4 \cdot 10^{51}$), the Milky Way ($N_{e, {\rm MW}} = N_{p, {\rm MW}} \approx N_{n, {\rm MW}} \sim 10^{67}$), and the cosmological matter ($N_{e,\mathrm{cos}} = N_{p,\mathrm{cos}} \sim 10^{79}$, $N_{n,\mathrm{cos}} \sim 10^{78}$) as continuous distributions. 

For the contribution of matter inside the Earth, we adopt the approximation of computing the average potential that acts on the neutrinos at their point of detection, as in \Refes~\cite{Bustamante:2018mzu, Agarwalla:2023sng, Singh:2023nek}.  We do not calculate the changing potential as the neutrinos traverse inside the Earth; see \Refes~\cite{ Smirnov:2019cae, Coloma:2020gfv} for such detailed treatment. Our approximation holds well for mediator mass below $10^{-14}$~eV, for which the interaction range is longer than the radius of the Earth (\figu{constraints_discovery_dune_t2hk-NMO}), so that all of the electrons, protons, and neutrons inside it contribute to the potential regardless of their position relative to the neutrino trajectory.

The above approximations allow us to simplify the calculation of the total potential, \equ{pot_total_general}.  First, for each celestial body, we compute the potential sourced by the electrons in it.  Then, for a choice of  symmetry, we compute the potential sourced by protons and neutrons by rescaling the electron potential by their abundance relative to electrons, \ie,
\begin{eqnarray}
 \label{equ:pot_total_simplified}
 V_{{\rm LRI}, \alpha}
 &=&
 b_\alpha
 \left[
 \left(
 \kappa_e
 +
 \kappa_p
 \frac{N_{p, \oplus}}{N_{e, \oplus}}
 +
 \kappa_n
 \frac{N_{n, \oplus}}{N_{e, \oplus}}
 \right)
 V_e^\oplus
 +
 (\oplus \to \leftmoon)
 +
 (\oplus \to \astrosun)
 +
 (\oplus \to {\rm MW})
 \right. 
 \nonumber
 \\
 &&
 \left.
 \qquad
 +~
 (\oplus \to {\rm cos})
 \right] \;.
\end{eqnarray}
By following this procedure, we need only compute explicitly the potential due to electrons --- which may be computationally taxing~\cite{Bustamante:2018mzu, Agarwalla:2023sng, Singh:2023nek} --- rather than the potential due to electrons, protons, and neutrons separately.  Table~\ref{tab:charges} shows \equ{pot_total_simplified} evaluated for each of our candidate symmetries.  Later, in section~\ref{sec:constraints_pot}, we use these expressions to convert the limits we place on the new matter potential into limits on $G^\prime$ as a function of $m_{Z^\prime}$.

%=================================================
\section{Neutrino oscillation probabilities and event rates}
\label{sec:probability_event_rates}
%=================================================

%=================================================
\subsection{Neutrino interaction Hamiltonian}
\label{sec:hamiltonians}
%=================================================

The Hamiltonian that describes neutrinos traveling through matter is, in the flavor basis,
\begin{equation}
 \label{equ:hamiltonian_tot}
 \mathbf{H}
 =
 \mathbf{H}_{\rm vac}
 +
 \mathbf{V}_{\rm mat}
 +
 \mathbf{V}_{\rm LRI} \;.
\end{equation}
The first term on the right-hand side is responsible for the oscillation of neutrinos in vacuum.  For neutrinos with energy $E$, it is
\begin{equation}
 \label{equ:hamiltonian_vac}
 \mathbf{H}_{\rm vac}
 =
 \frac{1}{2 E}
 \mathbf{U}~
 {\rm diag}(0, \Delta m^2_{21}, \Delta m^2_{31})
 ~\mathbf{U}^{\dagger} \;,
\end{equation}
where $\mathbf{U}$ is the Pontecorvo-Maki-Nakagawa-Sakata (PMNS) matrix, parametrized in terms of three mixing angles, $\theta_{23}$, $\theta_{13}$, and $\theta_{12}$, and one CP-violation phase, $\delta_{\rm CP}$, $\Delta m^2_{31} \equiv m^2_{3}-m^2_{1}$, and $\Delta m^2_{21} \equiv m^2_{2}-m^2_{1}$, with $m_i$ ($i = 1, 2, 3$) the mass of the neutrino mass eigenstate $\nu_i$.  

Table~\ref{tab:params_value1} shows the values of the oscillation parameters that we use in our analysis.  Later (section~\ref{sec:expt-details}), when producing mock event samples for DUNE and T2HK, we adopt as true values of the oscillation parameters their best-fit values of the recent NuFIT~5.1~\cite{Esteban:2020cvm, NuFIT} global fit to oscillation data.  When forecasting limits or discovery potential of long-range interactions, we allow their values to float as part of our statistical methods (section~\ref{sec:stat_methods}).  

%=================================================
\begin{table}[t!]
 \centering
 \begin{center} 
  \begin{tabular}{|c|c|c|c|}
   \hline
   Parameter &  Best-fit value & 3$\sigma$ range &  Statistical treatment \\
   \hline
   ${\theta_{12}}~[^{\circ}]$ & 33.45 & 31.27--35.87  & Fixed to best fit \\
   \hline
   \mr{2}{*}{${\theta_{13}}~[^{\circ}]$} & 8.62 & 8.25--8.98 & \mr{2}{*}{Fixed to best fit} \\ 
   & (8.61) & (8.24--9.02) & \\ 
   \hline
   \mr{2}{*}{${\theta_{23}}~[^{\circ}]$} & 42.1 & 39.7--50.9 & \mr{2}{*}{Minimized over $3\sigma$ range} \\
   & (49.0) & (39.8--51.6) & \\
   \hline
   \mr{2}{*}{$\delta_{\rm CP}~[^{\circ}]$} & 230 & 144--350 & \mr{2}{*}{Minimized over $3\sigma$ range} \\
   & (278) & (194--345) & \\
   \hline
   \mr{2}{*}{$\frac{\Delta{m^2_{21}}}{10^{-5} \, \rm{eV}^2}$} & \mr{2}{*}{7.42} & \mr{2}{*}{6.82--8.04}  & \mr{2}{*}{Fixed to best fit} \\
   & & & \\
   \hline
   \mr{2}{*}{$\frac{\Delta{m^2_{31}}}{10^{-3} \, \rm{eV}^2}$} & 2.51  & 2.430--2.593 & \mr{2}{*}{Minimized over $3\sigma$ range} \\
   & (-2.41)  & (-2.506--(-2.329)) & \\
   \hline
  \end{tabular}
  \caption{\textbf{\textit{Best-fit values and allowed ranges of the oscillation parameters used in our analysis.}}  The values are from the NuFIT~5.1 global fit to oscillation data~\cite{Esteban:2020cvm, NuFIT}.  Values outside parentheses are for normal neutrino mass ordering; values inside, for inverted mass ordering.  To produce the illustrative figs.~\ref{fig:dune_prob_events} and \ref{fig:t2hk_prob_events}, we fix all parameters to their best-fit values.}
  \label{tab:params_value1}
 \end{center}
\end{table}
%=================================================

The second term on the right-hand side of \equ{hamiltonian_tot} is the potential from the standard CC coherent forward $\nu_e$-$e$ scattering, \ie,
\begin{equation}
 \label{equ:v_mat}
 \mathbf{V}_{\rm mat}
 =
 {\rm diag}(V_{\rm CC}, 0, 0) \;,
\end{equation}
where $V_{\rm CC} = \sqrt{2} G_F n_e$, $G_F$ is the Fermi constant, and $n_e$ is the number density of electrons along the trajectory of the neutrinos.  This term contributes only during neutrino propagation inside Earth, where electron densities are high.  Since we do not compute the changing potential as the neutrino propagates (section~\ref{sec:yukawa_interaction}), and since the neutrino beams in DUNE and T2HK travel exclusively inside the crust of the Earth, where the matter density is fairly uniform, we use the average matter density along the neutrino trajectory from production to detection, $\rho_\textrm{avg}$, to approximate the potential, \ie,  $V_{\rm CC} \approx 7.6 \cdot Y_{e} \cdot 10^{-14} \left(\frac{\rho_{\text{avg}}}{\mathrm{g}~\mathrm{cm}^{-3}}\right) \; \mathrm{eV}$,where $Y_e \equiv n_e / (n_p + n_n)$ is the density of electrons relative to that of protons, $n_p$, and neutrons,  $n_n$.  We estimate $\rho_\textrm{avg}$ using the Preliminary Reference Earth Model~\cite{Dziewonski:1981xy}, which yields $2.848$~g~cm$^{-3}$ and 2.8~g~cm$^{-3}$ for DUNE and T2HK, respectively.  The potential above is for neutrinos; for antineutrinos, it flips sign, \ie, $\mathbf{V}_{\rm mat} \to - \mathbf{V}_{\rm mat}$.

The third term on the right-hand side of \equ{hamiltonian_tot} is the contribution from the new neutrino-matter interactions, \ie,
\begin{equation}
\label{equ:lri_pot}
 \mathbf{V}_{\rm LRI}
 =
 {\rm diag}(V_{{\rm LRI}, e}, V_{{\rm LRI},\mu}, V_{{\rm LRI},\tau}) \;,
\end{equation}
where, for a specific choice of $U(1)^\prime$ symmetry, and for given values of $G^\prime$ and $m_{Z^\prime}$, $V_{{\rm LRI}, \alpha}$ is computed using \equ{pot_total_simplified}.  When computing limits and discovery prospects of the new matter potential, we use for $\mathbf{V}_{\rm LRI}$ instead the textures in table~\ref{tab:charges}; see sections~\ref{sec:constraints_pot} and \ref{sec:discovery}.  The potential above is for neutrinos; for antineutrinos, it flips sign, \ie, $\mathbf{V}_{\rm LRI} \to - \mathbf{V}_{\rm LRI}$.

The relative sizes of the standard and new contributions to the total Hamiltonian, \equ{hamiltonian_tot}, determine the range of values of the new matter potential to which DUNE and T2HK are sensitive.  On the one hand, in the absence of new interactions, \ie, when $\mathbf{V}_{\rm LRI} = 0$, oscillations are driven by standard vacuum and matter effects.  Only the coherent forward scattering of $\nu_e$ on electrons inside the Earth modifies the oscillation parameters.  On the other hand, if the new matter potential is the dominant contribution, \ie, when $\mathbf{V}_{\rm LRI} \gg \mathbf{H}_{\rm vac} + \mathbf{V}_{\rm mat}$, oscillations are suppressed because $\mathbf{V}_{\rm LRI}$ is diagonal.  

In-between, when the new interactions contribute comparably to the standard contributions, \ie, when $\mathbf{V}_{\rm LRI} \approx \mathbf{H}_{\rm vac} + \mathbf{V}_{\rm mat}$, the new matter potential introduces a resonance that enhances the values of the oscillation parameters and affects the oscillation probabilities significantly.  For DUNE, the standard contribution to the Hamiltonian is roughly $10^{-13}$--$10^{-12}$~eV, depending on the specific neutrino energy; for T2HK, which has lower energies, it is slightly higher, roughly $10^{-12}$--$10^{-11}$~eV.  For the new matter potential to induce resonant flavor conversions --- and thus to boost the detectability of the new interactions --- it must be within this range.  Below, we show that this is indeed the case, by computing oscillation probabilities including the new interactions.

%=================================================
\begin{figure}[t!]
 \centering
 \includegraphics[width=\textwidth]{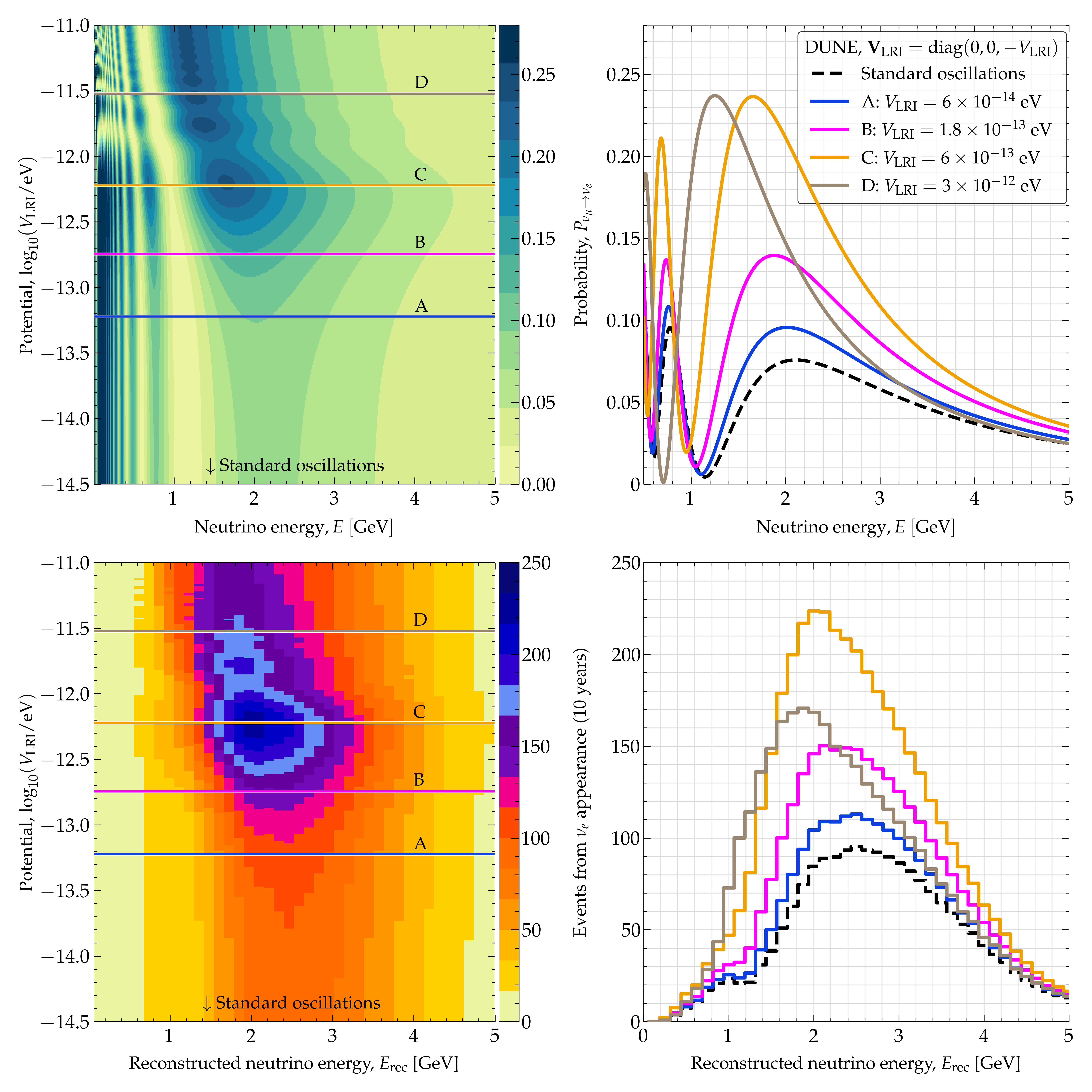}
 \caption{\textbf{\textit{Oscillation probabilities (top) and event distributions (bottom) in the presence of a new matter potential in DUNE.}}  In this figure, we show examples computed assuming a matter potential matrix of the form $\mathbf{V}_{\rm LRI} = \textrm{diag}(0,0,-V_{\rm LRI})$, with varying value of $V_{\rm LRI}$, as would be induced by the $B - 3L_\tau$ and $L - 3L_\tau$ symmetries (table~\ref{tab:charges}). \textit{Top left:} $\numu \to \nue$ probability as a function of the neutrino energy, $E$, and new matter potential, $V_{\rm LRI}$. \textit{Top right:} Oscillation probability computed for choices A--D of the potential, showing the change in amplitude and phase compared to standard oscillations.  \textit{Bottom left:}  Total number, \ie, signal plus background, of $\numu\to\nue$ appearance events after 10 years of run-time (5~yr in $\nu$ and $\bar{\nu}$ modes each), as a function of reconstructed neutrino energy, $E_{\rm rec}$, and $V_{\rm LRI}$.  \textit{Bottom right:}  Event spectra computed for choices A--D of the potential.  See section~\ref{sec:formalism} for details and \figu{t2hk_prob_events} for analogous results for T2HK.  \textit{In DUNE, resonant effects may appear if the new matter potential $V_{\rm LRI} \approx 10^{-13}$--$10^{-12}$~eV.}
 \label{fig:dune_prob_events}} 
\end{figure}
%=================================================

%=================================================
\begin{figure}[t!]
 \centering
 \includegraphics[width=\textwidth]{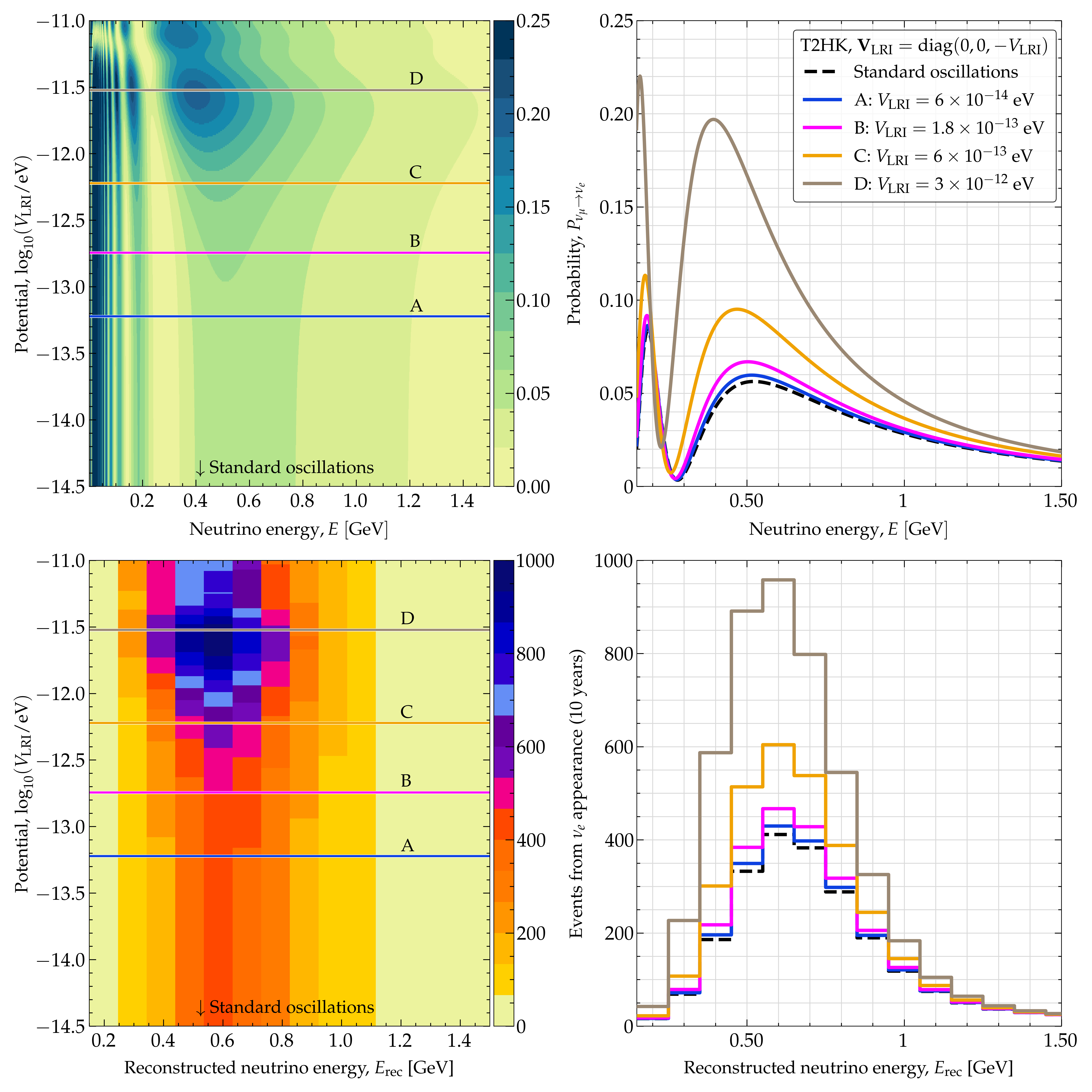}
 \caption{\textbf{\textit{Oscillation probabilities (top) and event distributions (bottom) in the presence of a new matter potential in T2HK.}}  Same as \figu{dune_prob_events}, but for T2HK.  The four illustrative choices of $V_{\rm LRI}$, A--D, are the same as in \figu{dune_prob_events}.  For the event rates, we use 10~years of run-time (2.5~yr in $\nu$ mode and 7.5~yr in $\bar{\nu}$ mode).  See section~\ref{sec:formalism} for details.  \textit{In T2HK, resonant effects may appear if the new matter potential $V_{\rm LRI} \approx 10^{-12}$--$10^{-11}$~eV.}
 \label{fig:t2hk_prob_events}} 
\end{figure}
%=================================================

%===============================================
\subsection{Neutrino oscillation probabilities}
\label{sec:prob_var_pot}
%===============================================

The $\nu_\alpha \to \nu_\beta$ transition probability under neutrino-matter interactions, including standard and new contributions, and governed by the Hamiltonian in \equ{hamiltonian_tot} is~\cite{Kimura:2002hb,Akhmedov:2004ny}
\begin{equation}
 \label{equ:osc_prob_gen}
 P_{\nu_{\alpha}\to \nu_{\beta}}
 =
 \left\vert
 \sum^3_{i=1}
 \tilde{U}_{\alpha i}
 \exp
 \left[
 -\frac{\Delta \tilde{m}^2_{i1} L}{2E}
 \right]
 \tilde{U}^{^{\ast}}_{\beta i}
 \right\vert^2 \;,
\end{equation}
where $L$ is the distance that the neutrino traverses from production to detection, $\tilde{m}^2_{i}/2E$ are the eigenvalues of the Hamiltonian, and $\Delta \tilde{m}^2_{ij} \equiv \tilde{m}_{i}^2 - \tilde{m}_{j}^2$. The matrix $\tilde{\mathbf{U}}$ diagonalizes the Hamiltonian; it is parameterized like the PMNS matrix but depends on the mixing parameters modified by matter effects, $\tilde{\theta}_{23}$, $\tilde{\theta}_{13}$, $\tilde{\theta}_{12}$, and $\tilde{\delta}_{\rm CP}$.  The values of the modified oscillation parameters deviate from their values in vacuum increasingly with rising neutrino energy; the magnitude of the deviation and its dependence with energy are different for the different symmetries.  Appendix~\ref{app:param_run} shows this explicitly.  We compute the oscillation parameters numerically (see below); \Refes~\cite{Barger:1980tf, Zaglauer:1988gz, Ohlsson:1999um, Akhmedov:2004ny, Agarwalla:2013tza, Chatterjee:2015gta, Agarwalla:2015cta, Khatun:2018lzs, Agarwalla:2021zfr} provide approximate analytical expressions for them, some of which we use in our discussion below.

To generate our results, including in figs.~\ref{fig:dune_prob_events} and \ref{fig:t2hk_prob_events}, we compute  the $\nu_\mu \to \nu_e$ and $\bar{\nu}_\mu \to \bar{\nu}_e$ \textit{appearance} probabilities and the $\nu_\mu \to \nu_\mu$ and $\bar{\nu}_\mu \to \bar{\nu}_\mu$ \textit{disappearance} probabilities to which DUNE and T2HK are sensitive.  We do this numerically using {\sc GLoBES}~\cite{Huber:2004ka, Huber:2007ji}, with a version of the {\sc snu} matrix-diagonalization library~\cite{Kopp:2006wp, Kopp:2007ne} modified by us to include the new matter potential.  Later (section~\ref{sec:expt-details}), we also use these tools to compute event rates. 

Figures~\ref{fig:dune_prob_events} and \ref{fig:t2hk_prob_events} show the $\nu_\mu \to \nu_e$ probability computed in the presence of a new matter potential, evaluated, respectively, for the baselines and energy ranges of DUNE and T2HK.  For the new matter potential matrix, we adopt the illustrative choice of $\mathbf{V}_{\rm LRI} = \mathrm{diag}(0, 0, -V_{\rm LRI})$, which has the texture of the potential induced by the $B - 3L_\tau$ and $L - 3L_\tau$ symmetries (table~\ref{tab:charges}), and vary the value of $V_{\rm LRI}$.  However, our  observations below hold also for other choices of the texture of the potential.  As anticipated above, the effects of the new interaction on the oscillation probabilities (and the event distributions) are significant when the new matter potential is comparable to the standard term in the Hamiltonian.  Broadly stated, the closer in size the new and standard contributions are, the closer to resonant are the effects induced by the new matter potential.  Appendix~\ref{app:prob_plots} shows oscillation probabilities for our other candidate symmetries. 

Exclusively for the purpose of understanding the effect of the new interactions on the probabilities, we use approximate analytical expressions for the $\nu_\mu \to \nu_e$ and $\nu_\mu \to \nu_\mu$ probabilities.  For the $\nu_\mu \to \nu_e$ probability, we expand  \equ{osc_prob_gen} under the approximation that $\tilde{\theta}_{12}$ saturates to $90^\circ$~\cite{Chatterjee:2015gta, Agarwalla:2021zfr}, which occurs early with rising energy in the presence of standard and new matter effects (\figu{theta_run}).  This yields~\cite{Agarwalla:2021zfr}
\begin{equation}
 \label{equ:app_mat}
 P_{\nu_{\mu} \to \nu_{e}}
 \approx
 \sin^{2}\tilde{\theta}_{23} ~ 
 \sin^{2}(2\tilde{\theta}_{13}) ~
 \sin^{2}
 \bigg[
 1.27~
 \frac{ (\Delta \tilde{m}^{2}_{32} / {\rm eV^2}) (L / {\rm km}) }
 {E / {\rm GeV}}
 \bigg] \,.
\end{equation}
For the $\nu_\mu \to \nu_\mu$ probability, we expand \equ{osc_prob_gen}, assuming $\tilde{\theta}_{12} = 90^{\circ}$, and keep the first two terms from Eq.~(3.27) of~\cite{Khatun:2018lzs} (see also \Refe~\cite{Agarwalla:2021zfr}).  This yields
\begin{equation}
 P_{\nu_{\mu}\to\nu_{\mu}}
 \approx
 1
 -
 \sin^2 (2\tilde{\theta}_{23})~
 \cos^2 \tilde\theta_{13}~
 \sin^2
 \bigg[
 1.27~
 \frac{(\Delta \tilde{m}^{2}_{31} / {\rm eV^2}) (L/{\rm km} )}
 {(E / {\rm GeV})}
 \bigg] \;.
\label{equ:surv_mat}
\end{equation}

In DUNE (\figu{dune_prob_events}), on account of its baseline, the leading vacuum contribution over most of the relevant energy range is $\propto \Delta m_{31}^2 / (2 E)$; see \equ{hamiltonian_vac}.  In the $\mu$-$\tau$ sector, which determines the modification of the $\tilde{\theta}_{23}$ and $\tilde{\theta}_{13}$  angles that drive the $\nu_\mu \to \nu_e$ probability, the vacuum contribution dominates over the matter potential at energies roughly below 6~GeV.  

From roughly 1~GeV to 2~GeV, resonant features induced by the new interaction on the probability are possible when the new matter potential is roughly of the same size as the standard contributions (see above).  The exact relation between these quantities, which we do not show explicitly but compute numerically and implicitly, stems from the conditions imposed on them in order to achieve the resonant enhancement of the probability.  In addition to increasing the probability amplitude, the new interaction shifts the position of the first oscillation maximum to slightly lower energies, due to the growth of $\Delta \tilde{m}^{2}_{31}$ and $\Delta \tilde{m}^{2}_{32}$ with energy (\figu{delm_run}).  Appendix~\ref{app:param_run} expands on this.  (Below about 0.5~GeV, the effects of the new interaction on the probability are driven instead by $\tilde{\theta}_{12}$, on account of its rapid growth with rising energy (\figu{theta_run}).  However, because of the paucity of the DUNE neutrino beam at these energies (section~\ref{sec:expt-details}), these effects contribute little to our analysis.)

Figure~\ref{fig:dune_prob_events} shows that, between 1~GeV and 2~GeV, for a given value of $V_{\rm LRI}$, the probability is enhanced at values of the energy for which the resonance condition is satisfied.  Higher values of $V_{\rm LRI}$ require larger matching energies to trigger the resonance, on account of the $\propto 1/E$ dependence of the vacuum term.  For a fixed value of $V_{\rm LRI}$, the modification with energy of the $\nu_\mu \to \nu_e$ probability is driven by the growth with energy of $\tilde{\theta}_{23}$ and $\tilde{\theta}_{13}$ (\figu{theta_run}).  Overall, for the illustrative symmetry in \figu{dune_prob_events}, this makes DUNE sensitive to $V_{\rm LRI} \approx (\mathbf{H}_{\rm vac})_{\tau\tau} \in [3.8 \cdot 10^{-14}, 1.4 
 \cdot10^{-12}]$~eV (see bottom panel of \figu{lrf_bounds_NH_selective}), given the reconstructed neutrino energy range of 0.5--18 GeV in DUNE (section~\ref{sec:expt-details}). Values of $V_{\rm LRI}$ significantly smaller than that are unable to match the standard contribution and, therefore, trigger no resonance. 

The above behavior is not limited to the illustrative symmetry in \figu{dune_prob_events}, but applies to all of the symmetries that we consider.  Indeed, although the specific elements of the standard contribution that the long-range potential must match are different for different symmetries, later we find that our upper limits on $V_{\rm LRI}$ are contained within the above range; most are within $10^{-14}$--$10^{-13}$~eV; see \figu{constraints_on_pot_dune_t2hk-NMO}.  

In T2HK (\figu{t2hk_prob_events}), the results are similar as in DUNE.  However, because the T2HK neutrino beam has lower energies, the vacuum contribution is larger than in DUNE and, therefore, the values of the new matter potential to which T2HK is sensitive are higher, \ie, $V_{\rm LRI} \in [2.3 \cdot 10^{-13}, 6.8 \cdot 10^{-12}]$~eV, corresponding to the reconstructed neutrino energy range of 0.1--3 GeV; see \figu{lrf_bounds_NH_selective}.

%=================================================
\subsection{Event rates in DUNE and T2HK}
\label{sec:expt-details}
%=================================================

Long-baseline neutrino experiments, with precisely characterized neutrino beams, are excellent platforms to perform precision tests of the standard oscillation paradigm and to search for physics beyond it.  Like \Refe~\cite{Singh:2023nek}, we gear our forecasts of the sensitivity to new neutrino-matter interactions to two of the leading long-baseline experiments under construction, DUNE and T2HK.  Both experiments plan to have near and far detectors; in our analysis, we focus exclusively on the latter, where the effects of oscillations are more apparent; however, the near detectors also have interesting probing capabilities~\cite{Melas:2023olz}.

The neutrino beams are produced as mainly $\nu_\mu$ or $\bar{\nu}_\mu$, with a small contamination of $\nu_e$ and $\bar{\nu}_e$.  The experiments will look for the appearance of $\nu_e$ and $\bar{\nu}_e$ and the disappearance of $\nu_\mu$ and $\bar{\nu}_\mu$. Hence, there are four oscillation channels that we use in our analysis:  $\nu_{\mu}\to\nu_e$, $\bar\nu_\mu\to \bar\nu_e$, $\nu_{\mu}\to\nu_\mu$, and $\bar\nu_\mu\to \bar\nu_\mu$.  Detection of a sought signal is primarily via CC neutrino interactions of $\nu_e$, $\bar{\nu}_e$, $\nu_\mu$, and $\bar{\nu}_\mu$.  While the majority of the background is contributed by the NC events triggered by neutrinos of all flavors, there is also a small contribution from CC events triggered by $\nu_{\tau}$ and $\bar{\nu}_{\tau}$ born from oscillations.  Following \Refes~\cite{Hyper-Kamiokande:2016srs, DUNE:2021cuw}, we assume 2\% and 5\% appearance and 5\% and 3.5\% disappearance systematic uncertainties when computing the signal event rates in DUNE and T2HK, respectively. For the background contribution, for T2HK we assume 10\% systematic uncertainties for all kinds of background events and, for DUNE, 5--20\% depending on the background channel.  For details, see table 7 of \Refe~\cite{Singh:2023nek}.  Since the far detectors cannot distinguish neutrinos from antineutrinos, we add the ``wrong-sign'' contamination events as part of the signal.

Just as for the oscillation probabilities, we compute the expected rate of neutrino-induced events detected by DUNE and T2HK using {\sc GLoBES}~\cite{Huber:2004ka, Huber:2007ji}, with the new neutrino-matter interactions included by modifying the {\sc snu} library~\cite{Kopp:2006wp, Kopp:2007ne}.  We compute events binned in reconstructed neutrino energy, $E_{\rm rec}$, \ie, the energy inferred by analyzing the properties of the particles created in the neutrino interaction.  

DUNE~\cite{DUNE:2020lwj, DUNE:2020jqi, DUNE:2021cuw, DUNE:2021mtg} will use a liquid-argon time-projection chamber detector with a net volume of 40~kton. Its neutrino beam will travel 1285~km from Fermilab to the Homestake Mine. Neutrinos will be produced by a 1.2-MW beam of 120-GeV protons delivering $1.1 \cdot 10^{21}$ protons-on-target (P.O.T.) per year.  It will produce a wide-band, on-axis beam of neutrinos with energies of 0.5--110~GeV and peaking around 2.5~GeV.  When simulating neutrino detection in DUNE, we follow the configuration details from \Refe~\cite{DUNE:2021cuw}.  DUNE will run in neutrino and antineutrino modes, 5 years in each, for a total run-time of 10 years.  To make our forecasts conservative, we use only the fiducial volume and beam power of the completed form of DUNE and ignore the smaller contribution from runs during its construction~\cite{DUNE:2021tad}.   We bin events between $E_{\rm rec} = 0.5$ and 8 GeV with a uniform bin width of 0.125~GeV; between 8 and 10~GeV, with a width of 1~GeV; and between 10 and 18~GeV, with a width of 2~GeV.

T2HK~\cite{Abe:2015zbg, Hyper-Kamiokande:2016srs, Hyper-Kamiokande:2018ofw} will use a water Cherenkov detector with a fiducial volume of 187~kton. Neutrinos will be produced at the J-PARC facility~\cite{McDonald:2001mc} by a 1.3-MW beam of 80-GeV protons delivering $2.7 \cdot 10^{22}$~P.O.T.~per year.  The ensuing neutrino beam will be narrow-band, will peak around 0.6~GeV, will travel 295~km to the detector at the Tochibara Mines in Japan, and arrive $2.5^\circ$ off-axis.  For our analysis, we follow the configuration details from \Refe~\cite{Hyper-Kamiokande:2016srs}.  T2HK will run 2.5 years in neutrino mode and 7.5 years in antineutrino mode, adhering to the proposed 1:3 ratio between the modes.  We bin events uniformly between $E_{\rm rec} = 0.1$ and 3~GeV with a bin width of 0.1~GeV.

Figures~\ref{fig:dune_prob_events} and \ref{fig:t2hk_prob_events} illustrate the event spectra from $\nu_e$ appearance expected in DUNE and T2HK, computed using the same illustrative potential of $\mathbf{V}_{\rm LRI} = \textrm{diag}(0, 0, -V_{\rm LRI})$ used for the probabilities in these figures.  The features in the event spectra reflect the features in the oscillation probabilities (section~\ref{sec:prob_var_pot}). The event distribution starts deviating from the standard-oscillation expectation at $V_{\rm LRI} \gtrsim 10^{-14}~{\rm eV}$, and the deviation grows with $V_{\rm LRI}$.  Because T2HK has a larger detector than DUNE, its event rate is 3--4 times higher.  Yet, because DUNE reaches higher energies than T2HK,  it is sensitive to smaller values of $V_{\rm LRI}$, since the sensitivity to $V_{\rm LRI}$ is $\propto 1/E$; see section~\ref{sec:prob_var_pot}.  Appendix~\ref{app:prob_plots} shows event spectra for our other candidate symmetries. 

The above interplay between the experiments has important consequences for their sensitivity to new neutrino interactions; they become apparent later in our analysis, \eg, in section~\ref{sec:constraints_pot}.  The constraints on and discovery potential of the new interactions are driven by DUNE, on account of it being sensitive to the smallest values of $V_{\rm LRI}$.  However, in order to reach high statistical significance in our claims --- \eg, for discovery or distinguishing between competing candidate symmetries --- the contribution of T2HK is key, since it can reach the larger values of $V_{\rm LRI}$ that are needed to make those claims.

%=================================================
\section{Limits and discovery potential}
\label{sec:results}
%=================================================

Based on the above calculation of oscillation probabilities and event rates, we forecast the sensitivity of T2HK and DUNE to the new neutrino-matter interactions induced by our candidate symmetries.  First, we forecast constraints on the new matter potential and then convert them into constraints on the mass and coupling strength of $Z^\prime$ (sections~\ref{sec:stat_methods} and \ref{sec:constraints_pot}).  Second, we forecast discovery prospects (sections~\ref{sec:stat_methods} and \ref{sec:discovery}).  Third, we forecast prospects of distinguishing between different symmetries (sections~\ref{sec:stat_methods} and \ref{sec:confusion-theory}).

In the main text, we show results as figures (figs.~\ref{fig:lrf_bounds_NH_selective}--\ref{fig:confusion-matrix}).  In the appendices, we show some of them in tables (tables~\ref{tab:upper_limits_potential_NMO_organized} and \ref{tab:discovery_strength}).   In \Refe~\cite{GitHub_lrf-repo}, we provide them as digitized files.

%=================================================
\subsection{Statistical methods}
\label{sec:stat_methods}
%=================================================

When computing the constraints and discovery prospects, we treat each symmetry individually.  We follow the same statistical methods as in \Refe~\cite{Singh:2023nek}; below, we sketch it, and defer to \Refe~\cite{Singh:2023nek} for details.  When computing prospects for distinguishing between symmetries, we extend these methods to make pairwise comparisons between symmetries.  Throughout, we fix $\theta_{13}$ and $\theta_{12}$ to their present-day best-fit values (table~\ref{tab:params_value1}).  For $\theta_{13}$, this is because the current precision on its value is already small, of 2.8\%~\cite{DayaBay:2022orm}.  For $\theta_{12}$, it is because it has only a small impact on our results (eqs.~(\ref{equ:app_mat}) and  (\ref{equ:surv_mat})).

The experiments are blind to the origin of the new matter potential.  They are only sensitive to how and by how much it influences neutrino oscillations, \ie, to the texture of the matter potential $\mathbf{V}_{\rm LRI}$, as shown in table~\ref{tab:charges}, and to the size of the parameter $V_{\rm LRI}$ on which it depends.  In our statistical analysis, for each choice of symmetry, we adopt its corresponding $\mathbf{V}_{\rm LRI}$ potential texture.  As a consequence, symmetries that have equal or similar  texture yield equal or similar sensitivity to $V_{\rm LRI}$; see, \eg, \figu{constraints_on_pot_dune_t2hk-NMO}.  Only afterwards, when we convert the resulting sensitivity to $V_{\rm LRI}$ into sensitivity on $G^\prime$ and $m_{Z^\prime}$, do the particular $U(1)^\prime$ charges of each symmetry and the knowledge of the matter content in celestial bodies play a role in yielding different sensitivity between equally or similarly textured symmetries; see, \eg, \figu{all_symmetry}.  We show this explicitly below.

\smallskip

\textbf{\textit{Constraints on new neutrino interactions.---}}  When forecasting constraints, we generate the \textit{true} event spectrum under standard oscillations by fixing $V_{\rm LRI}^{\rm true} = 0$.  We compare it to \textit{test} event spectra computed using non-zero values of $V_{\rm LRI}$.  In experiment $e = \{{\rm T2HK},~{\rm DUNE}\}$, we bin the spectra in $N_e$ bins of $E_{\rm rec}$; see section~\ref{sec:expt-details} for a description of the binning.  In the $i$-th bin, we compare the true {\it vs.}~test numbers of events from each detection channel $c = \{{\rm app}~\nu,~{\rm app}~\bar{\nu},~{\rm disapp}~\nu,~{\rm disapp}~\bar{\nu}\}$, \ie,  $N_{e,c,i}^\textrm{true}$ {\it vs.}~$N_{e,c,i}^\textrm{test}$, via the Poisson $\chi^2$ function~\cite{Baker:1983tu, Cowan:2010js, Blennow:2013oma, Singh:2023nek}
\begin{eqnarray}
 \chi_{e,c}^{2}
 (V_{\rm LRI}, \boldsymbol{\theta}, o)
 =
 &&
 \underset{\left\{\xi_{s}, \{\xi_{b, c, k}\}\right\}}{\mathrm{min}} 
 \left\{
 2\sum^{N_e}_{i=1}
 \left[
 N_{e, c, i}^{{\rm test}}
 (V_{\rm LRI}, \boldsymbol{\theta}, o, \xi_s, \{\xi_{b,c,k}\})
 \right. \right.
 \nonumber \\
 &&
 \left. \left.
 -
 N_{e, c, i}^{{\rm true}}
 \left( 
 1
 +
 \ln
 \frac{N_{e, c, i}^{{\rm test}}
 	(V_{\rm LRI}, \boldsymbol{\theta}, o, \xi_s, 
 	\{\xi_{b, c, k}\})}
 {N_{e, c, i}^{{\rm true}}}
 \right)
 \right]
 +
 \xi^{2}_{s}
 +
 \sum_k \xi^{2}_{b, c, k} 
 \right\} \;, \;\,\,\,\,\,\,\,\,\,\,
 \label{equ:chi2_per_channel}
\end{eqnarray}
where $\boldsymbol{\theta} \equiv \{ \sin^2\theta_{23}, \delta_{\rm CP}, \vert \Delta m_{31}^2\vert \}$ are the oscillation parameters that we vary (the other parameters are fixed, see above) and $o = \{ \text{NMO, IMO} \}$ is the choice of neutrino mass ordering, which can be normal (NMO) or inverted (IMO).  On the right-hand side of \equ{chi2_per_channel}, $\xi_{s}$ and $\xi_{b, c, k}$ represent, respectively, systematic uncertainties on the signal rate and the $k$-th background contribution to the detection channel $c$; these uncertainties are identical for neutrinos and antineutrinos.  We treat them as in \Refe~\cite{Singh:2023nek}.  The right-hand side of \equ{chi2_per_channel} is profiled over the systematic uncertainties; the last two terms are pull terms that keep the values of the systematic uncertainties under control when minimizing over them.  

In \equ{chi2_per_channel}, the event spectra contain both signal and background contributions.  The true number of events is computed using $V_{\rm LRI}^{\rm true} = 0$, and for choices of $\boldsymbol{\theta}^{\rm true}$ and $o^{\rm true}$, \ie,
\begin{equation}
 N_{e, c, i}^{{\rm true}}
 = 
 N_{e, c, i}^{s, {\rm true}}
 + 
 N_{e, c, i}^{b, {\rm true}} \;,
\end{equation}
where $N_{e, c, i}^{s, {\rm true}}$ and $N_{e, c, i}^{b, {\rm true}}$ are, respectively, the number of signal ($s$) and background ($b$) events; the latter is summed over all sources of background that affect this detection channel.  Similarly, the test number of events is
\begin{equation}
 \label{equ:num_test}
 N_{e, c, i}^{{\rm test}}
 (V_{\rm LRI}, \boldsymbol{\theta}, o, \xi_s, \left\{\xi_{b,c,k}\right\})
 =
 N^s_{e,c,i}(V_{\rm LRI}, \boldsymbol{\theta}, o)
 (1+\pi_{e,c}^s\xi_s)
 +
 \sum_k
 N^{b}_{e,c,k,i}(\boldsymbol{\theta}, o)
 \left(
 1+\pi_{e,c,k}^b\xi_{b,c,k}
 \right) \;,
\end{equation}
where $\pi_{e,c}^s$ and $\pi_{e,c,k}^b$ are normalization errors on the signal and background rates (refer to Table~\ref{tab:normalization_err}). See \Refe~\cite{Singh:2023nek} for more details.

%%%%%%%%%%%%%%%%%%%%%%%%%%%%%%%%%%%%%%%%%%
\begin{table}[t!]
 \resizebox{\columnwidth}{!}{%
  \centering
  \begin{tabular}{|c|c|c|c|c|c|c|c|c|}
   \hline
   \multirow{3}{*}{Experiment}  & \multicolumn{8}{c|}{Normalization errors~[\%]}  \\
   & \multicolumn{4}{c|}{Signal, $\pi_{e,c}^s$} & \multicolumn{4}{c|}{Background, $\pi_{e,c,k}^b$}\\
   &  App.~$\nu$ & App.~$\bar{\nu}$ & Disapp.~$\nu$ & Disapp.~$\bar{\nu}$ & $\nu_{e}$, $\bar{\nu}_{e}$ CC & $\nu_{\mu}$, $\bar{\nu}_{\mu}$ CC & $\nu_{\tau}$, $\bar{\nu}_{\tau}$ CC & NC \\ 
   \hline
   DUNE & 2 & 2 & 5 & 5 & 5 & 5 & 20 & 10\\
   T2HK & 5 & 5 & 3.5 & 3.5 & 10 & 10 & -- & 10\\
   \hline
  \end{tabular}}
 \caption{Normalization errors used in the calculation of event rates in DUNE and T2HK, including signal and background detection channels. We show them separately for the neutrino ($\nu$) and antineutrino ($\bar{\nu}$) modes, and for appearance (``App.'') and disappearance (``Disapp.'') channels. The errors, sourced from \cite{Hyper-Kamiokande:2016srs, DUNE:2021cuw}, are used in \equ{num_test}.}
 \label{tab:normalization_err}
\end{table}
%%%%%%%%%%%%%%%%%%%%%%%%%%%%%%%%%%%%%%%%%%

For T2HK or DUNE, separately and together, we compute the total $\chi^2$ by adding the contributions of all the channels $c$, \ie,
\begin{eqnarray}
 \label{equ:chi2_dune}
 \chi_{\rm DUNE}^{2}
 (V_{\rm{LRI}}, \boldsymbol{\theta}, o)
 &=&
 \sum_c \chi_{{\rm DUNE}, c}^{2}
 (V_{\rm{LRI}}, \boldsymbol{\theta}, o)
 \;, \\
 \label{equ:chi2_t2hk}
 \chi_{\rm T2HK}^{2}
 (V_{\rm{LRI}}, \boldsymbol{\theta}, o)
 &=&
 \sum_c \chi_{{\rm T2HK}, c}^{2}
 (V_{\rm{LRI}}, \boldsymbol{\theta}, o) 
 \;, \\
 \label{equ:chi2_dune_t2hk}
 \chi^2_{{\rm DUNE}+{\rm T2HK}}
 (V_{\rm{LRI}}, \boldsymbol{\theta}, o)
 &=&
 \chi^2_{\rm DUNE}
 (V_{\rm{LRI}}, \boldsymbol{\theta}, o)
 +
 \chi^2_{\rm T2HK}
 (V_{\rm{LRI}}, \boldsymbol{\theta}, o)
 \;.
\end{eqnarray}
We treat the contributions of different channels as uncorrelated.

We compute the sensitivity to $V_{\rm LRI}$ by comparing the minimum value of the above functions, $\chi^2_{e, {\rm min}}$, which is obtained when evaluating them at $V_{\rm LRI} = V_{\rm LRI}^{\rm true} = 0$, $\boldsymbol{\theta} = \boldsymbol{\theta}^{\rm true}$, and $o = o^{\rm true}$, against test values of these parameters.  In the main text, we fix $\boldsymbol{\theta}^{\rm true}$ to its best-fit value under normal ordering (table~\ref{tab:charges}) and $o^{\rm true}$ to NMO.  In appendix~\ref{app:other}, we fix them to inverted ordering instead; our conclusions do not change.  Since we are interested in obtaining limits only on $V_{\rm LRI}$, we profile over $\boldsymbol{\theta}$ and $o$.  This yields the test statistic that we use to place constraints on $V_{\rm LRI}$; \eg, for DUNE, it is
\begin{equation}
 \label{equ:delta_chi2_dune}
 \Delta\chi^2_{ {\rm DUNE}, {\rm con}}(V_{\rm LRI})
 =
 \underset
 {\{ \boldsymbol{\theta}, o\}}{\mathrm{min}} 
 \left[
 \chi_{\rm DUNE}^{2}
 (V_{\rm LRI}, \boldsymbol{\theta}, o) 
 -
 \chi_{{\rm DUNE}, {\rm min}}^{2}
 \right]
 \;,
\end{equation}
and similarly for T2HK and DUNE + T2HK.  When profiling, we follow the same procedure as in \Refe~\cite{Singh:2023nek}.  When profiling over $\sin^2 \theta_{23}$, $\delta_{\rm CP}$, and $\vert \Delta m_{31}^2 \vert$, we vary each of them within their present-day $3\sigma$ allowed ranges~\cite{Esteban:2020cvm}.  We assume no correlations between them, since these are expected to disappear in the near future; see, \eg, \Refe~\cite{Song:2020nfh}. 
In principle, varying the values of the oscillation parameters over ranges different than the ones we have used could change our results. However, we have found that the test statistic aligns closely with the chosen true values, so we do not expect significant changes were we to use wider ranges for the oscillation parameters. Using \equ{delta_chi2_dune}, we report upper limits on $V_{\rm LRI}$ with $2\sigma$ and $3\sigma$ significance, for 1 degree of freedom (d.o.f.).

\smallskip

\textbf{\textit{Discovery of new neutrino interactions.---}}  When forecasting discovery prospects, we follow a similar procedure as when forecasting constraints, with some changes.  In $\chi^2_{e,c}(V_{\rm LRI}, \boldsymbol{\theta}, o)$ in \equ{chi2_per_channel}, the true event spectrum is instead computed using the nonzero value $V_{\rm LRI}^{\rm true} = V_{\rm LRI}$ from the left-hand side and, like before, for choices of $\boldsymbol{\theta}^{\rm true}$ and $o^{\rm true}$ (NMO in the main text and IMO in appendix~\ref{app:other}), \ie, $N_{e,c,i}^{\rm true} \to N_{e,c,i}^{\rm true}(V_{\rm LRI}^{\rm true} = V_{\rm LRI})$ in \equ{chi2_per_channel}.  The test spectrum is instead computed under standard oscillations, \ie, $N_{e, c, i}^{{\rm test}} (V_{\rm LRI}, \boldsymbol{\theta}, o, \xi_s, \{\xi_{b,c,k}\}) \to N_{e, c, i}^{{\rm test}} (V_{\rm LRI} = 0, \boldsymbol{\theta}, o, \xi_s, \{\xi_{b,c,k}\})$ in \equ{chi2_per_channel}.  Like before we profile over $\boldsymbol{\theta}$ and $o$ to build the test statistic that we use to compute the significance with which oscillations with a new matter potential $V_{\rm LRI}$ would be discovered; \eg, for DUNE,
\begin{equation}
 \label{equ:delta_chi2_disc}
 \Delta\chi^2_{ {\rm DUNE}, \rm {disc}}(V_{\rm LRI})
 =
 \underset
 {\{ \boldsymbol{\theta}, o\}}{\mathrm{min}} 
 \left[\chi_{ {\rm DUNE}, {\rm min}}^{2}
 -
 \chi_{\rm DUNE}^{2}
 (V_{\rm LRI}, \boldsymbol{\theta}, o) 
 \right]
 \;,
\end{equation}
and similarly for T2HK and DUNE + T2HK.  This test statistic measures the separation between the observed event distribution, which includes the new matter potential, and standard oscillations.  Using \equ{delta_chi2_disc}, we report the values of $V_{\rm LRI}$ for which LRI would be discovered at $3\sigma$ and $5\sigma$, for 1 d.o.f.  Reference~\cite{Singh:2023nek} shows complementary results on jointly measuring the values of $V_{\rm LRI}$ and of the mixing parameters $\theta_{23}$ and $\delta_{\rm CP}$.

\smallskip

\textbf{\textit{Distinguishing between symmetries.---}}Symmetries that introduce new  matter potentials with different textures have qualitatively different effects on the oscillation probabilities (\figu{probability}).  Hence, we explore whether, in the event of discovery of evidence of a new neutrino interaction, we may identify which symmetry is responsible for it, or narrow down the possibilities to a subset of candidate symmetries.

We proceed similarly as before.  Out of the set of candidate symmetries  (table~\ref{tab:charges}), the true symmetry responsible for the new potential observed is SA, and SB is an alternative one.  We modify \equ{chi2_per_channel} by changing $N_{e,c,i}^{\rm true} \to  N_{e,c,i}^{\rm true}(V_{\rm LRI}) \vert_{\rm SA}$ and $N_{e, c, i}^{{\rm test}}
(V_{\rm LRI}, \boldsymbol{\theta}, o, \xi_s, \allowbreak \{\xi_{b,c,k}\}) \to N_{e, c, i}^{{\rm test}} (V_{\rm LRI}, \boldsymbol{\theta}, o, \xi_s, \{\xi_{b,c,k}\}) \vert_{\rm SB}$.  We show only results assuming NMO for $\boldsymbol{\theta}^{\rm true}$ and $o^{\rm true}$.  The test statistic that we use to distinguish SA from SB is
\begin{equation}
 \label{equ:delta_chi2_conf}
 \Delta\chi^2_{ {\rm DUNE}, {\rm dist}}(V_{\rm LRI}) \vert_{{\rm SA}, {\rm SB}}
 =
 \underset
 {\{ \boldsymbol{\theta}, o\}}{\rm{min}} 
 \left[\chi^{2}_{{\rm DUNE}, {\rm min}}(V_{\rm LRI}) \vert_{\rm SA}
 -
 \chi_{\rm DUNE}^{2}
 (V_{\rm LRI}, \boldsymbol{\theta}, o) \vert_{\rm SB}
 \right]
 \;,
\end{equation}
and similarly for T2HK and DUNE + T2HK.  Using \equ{delta_chi2_disc}, we report the significance, for 1~d.o.f., with which all pairs of SA and SB can be distinguished, via confusion matrices produced for illustrative values of $V_{\rm LRI}$.  

%=================================================
\begin{figure}[t!]
 \centering
 \includegraphics[width=0.75\textwidth]{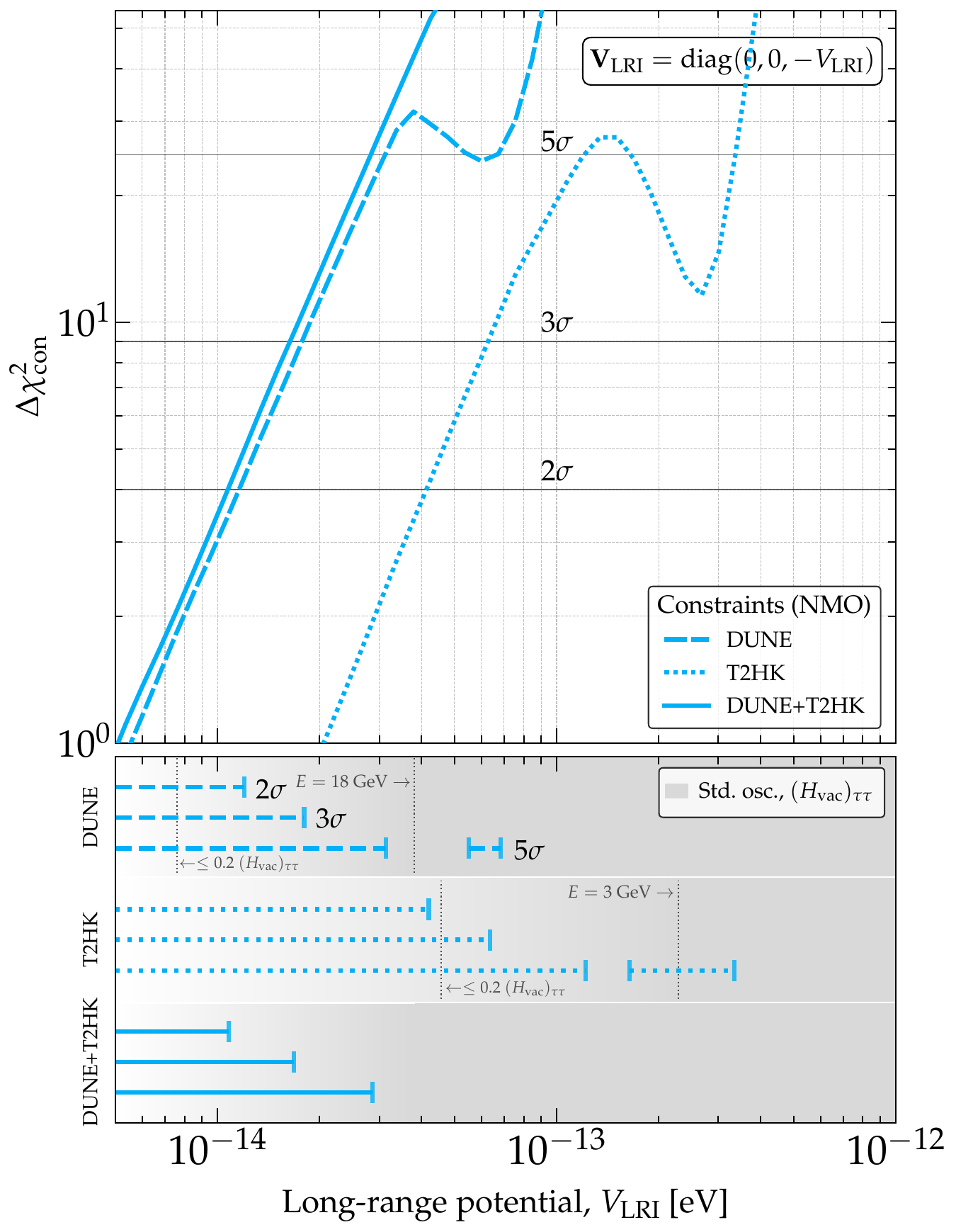}
 \caption{\textbf{\textit{Projected test statistic used to constrain the new matter potential induced by a $U(1)^\prime$ symmetry.}} For this plot, as illustration, we show limits on a potential of the form $\mathbf{V}_{\rm LRI} = \textrm{diag}(0, 0, -V_{\rm LRI})$ for neutrinos and $-\mathbf{V}_{\rm LRI}$ for antineutrinos, like in figs.~\ref{fig:dune_prob_events} and \ref{fig:t2hk_prob_events}, as would be introduced by symmetries $L - 3L_\tau$ or $B - 3L_\tau$ (table~\ref{tab:charges}); fig.~\ref{fig:lrf_bounds_NH} shows results for all the symmetries.  The test statistic is \equ{delta_chi2_dune}.  Results are for DUNE and T2HK separately and combined.  The true neutrino mass ordering is assumed to be normal; fig.~\ref{fig:lrf_bounds_IH} shows that results under inverted mass ordering are similar.  See sections~\ref{sec:stat_methods} and \ref{sec:constraints_pot} for details.  The experiments are sensitive to values of $V_{\rm LRI}$ that are comparable to the standard-oscillation terms in the Hamiltonian; for the choice of $\mathbf{V}_{\rm LRI}$ texture in this figure, this is $(\mathbf{H}_{\rm vac})_{\tau\tau}$, which is $\propto 1/E$. Constraints on $V_{\rm LRI}$ lie around 20\% of the value of $(\mathbf{H}_{\rm vac})_{\tau\tau}$ evaluated at the highest energy in each experiment.  \textit{Combining DUNE and T2HK strengthens the constraints, especially at high values of $V_{\rm LRI}$, by removing the degeneracies between $V_{\rm LRI}$ and $\theta_{23}$ and $\delta_{\rm CP}$ that plague each experiment individually (see also \Refe~\cite{Singh:2023nek}).}
 \label{fig:lrf_bounds_NH_selective}} 
\end{figure}
%=================================================

%=================================================
\subsection{Constraints on new neutrino interactions}
\label{sec:constraints_pot}
%=================================================

%=================================================
\begin{figure}[t!]
 \centering
 \includegraphics[width=\textwidth]{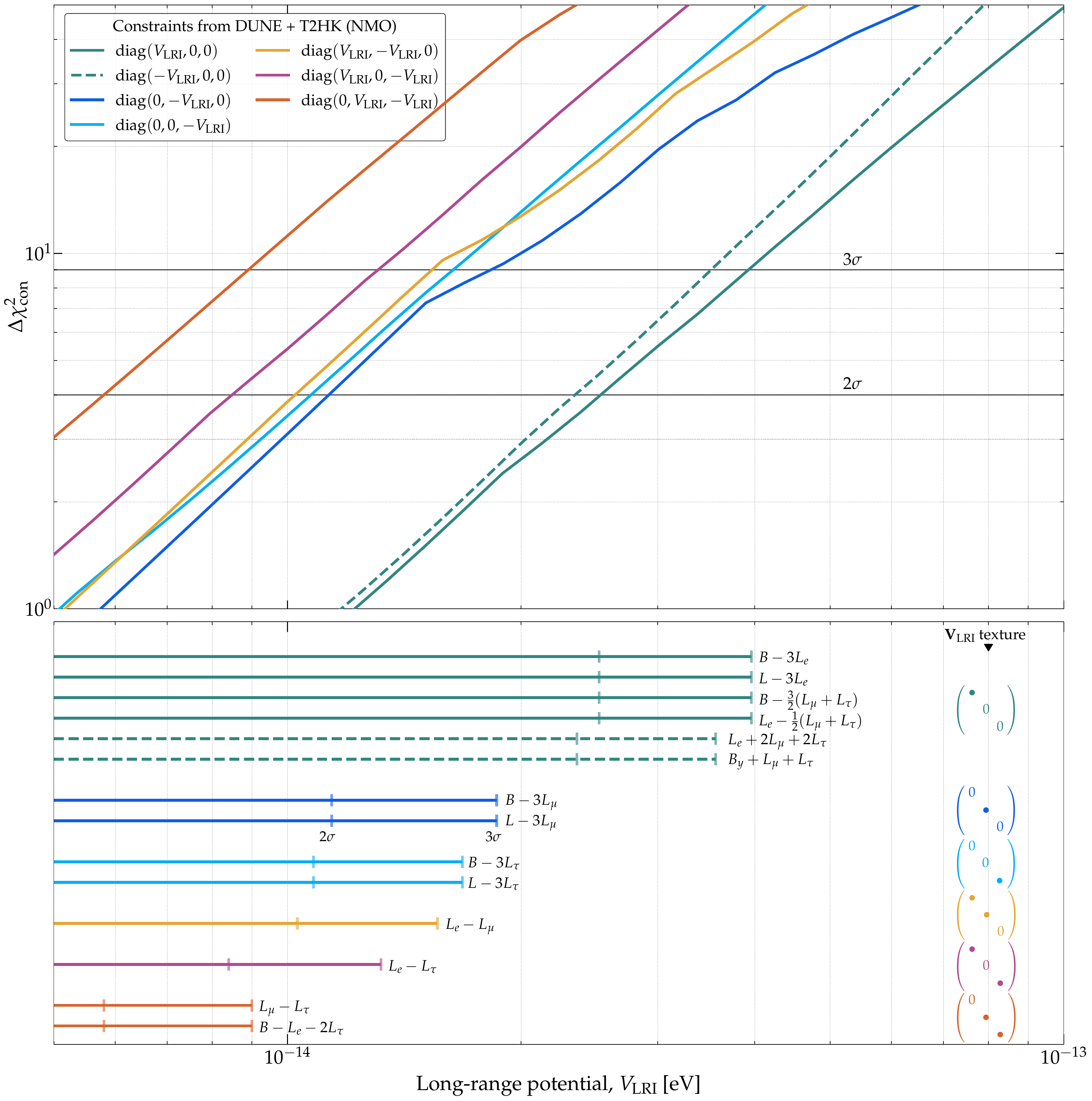}
 \caption{\textbf{\textit{Projected test statistic (top) used to place upper limits (bottom) on the new matter potential induced by our candidate $U(1)^\prime$ symmetries.}}  The true neutrino mass ordering is assumed to be normal; results under inverted ordering are similar (\figu{constraints_on_pot_dune_t2hk-IMO}).  Results are for DUNE and T2HK combined; the test statistic is \equ{delta_chi2_dune} computed for their combination.  See sections~\ref{sec:stat_methods} and \ref{sec:constraints_pot} for details.  The symmetries are grouped according to the texture of the new matter potential that they introduce, $\mathbf{V}_{\rm LRI}$ in table~\ref{tab:charges}.  The numerical values of the limits are in table~\ref{tab:upper_limits_potential_NMO_organized}.  \textit{Symmetries that induce equal or similar potential texture yield equal or similar upper limits on $V_{\rm LRI}$.  All limits lie near the value of the standard-oscillation terms in the Hamiltonian (see \figu{lrf_bounds_NH_selective}), since this triggers resonances in the oscillation probabilities (section~\ref{sec:hamiltonians}).}
 \label{fig:constraints_on_pot_dune_t2hk-NMO}} 
\end{figure}
%=================================================

Figure~\ref{fig:lrf_bounds_NH_selective} shows our resulting projected constraints on $V_{\rm LRI}$ assuming, for illustration, a matter potential with the texture $\mathbf{V}_{\rm LRI} = \textrm{diag}(0,0,-V_{\rm LRI})$, as introduced by the symmetries $L-3L_\tau$ and $B-3L_\tau$, just as in figs.~\ref{fig:dune_prob_events} and \ref{fig:t2hk_prob_events}.  (We show results for other symmetries later.)  Our findings reiterate one of the key results first reported by \Refe~\cite{Singh:2023nek}: DUNE and T2HK can each separately place upper limits on $V_{\rm LRI}$ --- with DUNE placing stronger constraints due to its higher energies ({\it cf.}~figs.~\ref{fig:dune_prob_events}~{\it vs.}~\ref{fig:t2hk_prob_events}), as anticipated in section~\ref{sec:expt-details}.   However, the limits that they can place individually weaken at high values of $V_{\rm LRI}$, due to degeneracies between $V_{\rm LRI}$, $\theta_{23}$, and $\delta_{\rm CP}$; in \figu{lrf_bounds_NH_selective}, this shows up as dips in the test statistic; see \Refe~\cite{Singh:2023nek} for details.  Combining DUNE and T2HK lifts these degeneracies: T2HK lifts the degeneracies due to $\theta_{23}$ and $\delta_{\rm CP}$, while DUNE fixes the neutrino mass ordering, \ie, the sign of $\Delta m_{31}^2$.  As anticipated (section~\ref{sec:prob_var_pot}), the limits on $V_{\rm LRI}$ are comparable to the size of standard-oscillation terms in the Hamiltonian.

Our results extend those of \Refe~\cite{Singh:2023nek}, which had shown the above interplay between DUNE and T2HK only for the symmetries $L_e-L_\mu$, $L_e-L_\tau$, and $L_\mu-L_\tau$.  We find that the same complementarity is present for all the other candidate symmetries that we consider (\figu{lrf_bounds_NH}), with some variation depending on whether the mass ordering is normal or inverted ({\it cf.}~figs.~\ref{fig:lrf_bounds_NH} {\it vs.}~\ref{fig:lrf_bounds_IH}), stemming from differences in the signs of the standard and new matter potentials for neutrinos and antineutrinos (section~\ref{sec:hamiltonians}), and in the run times for each in T2HK (section~\ref{sec:expt-details}).

Figure~\ref{fig:constraints_on_pot_dune_t2hk-NMO} (also, table~\ref{tab:upper_limits_potential_NMO_organized}) shows the upper limits on $V_{\rm LRI}$ for all the symmetries that we consider.  Like in table~\ref{tab:charges}, we group symmetries according to the texture of the matter potential, $\mathbf{V}_{\rm LRI}$, that they induce, since the effects of new interactions on the neutrino oscillation probabilities depend on the texture of the potential matrix $\boldsymbol{V}_{\rm LRI}$, and on the size of its elements, regardless of the source of the potential.  Because of this, the limits on $V_{\rm LRI}$ in \figu{constraints_on_pot_dune_t2hk-NMO} are equal or similar for symmetries that have equal or similar textures for $\mathbf{V}_{\rm LRI}$; {\it cf.}~table~\ref{tab:charges} and \figu{constraints_on_pot_dune_t2hk-NMO}.  Thus, our results broaden the perspectives first put forward by \Refe~\cite{Singh:2023nek}: \textbf{\textit{DUNE and T2HK may constrain the new matter potential to a level comparable to the standard-oscillation terms, roughly $10^{-14}$--$10^{-13}$~eV, regardless of what is the $U(1)^\prime$ symmetry responsible for inducing the new interaction.}}

%=================================================
\begin{figure}[t!]
 \centering
 \includegraphics[width=0.85\textwidth]{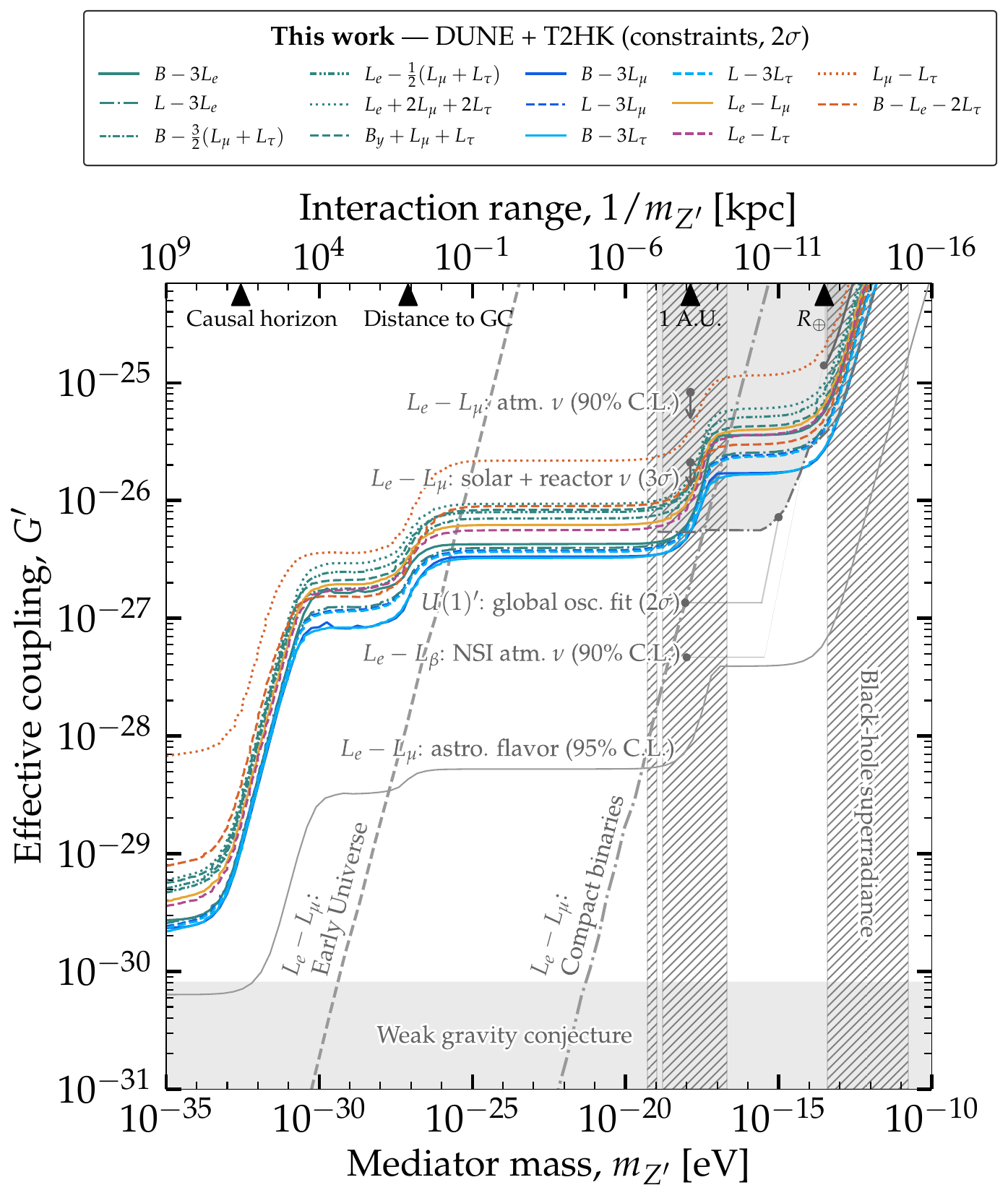}
 \caption{\textbf{\textit{Projected upper limits on the effective coupling of the new gauge boson, $Z^\prime$, that mediates flavor-dependent long-range neutrino interactions.}}  Results are for DUNE and T2HK, combined, after 10 years of operation, and for each of our candidate $U(1)^\prime$ symmetries (table~\ref{tab:charges}).  For this figure, we assume that the true neutrino mass ordering is normal. For each symmetry, the limits on the coupling, $G^\prime$, as a function of the mediator mass, $m_{Z^\prime}$, are converted from the limits on $V_{\rm LRI}$ in \figu{constraints_on_pot_dune_t2hk-NMO} (also in table~\ref{tab:upper_limits_potential_NMO_organized}) using the expressions for $V_{\rm LRI}$ in table~\ref{tab:charges}.  The existing limits are the same as in \figu{constraints_discovery_dune_t2hk-NMO}.  See sections~\ref{sec:stat_methods} and \ref{sec:constraints_pot} for details.  \textit{DUNE and T2HK may constrain long-range interactions more strongly than ever before, regardless of which $U(1)^\prime$ symmetry is responsible for inducing them, especially for mediators lighter than $10^{-18}$~ eV}.}
 \label{fig:all_symmetry}
\end{figure}
%=================================================

The strongest limits can be placed when the new matter potential affects primarily the $\mu$-$\tau$ sector, \ie, when $\mathbf{V}_{\rm LRI} = \textrm{diag}(0, V_{\rm LRI}, -V_{\rm LRI})$, as would be induced by symmetries $B-L_{e}-2L_{\tau}$ and $L_\mu - L_\tau$; see table~\ref{tab:charges}.  A potential of this form affects primarily the $\nu_\mu \to \nu_\mu$ and $\bar{\nu}_\mu \to \bar{\nu}_\mu$ disappearance probabilities. Because in DUNE and T2HK the disappearance channels have the highest event rates (\figu{event_spectra}), the effects of the new matter potential in this case can be detected more easily, leading to stronger limits on $V_{\rm LRI}$.  In contrast, the weakest limits can be placed when the new matter potential affects primarily the electron sector, \ie, when the only non-zero entry is $(\mathbf{V}_{\rm LRI})_{ee}$, as would be induced by symmetries $B - 3L_e$, $L - 3L_e$, $B-\frac{3}{2} (L_\mu + L_\tau)$, $B_y + L_\mu + L_\tau$, $L_{e} + 2L_\mu + 2L_\tau$, and $L_e - \frac{1}{2} (L_\mu + L_\tau)$.  A potential of this form affects primarily the $\nu_\mu \to \nu_e$ and $\bar{\nu}_\mu \to \bar{\nu}_e$ appearance probabilities.  Because in DUNE and T2HK the appearance channels have lower event rates, the limits on $V_{\rm LRI}$ in this case are weaker.

Figure~\ref{fig:all_symmetry} shows the limits on $V_{\rm LRI}$ converted into limits on $G^\prime$ as a function of $m_{Z^\prime}$.  To convert them, we use knowledge of the distribution of electrons, protons, and neutrons in the Earth, Moon, Sun, the Milky Way, and the cosmological matter distribution, and the long-range potential that they source, as introduced in section~\ref{sec:yukawa_interaction}.  In practice, for each symmetry, we take the limit on $V_{\rm LRI}$ from \figu{constraints_on_pot_dune_t2hk-NMO} and equate it to the expression for the simplified potential in table~\ref{tab:charges}, which depends on $m_{Z^\prime}$ and $G^\prime$, and which contains the contribution of the celestial bodies weighed by their  abundance of electrons, protons, and neutrons.  Then, for each value of $m_{Z^\prime}$, we find the upper limit on $G^\prime$ that we report in \figu{all_symmetry}.  In \figu{constraints_discovery_dune_t2hk-NMO}, the region constrained is the envelope of all the individual curves in \figu{all_symmetry}.

%=================================================
\begin{figure}[t!]
 \centering
 \includegraphics[width=1\textwidth]{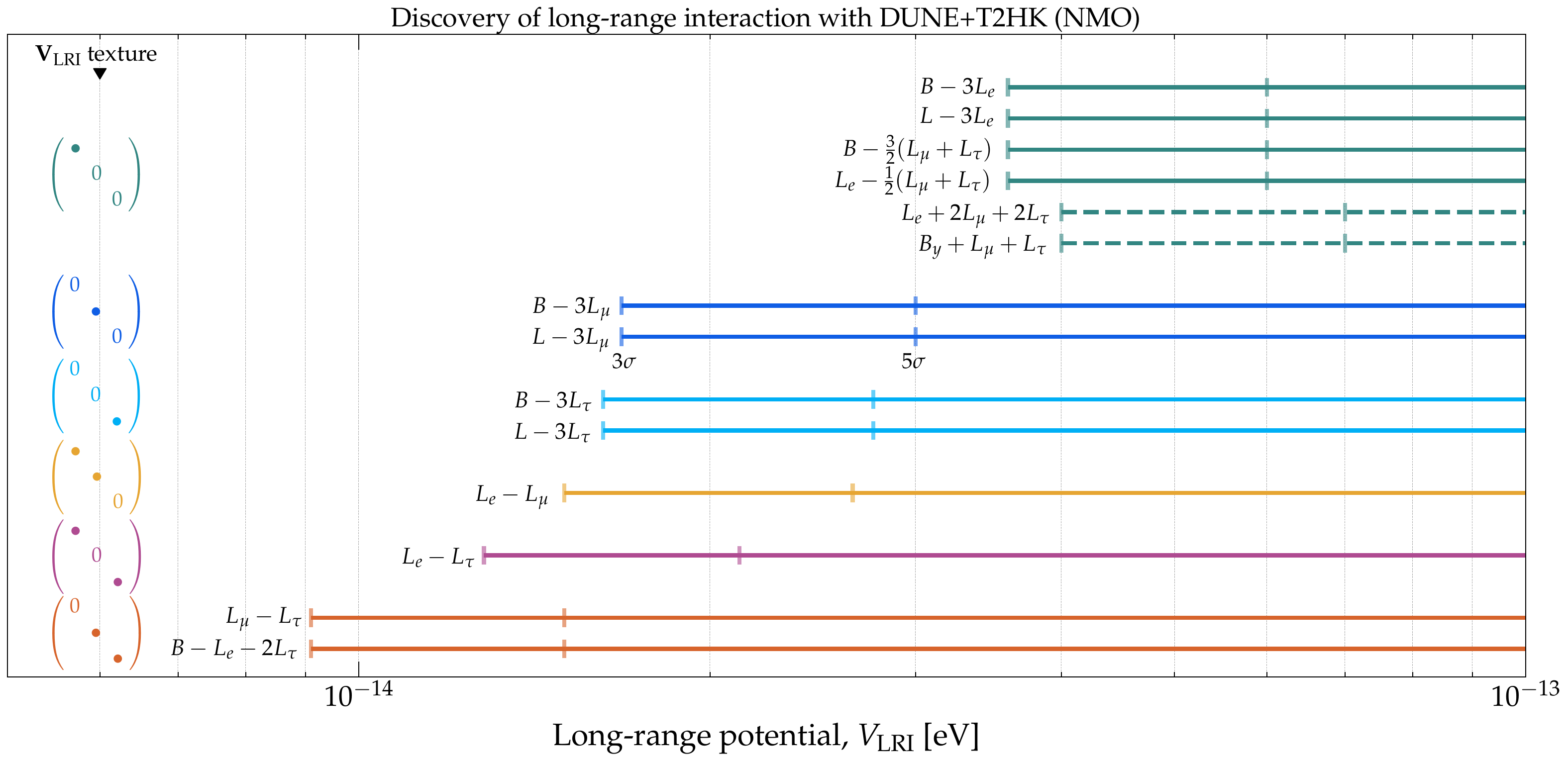}
 \caption{\textbf{\textit{Projected discovery prospects of the new matter potential induced by our candidate $U(1)^\prime$ symmetries.}}  The true neutrino mass ordering is assumed to be normal.  Results are for DUNE and T2HK combined; the test statistic is \equ{delta_chi2_disc} computed for their combination.   See sections~\ref{sec:stat_methods} and \ref{sec:discovery} for details.  The symmetries are grouped according to the texture of the new matter potential that they introduce, $\mathbf{V}_{\rm LRI}$ in table~\ref{tab:charges}.  The numerical values of the results are in table~\ref{tab:discovery_strength}.  \textit{Symmetries that induce equal or similar potential texture have equal or similar discovery prospects.  Discoverable ranges of $V_{\rm LRI}$ lie near the value of the standard-oscillation terms in the Hamiltonian, since this triggers resonances in the oscillation probabilities (section~\ref{sec:hamiltonians}).}}
 \label{fig:discovery_3s_5s_dune_t2hk}
\end{figure}
%=================================================

The limits in \figu{all_symmetry} exhibit step-like transitions occurring at different values of $m_{Z^\prime}$.  As explained in \Refe~\cite{Bustamante:2018mzu} (see also \Refes~\cite{Agarwalla:2023sng, Singh:2023nek} and section~\ref{sec:yukawa_interaction}), the transitions occur when the interaction range, $1/m_{Z^\prime}$, reaches the distance to a new celestial body.  Because different bodies have different abundances of electrons, protons, and neutrons, the tightest limits on $V_{\rm LRI}$ from \figu{constraints_on_pot_dune_t2hk-NMO} do not necessarily translate into the tightest limits on $G^\prime$ in \figu{all_symmetry}.  Once again, our results broaden the perspectives first put forward by \Refe~\cite{Singh:2023nek}: \textbf{\textit{regardless of which of our candidate $U(1)^\prime$ symmetries is responsible for inducing long-range neutrino interactions, DUNE and T2HK may outperform existing limits on the coupling strength of the new $Z^\prime$ mediator.}}  (See section~\ref{sec:intro} for an explanation of why the limits coming from flavor measurements of high-energy astrophysical neutrinos in \figu{all_symmetry} are in reality no match, for now, for DUNE and T2HK.)

References~\cite{Dolgov:1995hc, Blinnikov:1995kp, Joshipura:2003jh, Grifols:2003gy, Bustamante:2018mzu} indicated that if the relic neutrino background consists of equal numbers of $\nu_e$ and $\bar{\nu}_e$ it may partially screen out the long-range matter potential sourced by distant electrons by inducing corrections to the mass of the $Z^\prime$.  We have not considered this effect in our analysis, but, like \Refe~\cite{Bustamante:2018mzu}, we point out that it would affect the sensitivity to coupling strengths $G^\prime \lesssim 10^{-29}$, for which the Debye length of this effect, \ie, the distance at which it becomes appreciable, is about a factor-of-10 smaller~\cite{Joshipura:2003jh} than the interaction length to which we are sensitive in \figu{all_symmetry}.  A recent recalculation of the magnitude of the screening in \Refe~\cite{Chauhan:2024qew} suggests that it might have a stronger impact on the constraints on long-range interactions; however, a detailed assessment of this possibility within our analysis lies beyond the scope of the present work.

%=================================================
\begin{figure}[t!]
 \centering
 \includegraphics[width=0.85\textwidth]{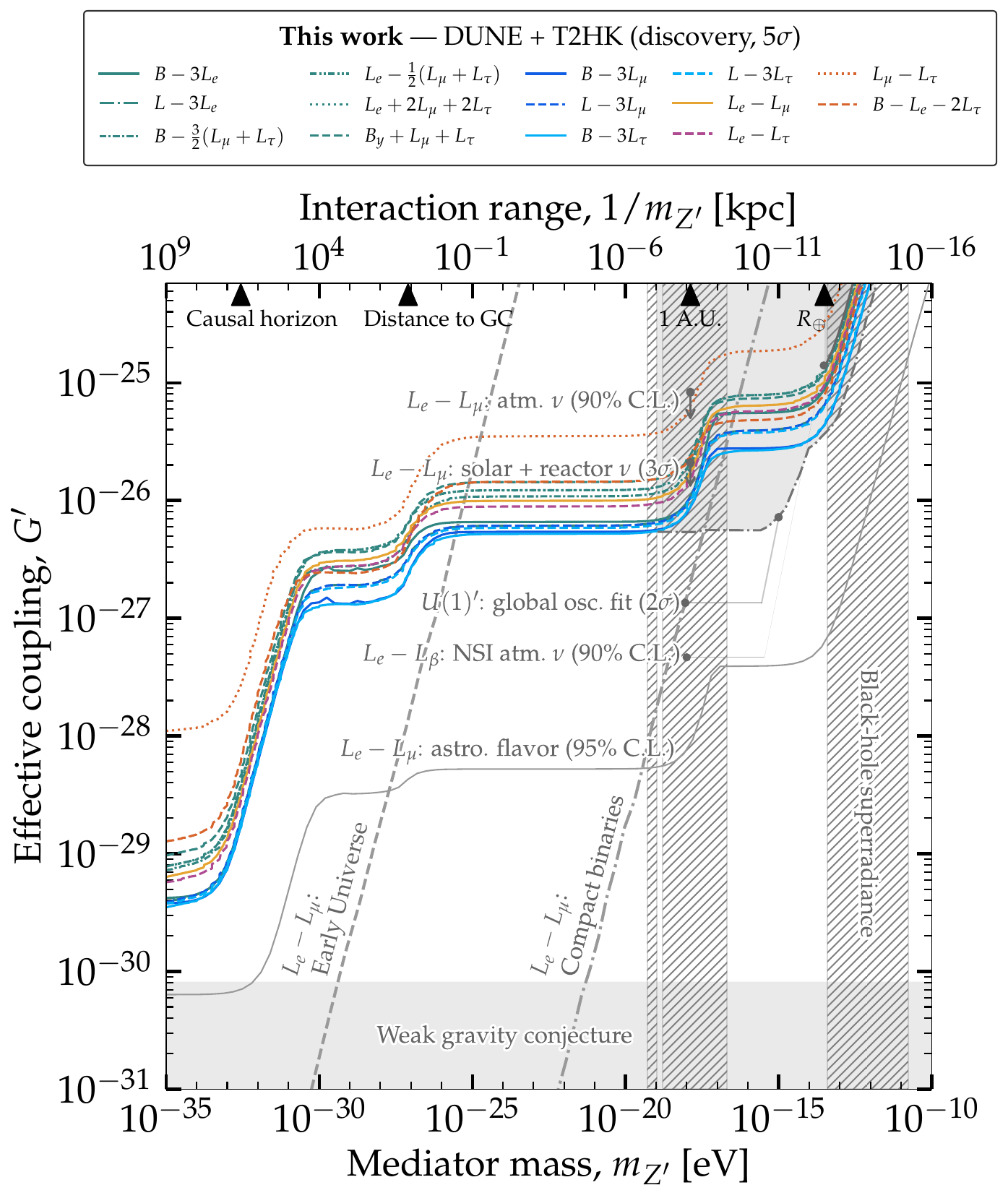}
 \caption{\textbf{\textit{Projected discovery prospects of the effective coupling of the new gauge boson, $Z^\prime$, that mediates flavor-dependent long-range neutrino interactions.}}  Results are for DUNE and T2HK, combined, after 10 years of operation, and for each of our candidate $U(1)^\prime$ symmetries (table~\ref{tab:charges}).  For this figure, we assume that the true neutrino mass ordering is normal. For each symmetry, the discovery prospects of the coupling, $G^\prime$, as a function of the mediator mass, $m_{Z^\prime}$, are converted from the discovery prospects on $V_{\rm LRI}$ in \figu{discovery_3s_5s_dune_t2hk} (also in table~\ref{tab:discovery_strength}) using the expressions for $V_{\rm LRI}$ in table~\ref{tab:charges}.  The existing limits are the same as in \figu{constraints_discovery_dune_t2hk-NMO}.  See sections~\ref{sec:stat_methods} and \ref{sec:discovery} for details.  \textit{DUNE and T2HK may discover long-range interactions, if they induce a matter potential comparable to the standard-oscillation terms of the Hamiltonian, regardless of which $U(1)^\prime$ symmetry is responsible for inducing them.}}
 \label{fig:discovery_all_DUNE+T2HK}
\end{figure}
%=================================================

%=================================================
\subsection{Discovery of new neutrino interactions}
\label{sec:discovery}
%=================================================

Figure~\ref{fig:discovery_3s_5s_dune_t2hk} (also, table~\ref{tab:discovery_strength}) shows, for each symmetry, the projected range of values of $V_{\rm LRI}$ that would result in the discovery of the presence of a new matter potential with a statistical significance of $3\sigma$ or $5\sigma$.  The ranges of values that can be discovered are similar to the ranges of values that can be constrained (\figu{constraints_on_pot_dune_t2hk-NMO}), since in both cases it is the size of the standard-oscillation term in the Hamiltonian that determines the sensitivity.  Like when placing constraints, symmetries whose matter potentials have equal or similar texture yield equal or similar discovery prospects.  Symmetries that affect primarily the $\nu_\mu \to \nu_\mu$ and $\bar{\nu}_\mu \to \bar{\nu}_\mu$ disappearance probabilities require smaller values of $V_{\rm LRI}$ to be discovered, due to the event rates being largest in the disappearance channels (\figu{event_spectra}).

Figure~\ref{fig:discovery_all_DUNE+T2HK} shows the associated discoverable regions of $G^\prime$ as a function of $m_{Z^\prime}$, converted from the discoverable intervals of $V_{\rm LRI}$ via the expressions for the potential sourced by celestial bodies in table~\ref{tab:charges}, just like we did for the constraints.  The results for discovery exhibit the same step-like transitions as for constraints, and the hierarchy of discoverability of the symmetries in \figu{discovery_all_DUNE+T2HK} is, as expected, the same as that of the constraints in \figu{all_symmetry}.  In \figu{constraints_discovery_dune_t2hk-NMO}, the discoverable region is the envelope of all the individual curves  in \figu{discovery_all_DUNE+T2HK}.

The results in figs.~\ref{fig:discovery_3s_5s_dune_t2hk} and \ref{fig:discovery_all_DUNE+T2HK} are the first reported discovery prospects in DUNE and T2HK of the full list of candidate $U(1)^\prime$ gauge symmetries in table~\ref{tab:charges}.  \textbf{\textit{DUNE and T2HK may discover long-range interactions, regardless of what is the $U(1)^\prime$ symmetry responsible for inducing them}}, provided the new matter potential is roughly within $10^{-14}$--$10^{-13}$~eV, \ie, comparable to the standard-oscillation terms.

%=================================================
\subsection{Distinguishing between symmetries}
\label{sec:confusion-theory}
%=================================================

Finally, we forecast how well, in the event of discovery of a new neutrino interaction, DUNE and T2HK could identify which of our candidate $U(1)^\prime$ symmetries is responsible for it.  

\begin{figure}[t!]
 \centering
 \includegraphics[width=1\textwidth]{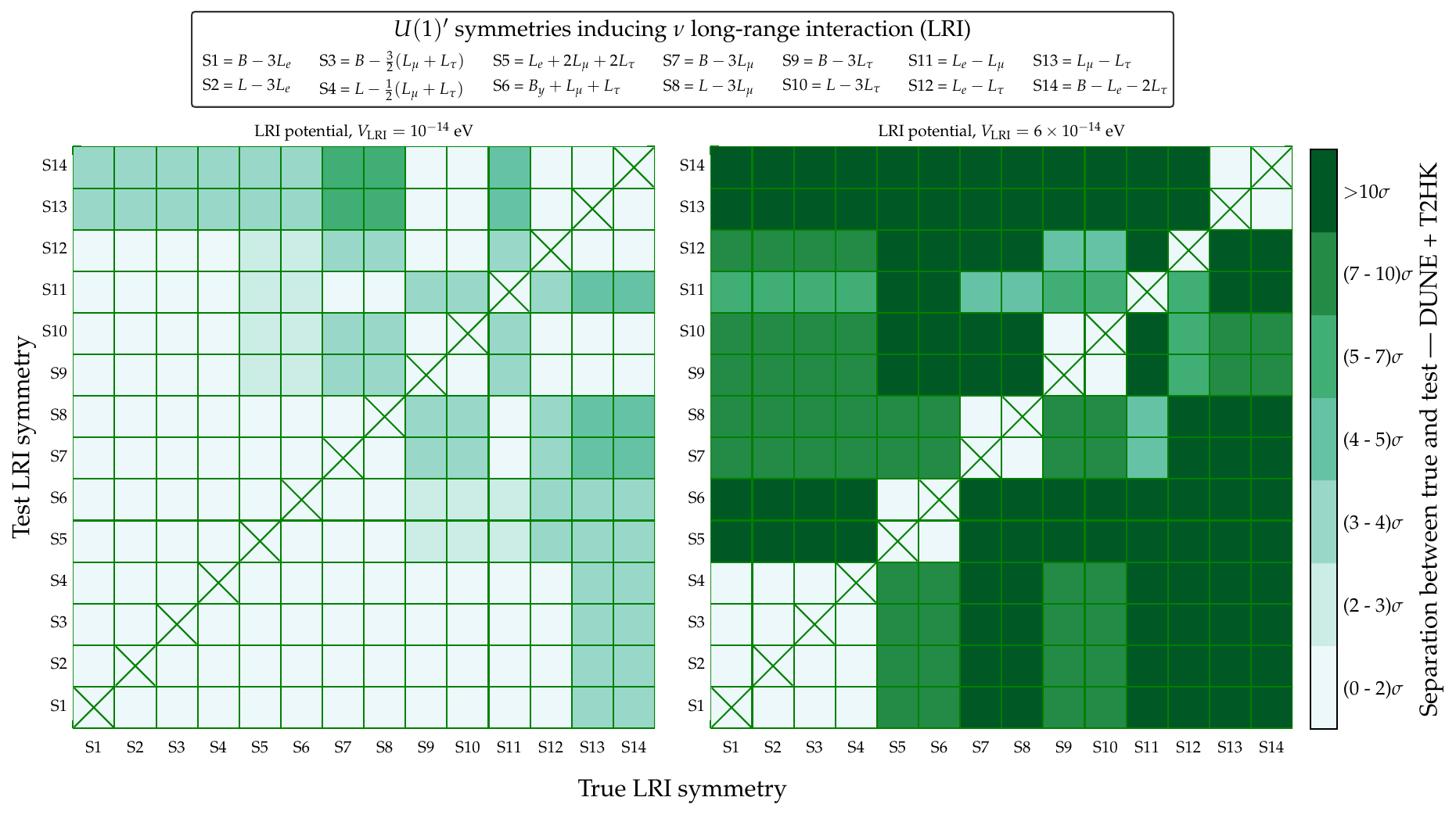}
 \caption{\textbf{\textit{Confusion matrices showing the degree of separation between true and test $U(1)^\prime$ symmetries.}}  The separation is evaluated using the test statistic in \equ{delta_chi2_conf}, and expressed in terms of number of standard deviations, $\sigma$, between the symmetries.  Results are for DUNE and T2HK combined, after 10 years of operation, assuming normal neutrino mass ordering.  We show results for two illustrative values of the new matter potential, $V_{\rm LRI} =  10^{-14}$~eV (\textit{left}) and $6 \cdot 10^{-14}$~eV (\textit{right}).  See sections~\ref{sec:stat_methods} and \ref{sec:confusion-theory} for details.  \textit{Distinguishing between competing symmetries may be feasible, especially for higher values of $V_{\rm LRI}$ and when the texture of the long-range matter interaction potential of each symmetry (table~\ref{tab:charges}) is different.}}
 \label{fig:confusion-matrix}
\end{figure}

Figure~\ref{fig:confusion-matrix} shows this via confusion matrices.  They depict the statistical separation between pairs of symmetries, one true and one test, computed using the test statistic in \equ{delta_chi2_conf} for the combination of DUNE and T2HK.  We show results assuming two illustrative values of the new matter potential, $V_{\rm LRI} = 10^{-14}$~eV and $6 \cdot 10^{-14}$~eV.  The higher the potential is, the more prominent the effects of the new interaction are, and the easier it becomes to contrast event distributions due to competing symmetries.  The confusion matrices are nearly, but not fully, symmetric, since the true and test event spectra are treated differently (section~\ref{sec:stat_methods}).

The separation is clearer between symmetries whose matter potential matrices, $\mathbf{V}_{\rm LRI}$, have different textures; see table~\ref{tab:charges}.  Conversely, the separation is blurred between symmetries whose matter potentials have similar texture, \eg, between $B - 3 L_e$ and $L_e + 2 L_\mu + 2 L_\tau$, and it is null between symmetries whose matter potentials have equal texture, \eg, between $B - 3 L_e$ and $L - 3 L_e$.  This is a fundamental limitation; it persists regardless of the value of $V_{\rm LRI}$.  As when constraining (section~\ref{sec:constraints_pot}) and discovering (section~\ref{sec:discovery}) a new interaction, symmetries that affect primarily the disappearance probabilities, \ie, $L_\mu-L_\tau$ and $B-L_e-2L_\tau$, introduce features into the event rate that may be more easily spotted due to higher event rates, and are thus more easily distinguished from alternative symmetries.  Conversely, symmetries that affect primarily the appearance probabilities are less easily distinguished from alternative symmetries.

%=================================================
%=================================================

%=================================================
\section{Summary and conclusions}
\label{sec:conclusion}
%=================================================

The growing precision achieved by neutrino oscillation experiments endows them with the capability to look for new neutrino interactions with matter that could modify the transitions between $\nu_e$, $\nu_\mu$, and $\nu_\tau$, revealing long-sought physics beyond the Standard Model.  The possibility that the interaction has a long range --- which we focus on --- is particularly compelling.  In this case, neutrinos on Earth could experience a large potential sourced by vast repositories of matter in the local and distant Universe --- the Earth, Moon, Sun, Milky Way, and the cosmological distribution of matter --- thereby enhancing our chances of probing the new interaction.

We have constructed the new interaction by gauging accidental global, anomaly-free $U(1)$ symmetries of the Standard Model that involve combinations of lepton and baryon numbers.  Doing this introduces a new neutral gauge boson, $Z^\prime$, that acts as mediator and induces a matter potential sourced by electrons, neutrons, or protons, depending on the specific symmetry considered.  The interaction range is inversely proportional to the mediator mass, which, along with its coupling strength, is a priori unknown; they are to be determined experimentally.   We have explored masses between $10^{-35}$~eV and $10^{-10}$~eV, corresponding to an interaction range between Gpc and hundreds of meters.

We have studied the prospects of constraining, discovering, and identifying the symmetry responsible for new neutrino interactions in two leading next-generation long-baseline oscillation experiments, DUNE and T2HK, expected to start operations within the next decade.  An initial study~\cite{Singh:2023nek} showed that they could outperform the present-day sensitivity to long-range interactions because of their large sizes, advanced detectors, and intense neutrino beams.  However, that study explored only three out of the many possible candidate symmetries that could induce new interactions, and ones involving exclusively lepton numbers.  Since different symmetries affect flavor transitions differently, it remained to be determined whether the sensitivity claimed in \Refe~\cite{Singh:2023nek} applies broadly.

We have addressed this, motivated by present-day comprehensive searches for long-range interactions~\cite{Coloma:2020gfv}, by exploring a plethora of possible candidate symmetries (table~\ref{tab:charges}) which induce a new matter potential that affects only $\nu_e$, $\nu_\mu$, or $\nu_\tau$, or combinations of them.  To make our forecasts realistic, we base them on detailed simulations of DUNE and T2HK, including accounting for multiple detection channels, energy resolution, backgrounds, and planned operation times of their neutrino beams in $\nu$ and $\bar{\nu}$ modes.

Our conclusions cement and broaden earlier promising perspectives. 
Although the different symmetries have diverse effects on oscillations, we find that \textbf{\textit{regardless of which symmetry is responsible for inducing new neutrino-matter interactions, including long-range ones, DUNE and T2HK may constrain them more strongly than ever before (figs.~\ref{fig:constraints_discovery_dune_t2hk-NMO}, \ref{fig:constraints_on_pot_dune_t2hk-NMO},  and \ref{fig:all_symmetry})  or may discover them (figs.~\ref{fig:constraints_discovery_dune_t2hk-NMO}, \ref{fig:discovery_3s_5s_dune_t2hk}, and \ref{fig:discovery_all_DUNE+T2HK}).}}  The experiments are predominantly sensitive to a new matter potential whose size is comparable to the standard-oscillation potential (\figu{lrf_bounds_NH_selective}), since this induces resonant effects on the oscillation probabilities.  Also, they are predominantly sensitive to new interactions that affect the disappearance channels, $\nu_\mu \to \nu_\mu$ and $\bar{\nu}_\mu \to \bar{\nu}_\mu$, since they have higher event rates, making it easier to spot subtle effects.  

In addition, for the first time, we report that \textbf{\textit{it may be possible to identify the symmetry responsible for the new interaction, or to narrow down the possibilities (\figu{confusion-matrix})}}, especially if the new matter potential is relatively large.  There is, however, an unavoidable limitation to disentangling the effects of two competing symmetries whose effects on the flavor transitions are equal or similar. Nevertheless, in all cases, combining events detected by DUNE and T2HK is key to lifting degeneracies between standard and new oscillation parameters that limit the sensitivity of each experiment individually.

Overall, our results demonstrate that the reach of DUNE and T2HK to probe new neutrino interactions is not only deep, but also broad in its scope.

%=================================================
\subsection*{Acknowledgments}
%=================================================

The authors thank Sudipta Das and Ashish Narang for helpful discussion. 
S.K.A. and P.S. acknowledge the support from the Department of Atomic Energy 
(DAE), Govt. of India, under the Project Identification Number RIO 4001. 
S.K.A. acknowledges the financial support from the Swarnajayanti Fellowship 
(sanction order no. DST/SJF/PSA-05/2019-20) provided by the Department of 
Science and Technology (DST), Govt. of India, and the Research Grant 
(sanction order no. SB/SJF/2020-21/21) provided by the Science and Engineering 
Research Board (SERB), Govt. of India, under the Swarnajayanti Fellowship project. M.B.~is supported by the {\sc Villum Fonden} under the project no. 29388. M.S. acknowledges the financial support from the DST, Govt. of India 
(DST/INSPIRE Fellowship/2018/IF180059). The numerical simulations 
are performed using the “SAMKHYA: High-Performance Computing Facility” 
provided by the Institute of Physics, Bhubaneswar, India.

%=================================================
\begin{appendix}
%=================================================

\newpage

%=================================================
\section{$U(1)^\prime$ charges of fermions}
\label{app:U1-charges}
%=================================================

\renewcommand\thefigure{A\arabic{figure}}
\renewcommand\theHfigure{A\arabic{figure}}
\renewcommand\thetable{A\arabic{table}}
\renewcommand\theequation{A\arabic{equation}}
\setcounter{figure}{0} 
\setcounter{table}{0}
\setcounter{equation}{0}

%-------------------------------------------------------------
Table~\ref{tab:charges2} shows the $U(1)^\prime$ charges of fermions for each of our candidate symmetries (table~\ref{tab:charges}).  The charges are used to compute the long-range matter potential, $V_{\rm LRI}$, using eqs.~(\ref{equ:pot_total_general}) and (\ref{equ:pot_total_simplified}) in the main text.

%=================================================
\begin{table}[t!]
 \centering
 \renewcommand{\arraystretch}{1.3}
 \begin{center}
  \begin{tabular}{|c|c|c|c|c|c|c|c|}
   \hline
   \mr{2}{*}{$U(1)^\prime$ symmetry} &  \mc{6}{c|}{$U(1)^\prime$ charge}
   \\
   &\mr{1}{*}{$a_{u}$} & \mr{1}{*}{$a_{d}$} & \mr{1}{*}{$a_{e}$} & \mr{1}{*}{$b_{e}$} & \mr{1}{*}{$b_{\mu}$} & \mr{1}{*}{$b_{\tau}$} 
   \\
   \hline
   \mr{1}{*}{$B-3L_{e}$}& \mr{1}{*}{$\frac{1}{3}$}&\mr{1}{*}{$\frac{1}{3}$} & \mr{1}{*}{$-3$} & \mr{1}{*} {$-3$} & \mr{1}{*}{0} & \mr{1}{*}{0} 
   \\
   \hline
   \mr{1}{*}{$L-3L_{e}$}& \mr{1}{*}{0}&\mr{1}{*}{0} & \mr{1}{*}{$-2$} & \mr{1}{*} {$-2$} & \mr{1}{*}{1} & \mr{1}{*}{1} 
   \\
   \hline
   \mr{1}{*}{$B-\frac{3}{2}(L_{\mu}+L_{\tau})$}& \mr{1}{*}{$\frac{1}{3}$}&\mr{1}{*}{$\frac{1}{3}$} & \mr{1}{*}{0} & \mr{1}{*} {0} & \mr{1}{*}{$-\frac{3}{2}$} & \mr{1}{*}{$-\frac{3}{2}$} 
   \\
   \hline
   \mr{1}{*}{$L_{e}-\frac{1}{2}(L_{\mu}+L_{\tau})$}& \mr{1}{*}{0}&\mr{1}{*}{0} & \mr{1}{*}{1} & \mr{1}{*} {1} & \mr{1}{*}{$-\frac{1}{2}$} & \mr{1}{*}{$-\frac{1}{2}$} 
   \\
   \hline
   \mr{1}{*}{$L_{e}+2L_{\mu}+2L_{\tau}$}& \mr{1}{*}{0}&\mr{1}{*}{0} & \mr{1}{*}{1} & \mr{1}{*} {1} & \mr{1}{*}{2} & \mr{1}{*}{2} 
   \\
   \hline
   \mr{1}{*}{$B_{y}+L_{\mu}+L_{\tau}$}& \mr{1}{*}{$\frac{1}{3}$}&\mr{1}{*}{$\frac{1}{3}$} & \mr{1}{*}{0} & \mr{1}{*} {0} & \mr{1}{*}{1} & \mr{1}{*}{1} 
   \\
   \hline
   \mr{1}{*}{$B-3L_{\mu}$}& \mr{1}{*}{$\frac{1}{3}$}&\mr{1}{*}{$\frac{1}{3}$} & \mr{1}{*}{0} & \mr{1}{*} {0} & \mr{1}{*}{$-3$} & 0 
   \\
   \hline
   \mr{1}{*}{$L-3L_{\mu}$}& \mr{1}{*}{0}&\mr{1}{*}{0} & \mr{1}{*}{1} & \mr{1}{*} {1} & \mr{1}{*}{$-2$} & \mr{1}{*}{1}
   \\
   \hline
   \mr{1}{*}{$B-3L_{\tau}$}& \mr{1}{*}{$\frac{1}{3}$}&\mr{1}{*}{$\frac{1}{3}$} & \mr{1}{*}{0} & \mr{1}{*} {0} & \mr{1}{*}{0} & \mr{1}{*}{$-3$}
   \\
   \hline
   \mr{1}{*}{$L-3L_{\tau}$}& \mr{1}{*}{0}&\mr{1}{*}{0} & \mr{1}{*}{1} & \mr{1}{*} {1} & \mr{1}{*}{1} & \mr{1}{*}{$-2$}
   \\
   \hline
   \mr{1}{*}{$L_{e}-L_{\mu}$}& \mr{1}{*}{0}&\mr{1}{*}{0} & \mr{1}{*}{1} & \mr{1}{*} {1} & \mr{1}{*}{$-1$} & \mr{1}{*}{0} 
   \\
   \hline
   \mr{1}{*}{$L_{e}-L_{\tau}$}& \mr{1}{*}{0}&\mr{1}{*}{0} & \mr{1}{*}{1} & \mr{1}{*} {1} & \mr{1}{*}{0} & \mr{1}{*}{$-1$} 
   \\
   \hline
   \mr{1}{*}{$L_{\mu}-L_{\tau}$}& \mr{1}{*}{0}&\mr{1}{*}{0} & \mr{1}{*}{0} & \mr{1}{*} {0} & \mr{1}{*}{1} & \mr{1}{*}{$-1$} 
   \\
   \hline					
   \mr{1}{*}{$B-L_{e}-2L_{\tau}$}& \mr{1}{*}{$\frac{1}{3}$}&\mr{1}{*}{$\frac{1}{3}$} & \mr{1}{*}{$-1$} & \mr{1}{*} {$-1$} & \mr{1}{*} {0} & \mr{1}{*} {$-2$}
   \\
   \hline
  \end{tabular}
  \caption{\textbf{\textit{$U(1)^{\prime}$ charges of the fermions for the candidate symmetries.}}  Charges $a_u$, $a_d$, and $a_e$ are, respectively, of the up quark, down quark, and the electron; and $b_e$, $b_\mu$, and $b_\tau$ are, respectively, of $\nu_e$, $\nu_\mu$, and $\nu_\tau$.  For protons, the charge is $a_p = 2 a_u + a_d$; for neutrons, it is $a_n = 2 a_d + a_u$.  The charges are used to compute the long-range potential, $V_{\rm LRI}$, using eqs.~(\ref{equ:pot_total_general}) and (\ref{equ:pot_total_simplified}) in the main text.
  \label{tab:charges2}}
 \end{center}
\end{table}
%=================================================

%=================================================
\section{Effect of long-range interactions on neutrino oscillation parameters}
\label{app:param_run}
%=================================================

\renewcommand\thefigure{B\arabic{figure}}
\renewcommand\theHfigure{B\arabic{figure}}
\renewcommand\thetable{B\arabic{table}}
\renewcommand\theequation{B\arabic{equation}}
\setcounter{figure}{0} 
\setcounter{table}{0}
\setcounter{equation}{0}

\begin{figure}[t!]
 \centering
 \includegraphics[width=0.82\textwidth]{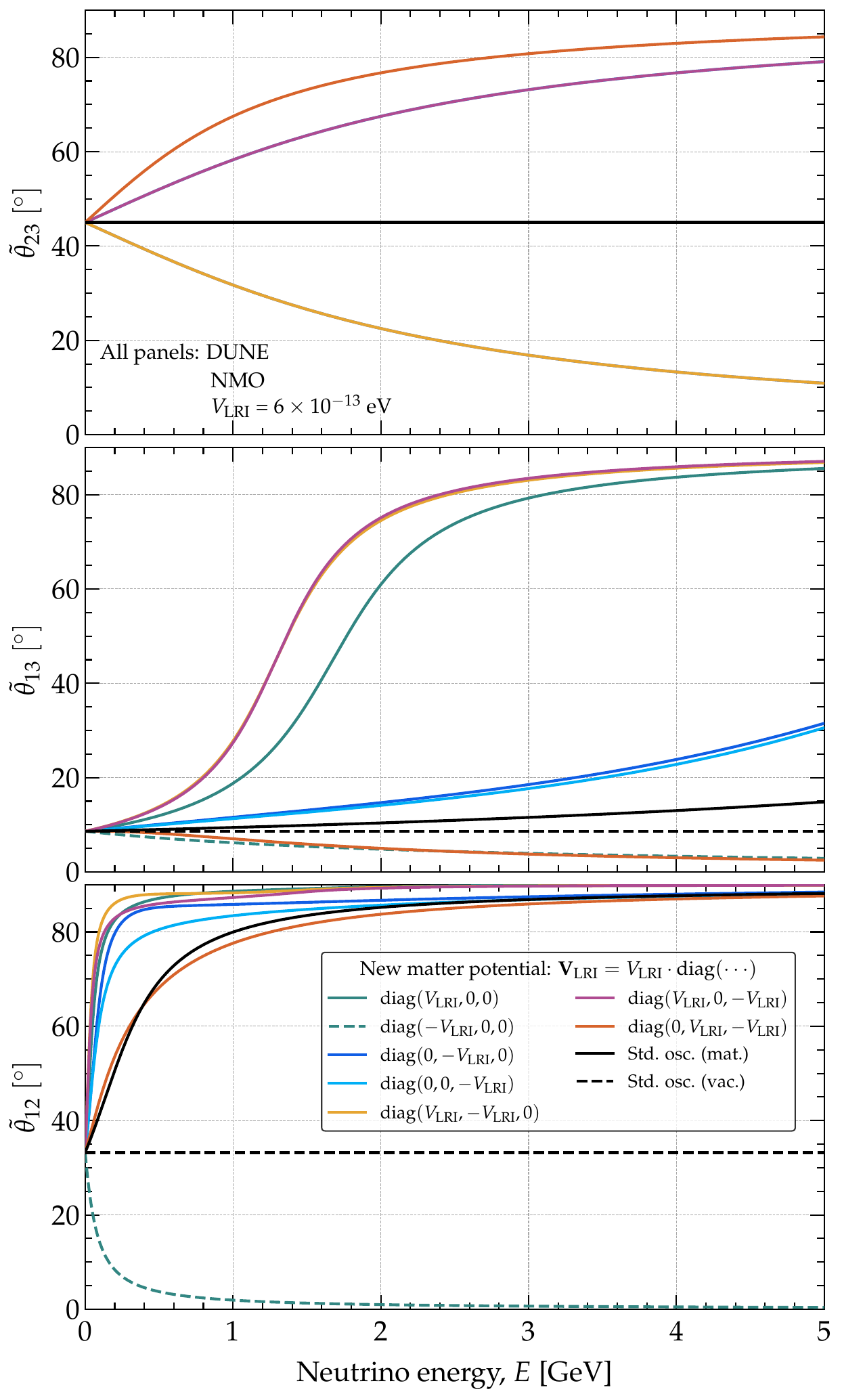}
 \caption{\textbf{\textit{Modification of the mixing angles with energy.}}  We compare their modification in the presence of the new matter potential induced by our candidate $U(1)^\prime$ symmetries {\it vs.}~their standard values in vacuum and modified by matter inside Earth.  We assume the DUNE baseline,  an illustrative value of the new matter potential, of $V_{\rm LRI} = 6 \cdot 10^{-13}$~eV, and the values of the oscillation parameters from table~\ref{tab:params_value1}, except for $\theta_{23}$, which we set to $45^\circ$.  See \figu{delm_run} for the modification of the neutrino mass-squared differences and appendix~\ref{app:param_run} for details.}
 \label{fig:theta_run}
\end{figure}

\begin{figure}[t!]
 \centering
 \includegraphics[width=0.82\textwidth]{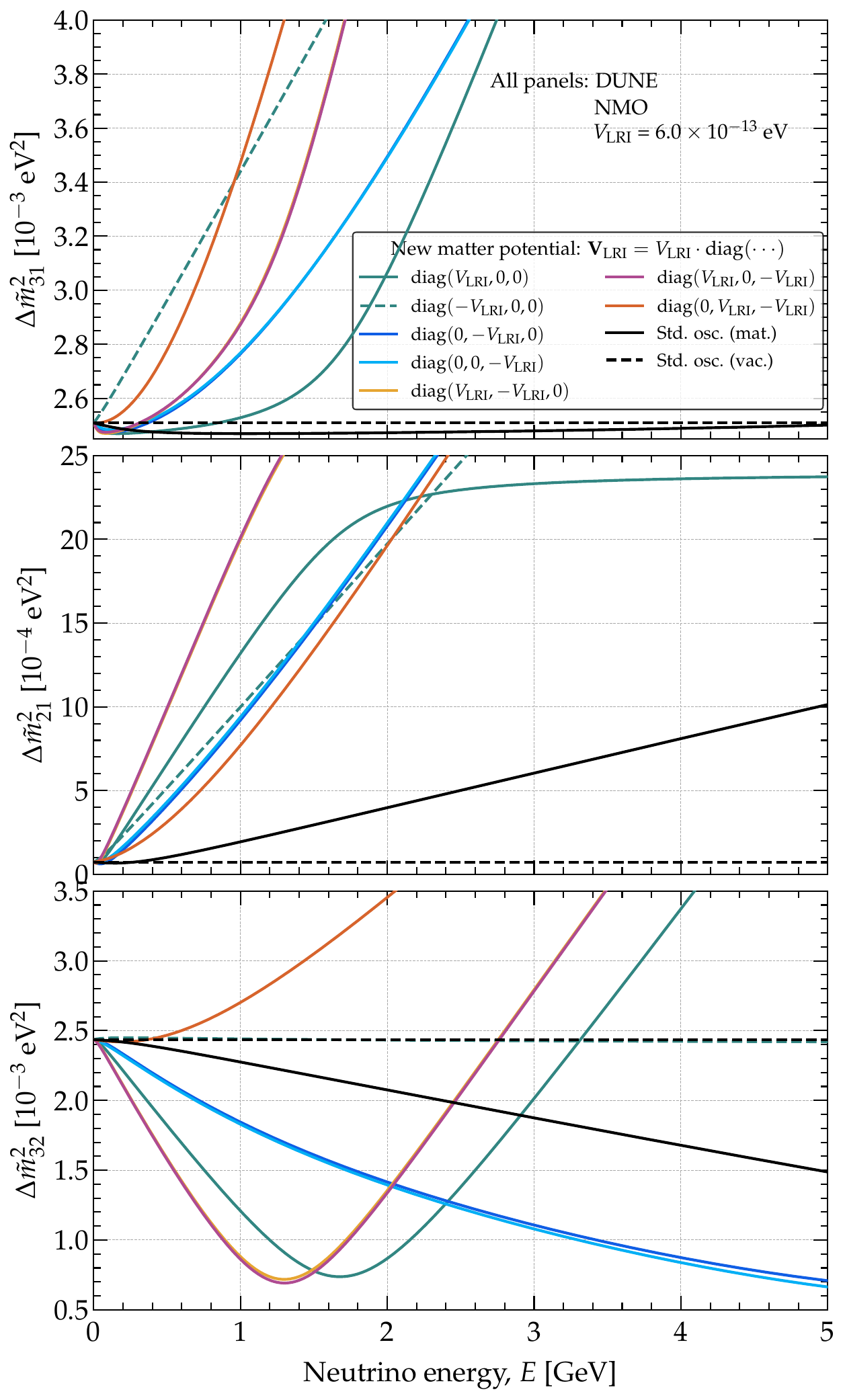}
 \caption{\textbf{\textit{Modification of the neutrino mass-squared differences with energy.}}  We compare their modification in the presence of the new matter potential induced by our candidate $U(1)^\prime$ symmetries {\it vs.}~their standard values in vacuum and modified by matter inside Earth.  We assume the DUNE baseline,  an illustrative value of the new matter potential, of $V_{\rm LRI} = 6 \cdot 10^{-13}$~eV, and the values of the oscillation parameters from table~\ref{tab:params_value1}, except for $\theta_{23}$, which we set to $45^\circ$.  See \figu{theta_run} for the modification of the mixing angles and appendix~\ref{app:param_run} for details.}
 \label{fig:delm_run}
\end{figure}

Figures~\ref{fig:theta_run} and~\ref{fig:delm_run} show the modification with energy of the mixing angles and mass-squared differences modified under the new matter potential introduced by our candidate $U(1)^\prime$ symmetries (table~\ref{tab:charges}).  To produce these figures, we compute the modified oscillation parameters using the approximate expressions from \Refe~\cite{Agarwalla:2021zfr}; to produce all other results, we compute them numerically and implicitly as part of the calculation of the oscillation probabilities.  We show results for DUNE; the results for T2HK, not shown, are analogous.

%=================================================
%=================================================

\begin{figure}[t!]
	\centering
	\includegraphics[width=0.82\textwidth]{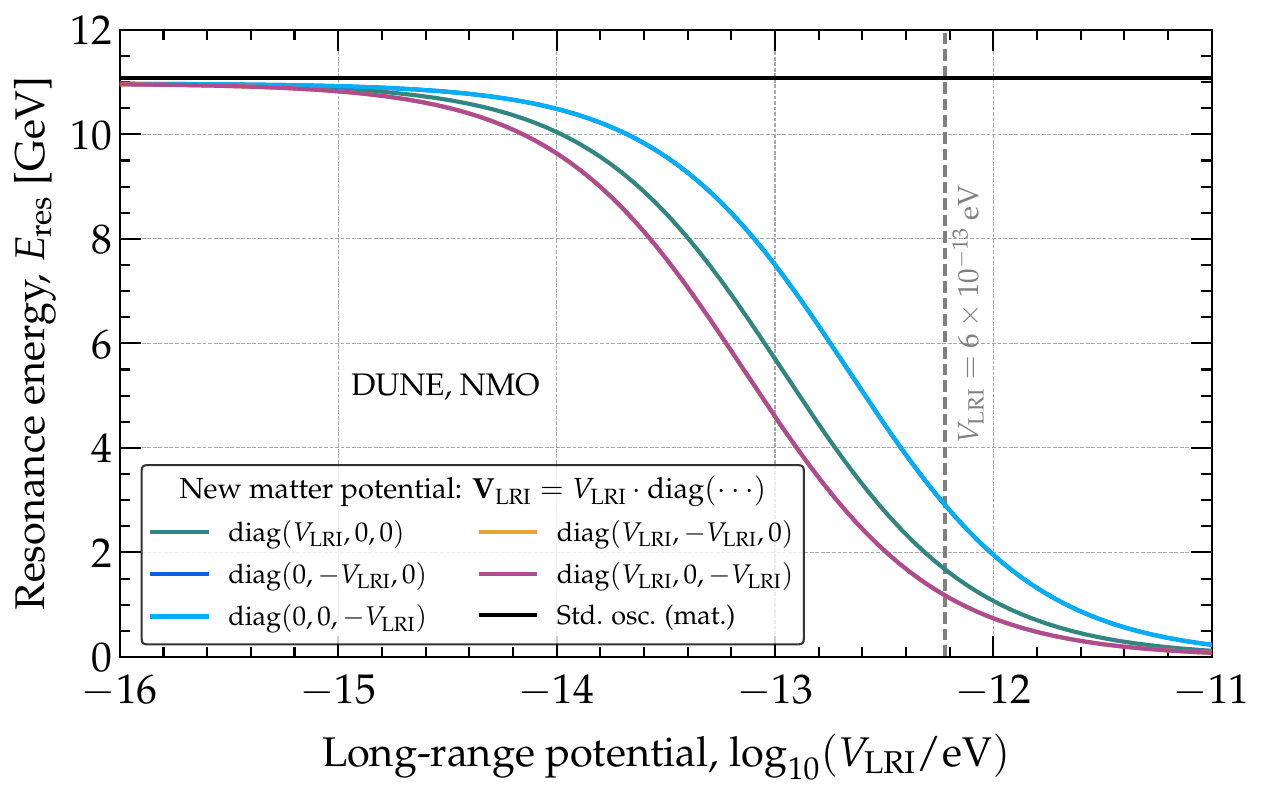}
	\caption{\textbf{\textit{Resonance neutrino energy as a function of long-range interaction potential.}} The energies are calculated for the DUNE baseline assuming normal neutrino mass ordering, and the values of the oscillation parameters are from table~\ref{tab:params_value1}, except for $\theta_{23}$, which we set to $45^\circ$. The grey vertical line corresponds to the illustrative value of long-range potential taken when we showcase its impact on the modification of mixing parameters in figs.~\ref{fig:theta_run}, \ref{fig:delm_run}, and \ref{fig:osc_length_run}, the oscillation probabilities and the event spectra in figs.~\ref{fig:dune_prob_events}, \ref{fig:t2hk_prob_events}, \ref{fig:probability}, and \ref{fig:event_spectra}. It reaffirms that $\tilde\theta_{13}$ attains the resonance value between 1-2 GeV for some symmetries, as shown in the middle panel of \figu{theta_run} and validates our approximate expression, \equ{resonance_lri}.}
	\label{fig:res_energy}
\end{figure}

We group symmetries according to the texture of the new matter potential that they induce, $\mathbf{V}_{\rm LRI}$ in table~\ref{tab:charges}.  Symmetries with equal or similar potential texture yield equal or similar modification of the oscillation parameters, which, in turn yields equal or similar oscillation probabilities (figs.~\ref{fig:probability} and \ref{fig:event_spectra}).  Below, we point out the salient features in the modification of the oscillation parameters:

%=================================================
\begin{description}
 %%%
 \item[Modification of $\tilde\theta_{23}$ (\figu{theta_run}):]
  The value of $\tilde\theta_{23}$ drives both the transition and survival probabilities, eqs.~(\ref{equ:app_mat}) and (\ref{equ:surv_mat}).  Symmetries that induce a new potential in the muon sector, tau sector, or both, of the Hamiltonian, \equ{hamiltonian_tot} in the main text, affect the modification of $\tilde\theta_{23}$.  This includes symmetries that induce a matter potential, $\mathbf{V}_{\rm LRI}$ of the form $\textrm{diag}(0,\bullet,0)$ (\ie, $B-3L_\mu$ and $L-3L_\mu$), $\textrm{diag}(0,0,\bullet)$ (\ie, $B-3L_\tau$, $L-3L_\tau$), $\textrm{diag}(\bullet,\bullet,0)$ (\ie, $L_e-L_\mu$), $\textrm{diag}(\bullet,0,\bullet)$ (\ie, $L_e-L_\tau$), and, especially, $\textrm{diag}(0,\bullet,\bullet)$ (\ie, $L_\mu-L_\tau$ and $B-L_e-2L_\tau$).  Depending on the signs of the nonzero elements of $\mathbf{V}_{\rm LRI}$, the value of $\tilde\theta_{23}$ either increases or decreases {\it vs.}~its value in vacuum, $\theta_{23}$.  Symmetries that induce a new potential only in the electron sector, \ie, with texture $\textrm{diag}(\bullet,0,0)$, do not affect the modification of $\tilde\theta_{23}$.  This encompasses  symmetries  $B-3L_e$, $L-3L_e$, $B-\frac{3}{2}(L_\mu+L_\tau)$, $L_e-\frac{1}{2}(L_\mu+L_\tau)$, $L_e + 2L_\mu + 2L_\tau$, and $B_y + L_\mu + L_\tau$.
 %%%
 \item[Modification of $\tilde\theta_{13}$ (\figu{theta_run}):]
  The value of $\tilde\theta_{13}$ also drives both probabilities, eqs.~(\ref{equ:app_mat}) and (\ref{equ:surv_mat}).  Most of our candidate symmetries induce a new potential in either the electron sector, tau sector, or both, and so directly affect the modification of $\tilde\theta_{13}$.  The exceptions are the symmetries that induce a potential only in the muon sector, of the form $\textrm{diag}(0,\bullet,0)$ (\ie, $B-3L_\mu$ and $L-3L_\mu$), which, however, still affect $\tilde\theta_{13}$ indirectly due to the mixing of the muon sector with the electron and tau sectors via $\mathbf{H}_{\rm vac}$.  Like for $\tilde\theta_{23}$ above, depending on the signs of the nonzero elements of $\mathbf{V}_{\rm LRI}$, the value of $\tilde\theta_{13}$ either increases or decreases {\it vs.}~its value in vacuum, $\theta_{13}$.  The largest deviations from the vacuum value occur under symmetries that induce a potential that affects both sectors: $\textrm{diag}(\bullet,0,\bullet)$ (\ie, $L_e-L_\tau$) and $\textrm{diag}(\bullet,\bullet,0)$ (\ie, $L_e-L_\mu$, which affects the tau sector via the standard mixing between it and the muon sector). In the presence of new matter potentials arising from most of our candidate symmetries, $\tilde\theta_{13}$ reaches $45^\circ$ at the resonance energy and continues to increase as the energy rises. 
  From \equ{app_mat}, it is clear that the probability approaches maximum as soon as $\tilde\theta_{13}$ attains resonance. 
  
  The resonance energy under LRI in the one-mass-scale-dominance (OMSD, $\Delta m^2_{31}L/4E$ $>>\Delta m^2_{21}L/4E$) approximation and assuming $\theta_{23} = 45^\circ$, is given by~\cite{Agarwalla:2021zfr},
  \begin{equation}
  \label{equ:resonance_lri}
  E_{\text{res}}^{\text{LRI}} \simeq  
  \left[E_{\text{res}}^{\text{SI}}\right]_\text{OMSD} \cdot V_{\rm CC} \cdot \bigg[\frac{1-(\alpha s^2_{12} c_{13}^{2} /\cos 2\theta_{13})}{V_{\rm CC} - \frac{1}{2}(V_{{\rm LRI}, \mu}+V_{{\rm LRI}, \tau}-2V_{{\rm LRI}, e})}\bigg] ,
  \end{equation}  
  where 
  \begin{equation}
  \label{equ:resonance_si}
  \left[E_{\text{res}}^{\text{SI}}\right]_\text{OMSD} = \frac{\Delta m^2_{31}\cos2\theta_{13}}{2V_{CC}} , 
  \end{equation}
  is the resonance energy in presence of standard charged-current interactions.
  Figure~\ref{fig:res_energy} shows the resonance energy as a function of $V_{\rm LRI}$ for the symmetry textures that are expected to give $\tilde\theta_{13}$ resonance. The resonance energies decrease from that of the standard oscillation value as $V_{\rm LRI}$ grows. The resonance is attained at lower energies for the symmetries with (i) ${\bf V}_{\rm LRI}={\rm diag}(V_{\rm LRI}, 0, -V_{\rm LRI})$ than (ii) ${\bf V}_{\rm LRI}={\rm diag}(V_{\rm LRI}, 0, 0)$ than (iii) ${\bf V}_{\rm LRI}={\rm diag}(0, 0, -V_{\rm LRI})$. This can be understood from \equ{resonance_lri}, as the denominator of the term inside the bracket increases more for (i) than (ii) than (iii) leading to the decrease in the energy needed to attain the resonance. The resonance energies are equivalent for symmetries with ${\bf V}_{\rm LRI}$ of the form ${\rm diag}(V_{\rm LRI}, -V_{\rm LRI}, 0)$ \& ${\rm diag}(V_{\rm LRI}, 0, -V_{\rm LRI})$ and ${\rm diag}(0, -V_{\rm LRI}, 0)$ \& ${\rm diag}(0, 0, -V_{\rm LRI})$. For all these symmetries, the $\tilde\theta_{13}$ resonance for DUNE baseline happens below 2 GeV when the strength of the new potential reaches around $10^{-12}-10^{-11}$ eV, which is clearly seen in \figu{dune_prob_events}, where we illustrate one particular texture, ${\rm diag}(0, 0, -V_{\rm LRI})$. For the textures, ${\rm diag}(-V_{\rm LRI}, 0, 0)$ and ${\rm diag}(0, V_{\rm LRI}, -V_{\rm LRI})$, $\tilde\theta_{13}$ decreases with energy even from its vacuum and will never achieve resonance. 
 %%%
 \item[Modification of $\tilde\theta_{12}$ (\figu{theta_run}):]
  Most of our candidate symmetries induce a new potential in either the electron sector, muon sector, or both, and so directly affect the modification of $\tilde\theta_{12}$.  The exceptions are the symmetries that induce a potential only in the tau sector, of the form $\textrm{diag}(0,0,\bullet)$ (\ie, $B-3L_\tau$ and $L-3L_\tau$), which, however, still affect $\tilde\theta_{12}$ indirectly due to the mixing of the tau sector with the electron and muon sectors via $\mathbf{H}_{\rm vac}$.  In nearly all cases, the value of $\tilde\theta_{12}$ saturates to $90^\circ$ early in its energy modification, which justifies our use of the approximate expression for the $\nu_\mu \to \nu_e$ probability, \equ{app_mat}, to interpret our results in the main text.  The exceptions are the symmetries that induce a potential of the form $\textrm{diag}(-V_{\rm LRI}, 0, 0)$ (\ie, $L_e+2L_\mu+2L_\tau$ and $B_y+L_\mu+L_\tau$), which instead quickly drives $\tilde\theta_{12}$ to zero; however, in this case, $\tilde\theta_{12}$ instead saturates early to $90^\circ$ for antineutrinos, since they are affected by the potential $-\mathbf{V}_{\rm LRI}$.
  %%%
  \item[Modification of $\Delta \tilde m^2_{31}$ (\figu{delm_run}):] 
   The modification of $\Delta \tilde m^2_{31}$ affects the oscillation phase of the $\nu_\mu \to \nu_\mu$ probability, \equ{surv_mat}.  Figure~\ref{fig:osc_length_run} shows the modification with energy of the oscillation length associated to $\Delta \tilde m^2_{31}$, \ie, $L_{\rm osc}^{31} \equiv 2.47~{\rm km}~(E/{\rm GeV}) / (\Delta \tilde{m}_{31}^2/{\rm eV}^2)$, which helps understand the impact of the modification of $\Delta \tilde m^2_{31}$ on the $\nu_\mu \to \nu_\mu$ probability.  At low energies, below about 0.8~GeV, the value of $\Delta \tilde m^2_{31}$ decreases slightly below its vacuum value, $\Delta m^2_{31}$, for most of our candidate symmetries.  However, this is overcome by the low energies and, as a result, the $\nu_\mu \to \nu_\mu$ oscillation length is shorter than in vacuum (\figu{osc_length_run}) and grows more slowly than in vacuum, and so the first oscillation maximum of the probability is shifted to slightly higher energies to compensate for the slower growth in $L_{\rm osc}^{31}$; see \figu{probability}.  At higher energies, \figu{delm_run} shows that the value of $\Delta \tilde m^2_{31}$ increases quickly, but because the energy is also growing, the net effect is to first stall and then overturn the growth of $L_{\rm osc}^{31}$, which, again, shifts the position of the second maximum of the probability further to higher energies.
  %%%
  \item[Modification of $\Delta \tilde m^2_{21}$ (\figu{delm_run}):]
   The value of $\Delta \tilde m^2_{21}$ grows with energy under all of our candidate symmetries.  However, this has only a mild impact on our results, since the transition and survival probabilities for DUNE and T2HK, eqs.~(\ref{equ:app_mat}) and (\ref{equ:surv_mat}), are driven by $\Delta \tilde m^2_{31}$ and $\Delta \tilde m^2_{32}$.
  %%%
  \item[Modification of $\Delta \tilde m^2_{32}$ (\figu{delm_run}):]
   The modification of $\Delta \tilde m^2_{32} \equiv \Delta \tilde m^2_{31}-\Delta \tilde m^2_{21}$ affects the oscillation phase of the $\nu_\mu \to \nu_e$ probability, \equ{app_mat}.  For most of our candidate symmetries, the value of $\Delta \tilde m^2_{32}$ is smaller than the vacuum value, $\Delta m^2_{32}$, across most of the energy range in \figu{delm_run}, roughly below 3~GeV.  As a result, in this range, the oscillation length associated to $\Delta \tilde m^2_{32}$, \ie, $L_{\rm osc}^{32} \equiv 2.47~{\rm km}~(E/{\rm GeV}) / (\Delta \tilde{m}_{32}^2/{\rm eV}^2)$, grows with energy faster than in vacuum (\figu{osc_length_run}) which, in turn, shifts the first and second oscillation maxima in the $\nu_\mu \to \nu_e$ probability to lower energies; see \figu{probability}.
\end{description}

%=================================================
%=================================================

%=================================================
%=================================================

%=================================================
\begin{figure}[t!]
 \centering
 \includegraphics[width=0.82\textwidth]{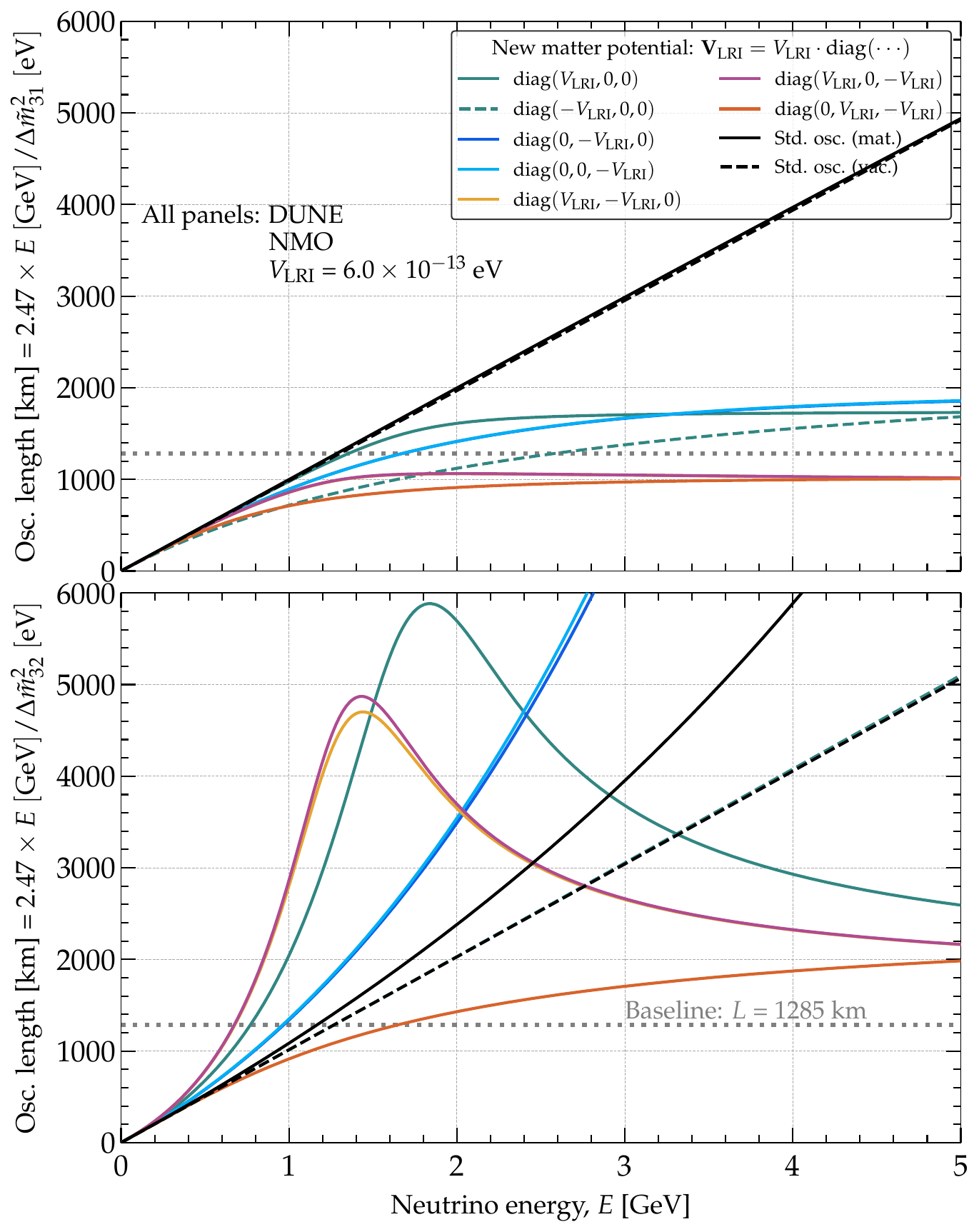}
 \caption{\textbf{\textit{Modification of the neutrino oscillation length with energy.}}  We show the modification of the oscillation length associated with the squared-mass difference modified by the new matter, $\Delta \tilde{m}_{31}^2$ ({\it top}) and $\Delta \tilde{m}_{32}^2$ ({\it bottom}), for our candidate $U(1)^\prime$ symmetries.  We compare them against the standard values in vacuum and modified by matter inside Earth.  We assume the DUNE baseline, an illustrative value of the new matter potential, of $V_{\rm LRI} = 6 \cdot 10^{-13}$~eV, and the values of the oscillation parameters from table~\ref{tab:params_value1}, except for $\theta_{23}$, which we set to $45^\circ$.  See \figu{delm_run} for the modification of the mass-squared differences and appendix~\ref{app:param_run} for details.}
 \label{fig:osc_length_run}
\end{figure}
%=================================================

\renewcommand\thefigure{C\arabic{figure}}
\renewcommand\theHfigure{C\arabic{figure}}
\renewcommand\thetable{C\arabic{table}}
\renewcommand\theequation{C\arabic{equation}}
\setcounter{figure}{0} 
\setcounter{table}{0}
\setcounter{equation}{0}

%=================================================
\begin{figure}[t!]
 \centering
 \includegraphics[width=\textwidth]{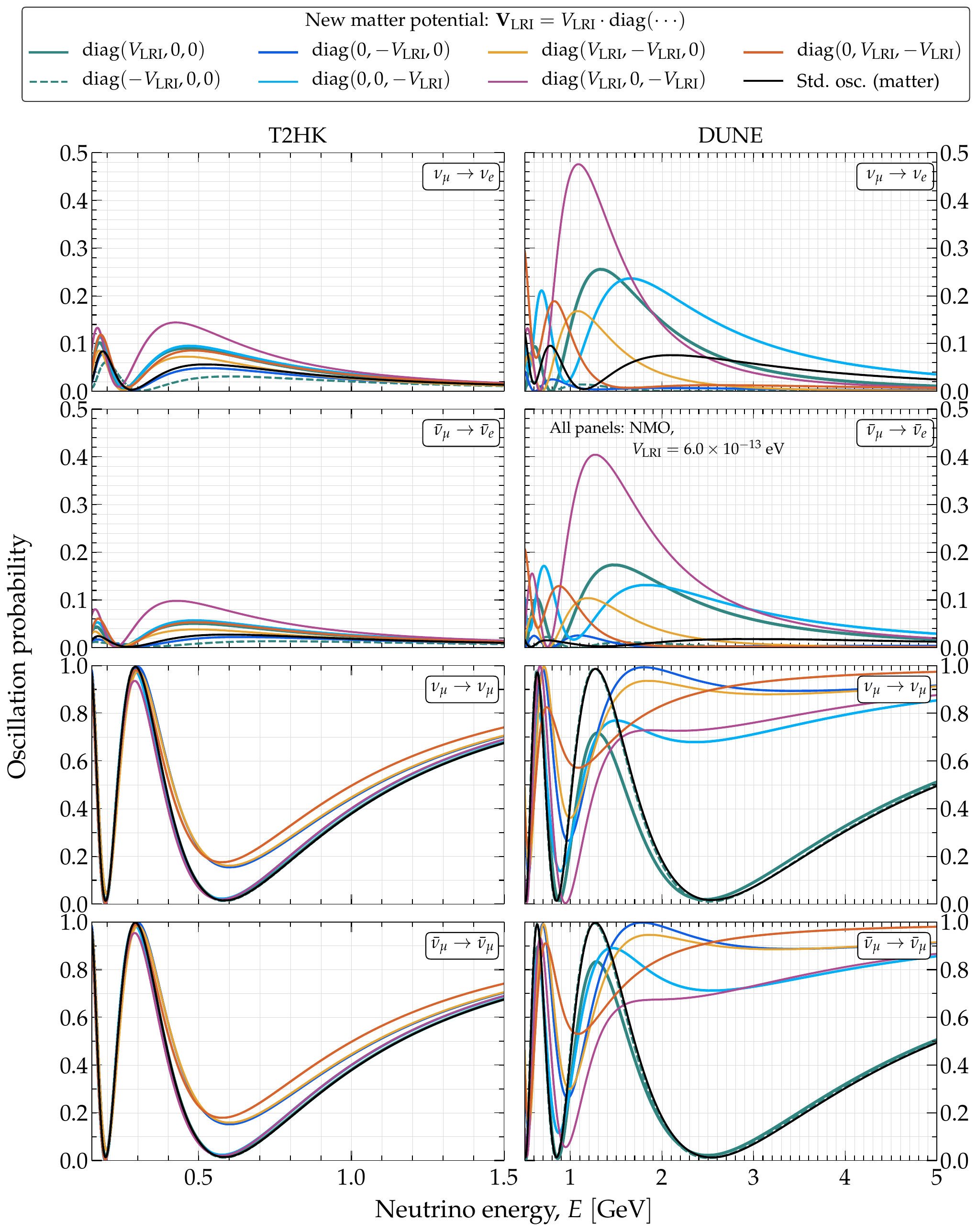}
 \caption{\textbf{\textit{Neutrino oscillation probabilities in the presence of a new matter potential.}}  The new matter potential is induced by each of our candidate $U(1)^\prime$ symmetries (table~\ref{tab:charges}).  In this figure, the neutrino mass ordering is normal, the values of the standard oscillation parameters are the best-fit values from table~\ref{tab:params_value1}, and we pick an illustrative value of the potential, of $V_{\rm LRI} = 6 \cdot 10^{-13}$~eV.  The probabilities are for T2HK ({\it left column}) and DUNE ({\it right column}), and for all the detection channels that we consider in our analysis: $\nu_\mu \to \nu_e$, $\bar{\nu}_\mu \to \bar{\nu}_e$, $\nu_\mu \to \nu_\mu$, and $\bar{\nu}_\mu \to \bar{\nu}_\mu$.  This figure extends the results shown in figs.~\ref{fig:dune_prob_events} and \ref{fig:t2hk_prob_events}.  See section~\ref{sec:prob_var_pot} for details and \figu{event_spectra} for corresponding results for the distribution of detected events.
 \label{fig:probability}} 
\end{figure}
%=================================================

%=================================================
\begin{figure}[t!]
 \centering
 \includegraphics[width=1\textwidth]{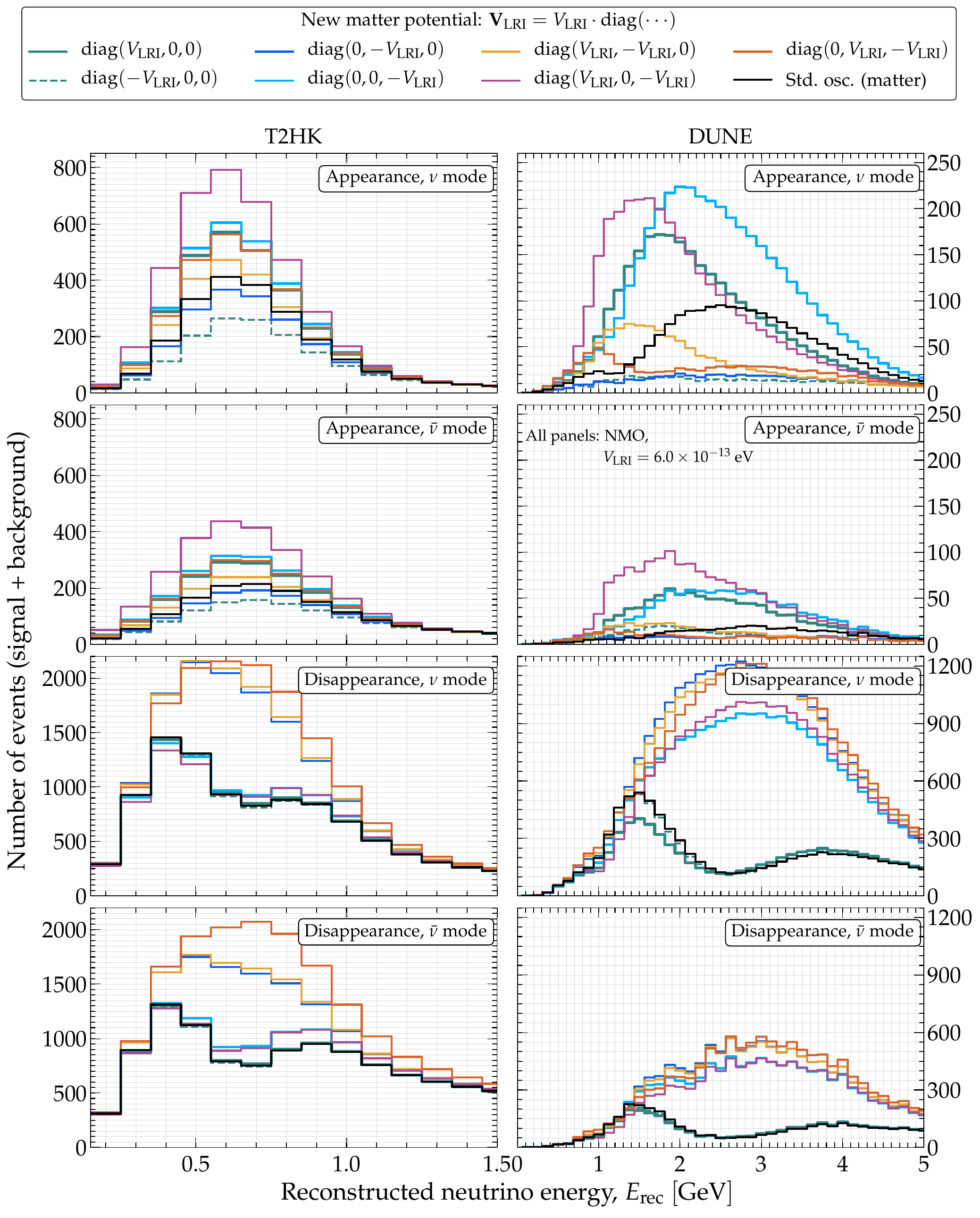}
 \caption{\textbf{\textit{Spectra of detected neutrino-initiated events in the presence of a new matter potential.}}  The new matter potential is induced by each of our candidate $U(1)^\prime$ symmetries (table~\ref{tab:charges}).  In this figure, the neutrino mass ordering is normal, the values of the standard oscillation parameters are the best-fit values from table~\ref{tab:params_value1}, and we pick an illustrative value of the potential, of $V_{\rm LRI} = 6 \cdot 10^{-13}$~eV.  The spectra are for T2HK ({\it left column}) and DUNE ({\it right column}), and for all the detection channels that we consider in our analysis: appearance and disappearance in neutrino and antineutrino models.  This figure extends the results shown in figs.~\ref{fig:dune_prob_events} and \ref{fig:t2hk_prob_events}.  See section~\ref{sec:expt-details} for details and \figu{probability} for corresponding results for the neutrino oscillation probabilities.
 \label{fig:event_spectra}} 
\end{figure}
%=================================================

%=================================================
\section{Effect of a new matter potential on oscillations and event rates}
\label{app:prob_plots}
%=================================================

Figures~\ref{fig:probability} and~\ref{fig:event_spectra} show, respectively, the oscillation probabilities and event spectra across all the detection channels of DUNE and T2HK, for all our candidate $U(1)^\prime$ symmetries (table~\ref{tab:charges}).  They extend the illustrative case for a single choice of symmetry shown in figs.~\ref{fig:dune_prob_events} and \ref{fig:t2hk_prob_events} in the main text.  The features of the event spectra reflect the features of the oscillation probabilities.  The behavior of the latter results from the running of the oscillation parameters in the presence of the new matter potential (figs.~\ref{fig:theta_run} and~\ref{fig:delm_run}).

%=================================================
\section{Detailed results on constraints}
\label{app:other}
%=================================================

\renewcommand\thefigure{D\arabic{figure}}
\renewcommand\theHfigure{D\arabic{figure}}
\renewcommand\thetable{D\arabic{table}}
\renewcommand\theequation{D\arabic{equation}}
\setcounter{figure}{0} 
\setcounter{table}{0}
\setcounter{equation}{0}

\begin{figure}[t!]
 \centering
 \includegraphics[width=1\textwidth]{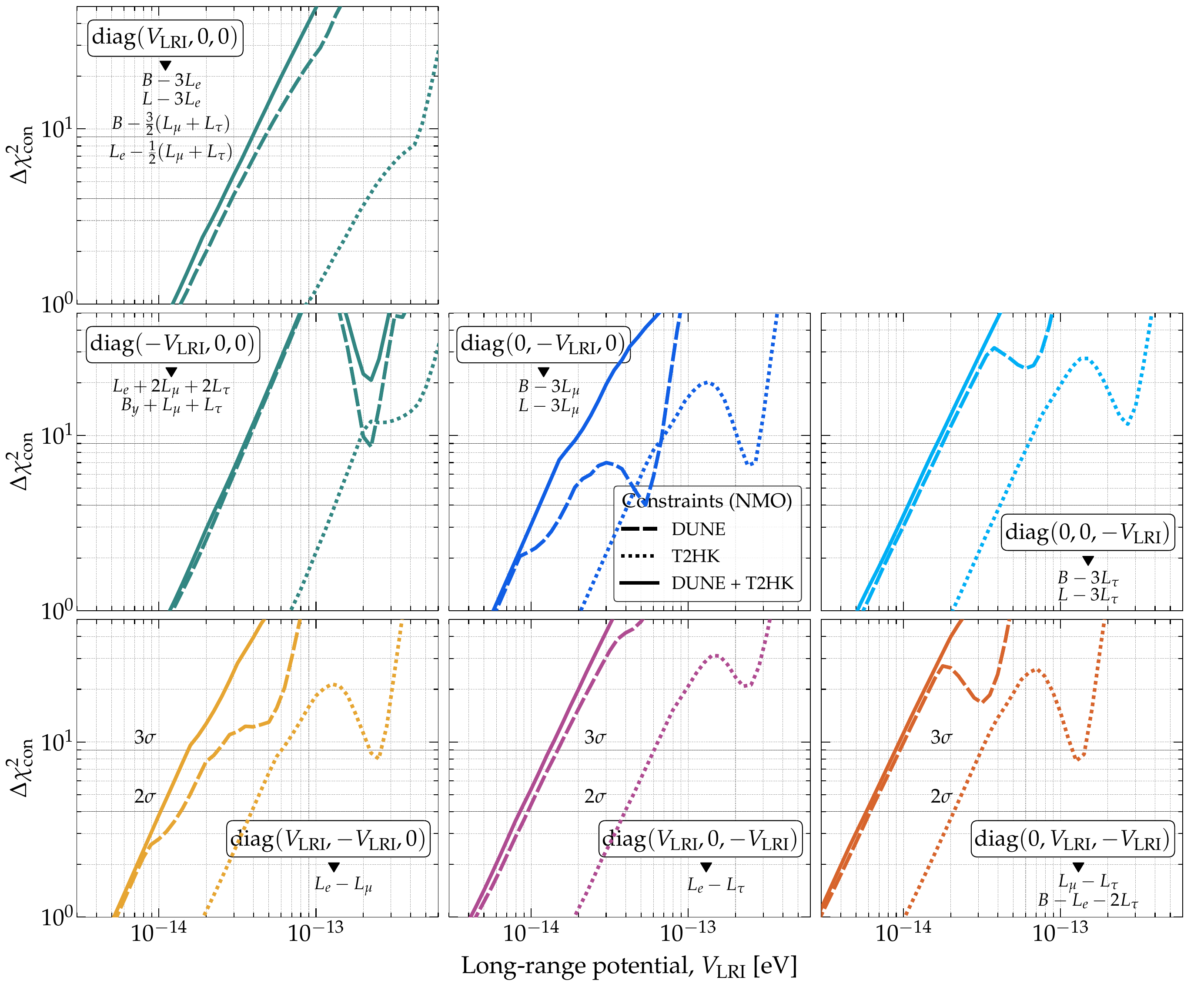}
 \caption{\textbf{\textit{Projected test statistic used to constrain the new matter potential induced by our candidate $U(1)^\prime$ symmetries, assuming normal mass ordering}}  This figure extends the illustrative case shown in \figu{lrf_bounds_NH_selective}.  The test statistic is \equ{delta_chi2_dune}, computed for DUNE and T2HK separately and combined.  Figure~\ref{fig:lrf_bounds_IH} shows results under inverted mass ordering.  See sections~\ref{sec:stat_methods} and \ref{sec:constraints_pot} for details.
 \label{fig:lrf_bounds_NH}} 
\end{figure}

Figure~\ref{fig:lrf_bounds_NH} shows the test statistic that we use to place constraints on the new matter potential, \equ{delta_chi2_dune} for DUNE and analogous expressions for T2HK and DUNE + T2HK, for all our candidate symmetries, assuming the true neutrino mass ordering is normal.  This figure includes and extends \figu{lrf_bounds_NH_selective} in the main text, where we showed a single illustrative case.  In \figu{lrf_bounds_NH}, the symmetries are grouped according to the texture of the matter potential they induce, $\mathbf{V}_{\rm LRI}$ in table~\ref{tab:charges}, since symmetries with equal or similar potential texture yield equal or similar constraints on $V_{\rm LRI}$ (section~\ref{sec:constraints_pot}).  The results in \figu{lrf_bounds_NH} reaffirm and extend those in \figu{lrf_bounds_NH_selective}: while the constraints are driven by DUNE, it is only by combining it with T2HK that the parameter degeneracies that plague each experiment separately  --- the ``dips'' in their individual test statistics --- are lifted (section~\ref{sec:constraints_pot}).  

In line with section~\ref{sec:constraints_pot}, \figu{lrf_bounds_NH} shows that the tightest limits on $V_{\rm LRI}$ are obtained for symmetries that affect primarily $\tilde\theta_{23}$, since this is the mixing angle that drives the amplitude of the oscillation probabilities, eqs.~(\ref{equ:app_mat}) and (\ref{equ:surv_mat}).  Among those symmetries, the ones that induce $\mathbf{V}_{\rm LRI}$ with texture of the form $\textrm{diag}(0,\bullet,\bullet)$ (\ie, $L_\mu-L_\tau$ and $B-L_e-2L_\tau$), and, therefore, affect primarily the muon and tau sector of the Hamiltonian, yield the best limits on $V_{\rm LRI}$ (table~\ref{tab:upper_limits_potential_NMO_organized}); see also appendix~\ref{app:param_run}.  Conversely, symmetries that induce a potential texture of the form $\textrm{diag}(\bullet,0,0)$ and that, therefore, affect predominantly the electron sector, yield the weakest limits, since they do not modify $\tilde\theta_{23}$; see appendix~\ref{app:param_run}.

In \figu{lrf_bounds_NH}, the degeneracies in the test statistic are larger for symmetries whose matter potential contains negative entries; see table~\ref{tab:charges}.  These negative entries partially cancel the standard matter potential $\mathbf{V}_{\rm mat}$ (section~\ref{sec:hamiltonians}), hindering the capability of DUNE to single out the neutrino mass ordering, and resulting in the large dips in some of the test statistics seen in \figu{lrf_bounds_NH}.  Because T2HK has a shorter baseline than DUNE, it is less affected by standard matter effects, and therefore less impacted by the above issue.  This is why combining DUNE and T2HK strengthens the resulting constraints on $V_{\rm LRI}$.

\begin{figure}[t!]
 \centering
 \includegraphics[width=1\textwidth]{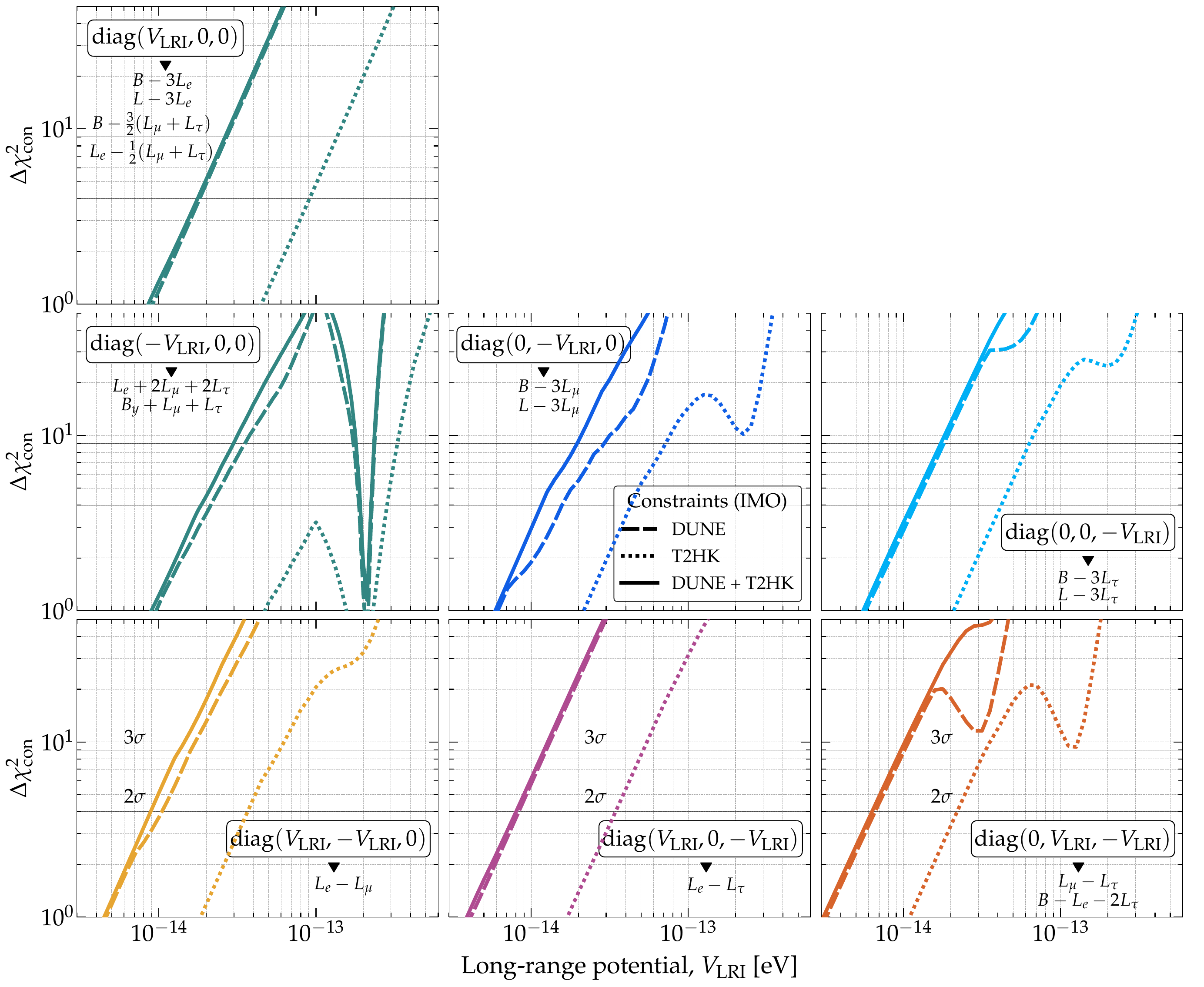}
 \caption{\textbf{\textit{Projected test statistic used to constrain the new matter potential induced by our candidate $U(1)^\prime$ symmetries, assuming inverted mass ordering.}}  Same as \figu{lrf_bounds_NH}, but for inverted mass ordering.  See sections~\ref{sec:stat_methods} and \ref{sec:constraints_pot} for details.
 \label{fig:lrf_bounds_IH}} 
\end{figure}

Figure~\ref{fig:lrf_bounds_IH} shows the same test statistic as \figu{lrf_bounds_NH}, but computed assuming that the true neutrino mass ordering is inverted.  Compared to \figu{lrf_bounds_NH}, the impact of the parameter degeneracies on the test statistic is milder (except for symmetries with a potential of the form $\mathbf{V}_{\rm LRI} = (-V_{\rm LRI}, 0, 0)$, which we explain below).  This is related to the degeneracy between $\theta_{23}$ and $\delta_{\rm CP}$, and the need to detect comparable event rates of neutrinos and antineutrinos in order to resolve it~\cite{Agarwalla:2013ju}.  Under normal mass ordering, the antineutrino rates are suppressed by the smaller interaction cross section and flux, which makes the rates of events due to neutrinos and antineutrinos uneven.  Under inverted ordering, the antineutrino rates (not shown) are significantly enhanced due to the matter effects, thereby making them comparable to those of neutrinos.  This helps to break the degeneracy between $\theta_{23}$ and $\delta_{\rm CP}$~\cite{Ballett:2016daj, Bernabeu:2018twl, Kelly:2018kmb, King:2020ydu, Singh:2021kov, Agarwalla:2022xdo}, and to remove the dips in the test statistic in \figu{lrf_bounds_IH}. However, for the texture ${\textbf V_{\rm LRI}}=(-V_{\rm LRI}, 0, 0)$, it is instead the degeneracy between $V_{\rm LRI}$ and the mass ordering that affects the test statistic, which leads its exhibiting a deeper dip under inverted mass ordering, where the determination of the mass ordering is impaired, than under normal ordering ({\it cf.}~\figu{lrf_bounds_NH} and \figu{lrf_bounds_IH}).

\begin{figure}[t!]
 \centering
 \includegraphics[width=1\textwidth]{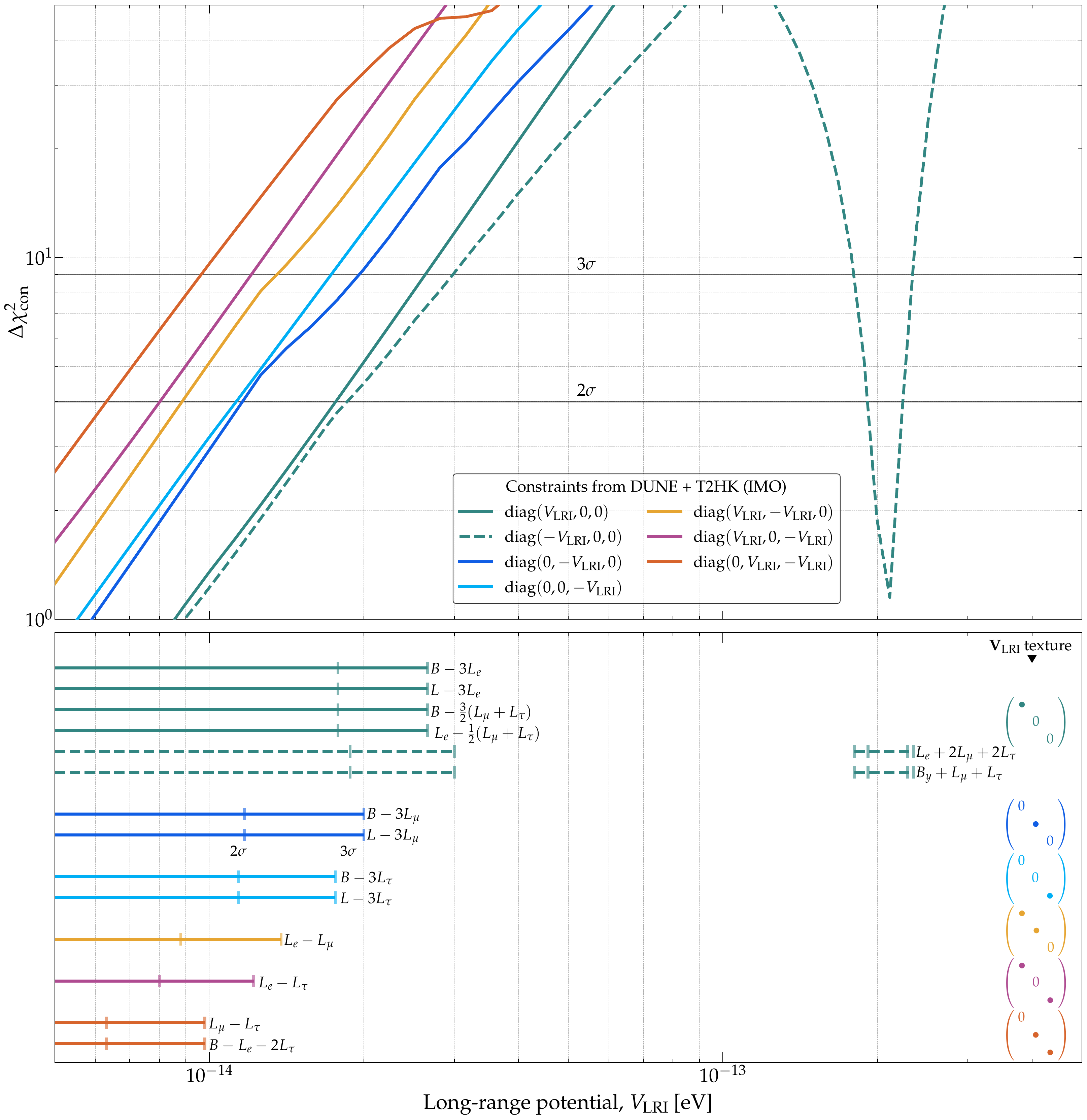}
 \caption{ \textbf{\textit{Projected test statistic (top) used to place upper limits (bottom) on the new matter potential induced by our candidate $U(1)^\prime$ symmetries.}}  Same as \figu{constraints_on_pot_dune_t2hk-NMO}, but assuming that the true neutrino mass ordering is inverted.
 \label{fig:constraints_on_pot_dune_t2hk-IMO}} 
\end{figure}

Figure~\ref{fig:constraints_on_pot_dune_t2hk-IMO} shows the upper limits on $V_{\rm LRI}$ obtained from \figu{lrf_bounds_IH}.  Because of the above explanation, the limits on symmetries that induce the new matter potential in the electron sector, \ie, those that have $\mathbf{V}_{\rm LRI} = \textrm{diag}(\bullet,0,0)$, improve compared to assuming normal ordering, except for the case ${\textbf V_{\rm LRI}}=(-V_{\rm LRI}, 0, 0)$, {\it cf.}~\figu{constraints_on_pot_dune_t2hk-IMO} and \figu{constraints_on_pot_dune_t2hk-NMO} in the main text.

Table~\ref{tab:upper_limits_potential_NMO_organized} shows the numerical values of the upper limits on $V_{\rm LRI}$ from figs.~\ref{fig:constraints_on_pot_dune_t2hk-NMO} and \ref{fig:constraints_on_pot_dune_t2hk-IMO}, for all our candidate symmetries, for normal and inverted mass ordering, and for DUNE and T2HK, separate and together. 

%=================================================
%=================================================

%=================================================
%=================================================

%=================================================
%=================================================

%=================================================
\begin{table}[t!]
 \centering
 \renewcommand{\arraystretch}{1.3}
 \resizebox{\columnwidth}{!}{%
  \begin{tabular}{|c|c|c|c|c|c|c|c|c|c|c|c|c|}
   \hline
   \multirow{4}{*}{$U(1)^{\prime}$ symmetry} &
   \multicolumn{12}{c|}{Upper limit on the new matter potential, $V_{\mathrm{LRI}}$ [$10^{-14}$~eV]} 
   \\ 
   & \mc{6}{c|}{Normal mass ordering (NMO)} & \mc{6}{c|}{Inverted mass ordering (IMO)}
   \\
   & \mc{2}{c|}{DUNE} & \mc{2}{c|}{T2HK} &\mc{2}{c|}{ DUNE+T2HK} & \mc{2}{c|}{DUNE} & \mc{2}{c|}{T2HK} &\mc{2}{c|}{ DUNE+T2HK} 
   \\
   & 2$\sigma$ & 3$\sigma$ & 2$\sigma$ & 3$\sigma$ & 2$\sigma$ & 3$\sigma$ & 2$\sigma$ & 3$\sigma$ & 2$\sigma$ & 3$\sigma$ & 2$\sigma$ & 3$\sigma$
   \\ 
   \hline
   $B-3L_e$  & 3.0 & 4.80 & 21.60 & 45.0 & 2.52 & 3.96 & 1.82 & 2.66 & 9.13 & 13.51 & 1.78 & 2.66
   \\
   $L-3L_e$ & $\vert$ & $\vert$ & $\vert$ & $\vert$ & $\vert$ & $\vert$ & $\vert$ & $\vert$ & $\vert$ & $\vert$ & $\vert$ & $\vert$  
   \\
   $B-\frac{3}{2}(L_{\mu}+L_{\tau})$ & $\vert$ & $\vert$ & $\vert$ & $\vert$ & $\vert$ & $\vert$ & $\vert$ & $\vert$ & $\vert$ & $\vert$ & $\vert$ & $\vert$  
   \\
   $L_e-\frac{1}{2}(L_\mu+L_\tau)$ & $\vert$ & $\vert$ & $\vert$ & $\vert$ & $\vert$ & $\vert$ & $\vert$ & $\vert$ & $\vert$ & $\vert$ & $\vert$ & $\vert$  
   \\ 
   $L_e+2L_\mu+2L_\tau$ & 2.50 & 22.50 & 13.60 & 18.60 & 2.36 & 3.56 & 22.68 & 23.52 & 28.62 & 33.35 & 22.62 & 23.52 
   \\
   $B_y+L_\mu+L_\tau$ & $\vert$ & $\vert$ & $\vert$ & $\vert$ & $\vert$ & $\vert$ & $\vert$ & $\vert$ & $\vert$ & $\vert$ & $\vert$ & $\vert$  
   \\
   \hline
   $B-3L_\mu$ & 5.40 & 6.72 & 4.20 & 28.20 & 1.14 & 1.86 & 1.58 & 2.95 & 4.37 & 7.22 & 1.17 & 2.0 
   \\
   $L-3L_\mu$ & $\vert$ & $\vert$ & $\vert$ & $\vert$ & $\vert$ & $\vert$ & $\vert$ & $\vert$ & $\vert$ & $\vert$ & $\vert$ & $\vert$  
   \\ 
   \hline
   $B-3L_\tau$ & 1.20 & 1.80 & 4.20 & 6.36 & 1.08 & 1.68 & 1.21 & 1.82 & 4.28 & 6.31 & 1.14 & 1.76 
   \\
   $L-3L_\tau$ & $\vert$ & $\vert$ & $\vert$ & $\vert$ & $\vert$ & $\vert$ & $\vert$ & $\vert$ & $\vert$ & $\vert$ & $\vert$ & $\vert$  
   \\ 
   \hline
   $L_e-L_\mu$ &  1.40 & 2.50 & 4.10 & 25.70 & 1.03 & 1.56 & 1.05 & 1.61 & 3.74 & 6.06 & 0.88 & 1.38
   \\
   \hline
   $L_e-L_\tau$ & 0.98 & 1.50 & 4 & 6.10 & 0.84 & 1.32 & 0.84 & 1.26 & 3.49 & 5.21 & 0.8 & 1.22 
   \\
   \hline
   $L_\mu-L_\tau$ & 0.62 & 0.95 & 2.12 & 14.30 & 0.58 & 0.9 & 0.66 & 1.03 & 2.25 & 3.41 & 0.63 & 0.98 
   \\
   $B-L_{e}-2L_{\tau}$ & $\vert$ & $\vert$ & $\vert$ & $\vert$ & $\vert$ & $\vert$ & $\vert$ & $\vert$ & $\vert$ & $\vert$ & $\vert$ & $\vert$
   \\
   \hline
  \end{tabular}
 }
 \caption{\textbf{\textit{Projected upper limits on the new matter potential, $V_{\rm LRI}$, induced by our candidate $U(1)^\prime$ symmetries.}}  As in table~\ref{tab:charges}, the symmetries are grouped according to the texture of the matter potential, $\mathbf{V}_{\rm LRI}$, that they induce.  Symmetries with equal or similar potential texture yield equal or similar upper limits.  See figs.~\ref{fig:constraints_on_pot_dune_t2hk-NMO} and \ref{fig:constraints_on_pot_dune_t2hk-IMO} for a graphical representation of the limits in this table, and sections~\ref{sec:stat_methods} and \ref{sec:constraints_pot} for details.  The contents of this table are available in \Refe~\cite{GitHub_lrf-repo}.}
 \label{tab:upper_limits_potential_NMO_organized}
\end{table}
%=======f==========================================

%=================================================
\section{Detailed results on discovery prospects}
\label{app:discovery}
%=================================================

\renewcommand\thefigure{E\arabic{figure}}
\renewcommand\theHfigure{E\arabic{figure}}
\renewcommand\thetable{E\arabic{table}}
\renewcommand\theequation{E\arabic{equation}}
\setcounter{figure}{0} 
\setcounter{table}{0}
\setcounter{equation}{0}

\begin{figure}[t!]
 \centering
 \includegraphics[width=0.85\textwidth]{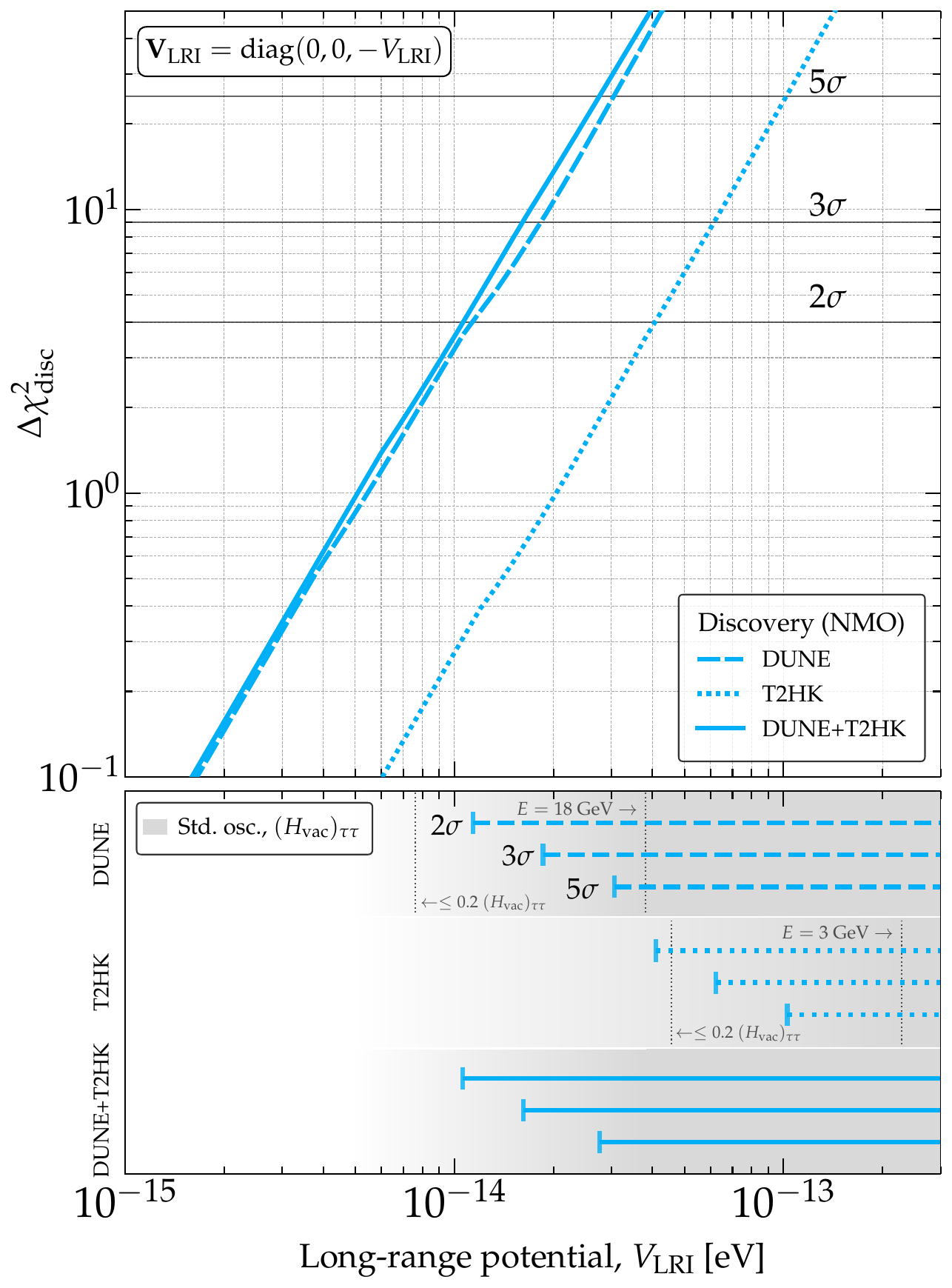}
 \caption{\textbf{\textit{Projected test statistic used to compute discovery prospects on the new matter potential induced by a $U(1)^\prime$ symmetry.}} For this plot, as illustration, we show limits on a potential of the form $\mathbf{V}_{\rm LRI} = \textrm{diag}(0, 0, -V_{\rm LRI})$ for neutrinos and $-\mathbf{V}_{\rm LRI}$ for antineutrinos, as would be introduced by symmetries $L - 3L_\tau$ or $B - 3L_\tau$ (table~\ref{tab:charges}).  The test statistic is \equ{delta_chi2_disc}.  Results are for DUNE and T2HK separately and combined.  The true neutrino mass ordering is assumed to be normal.  See sections~\ref{sec:stat_methods} and \ref{sec:discovery} for details.  Like when placing constraints (\figu{lrf_bounds_NH_selective}), the experiments are sensitive to values of $V_{\rm LRI}$ that are comparable to the standard-oscillation terms in the Hamiltonian; for the choice of $\mathbf{V}_{\rm LRI}$ texture in this figure, this is $(\mathbf{H}_{\rm vac})_{\tau\tau}$.  Figure~\ref{fig:discovery_3s_5s_dune_t2hk} shows the discovery prospects for all of our candidate symmetries.} 
 \label{fig:discovery_b-3ltau}
\end{figure}

Figure~\ref{fig:discovery_b-3ltau} shows the test statistic, \equ{delta_chi2_disc}, that we use to forecast discovery prospects of $V_{\rm LRI}$, assuming, for illustration, a matter potential with the texture $\mathbf{V}_{\rm LRI} = {\rm diag}(0,0, -V_{\rm LRI})$, as in figs.~\ref{fig:dune_prob_events}, \ref{fig:t2hk_prob_events}, and \ref{fig:lrf_bounds_NH_selective}.  Unlike in the test statistic that we use to place constraints (figs.~\ref{fig:lrf_bounds_NH_selective}, \ref{fig:lrf_bounds_NH}, and \ref{fig:lrf_bounds_IH}), there are no large degeneracies between $V_{\rm LRI}$ and the standard oscillation parameters since, when computing the test statistic, we fix the test value of $V_{\rm LRI}$ to zero while varying the standard oscillation parameters.

Table~\ref{tab:discovery_strength} gives the numerical values of $V_{\rm LRI}$ that lead to discovery for all our candidate symmetries, as shown in \figu{discovery_3s_5s_dune_t2hk} in the main text.  

%=================================================
%=================================================

%=================================================
\begin{table}[t!]
 \centering
 \renewcommand{\arraystretch}{1.3}
 \begin{tabular}{|c|c|c|c|c|c|c|}
  \hline
  \multirow{4}{*}{$U(1)^{\prime}$ symmetry} & 
  \multicolumn{6}{c|}{\makecell{Discovery strength of LRI potential}} 
  \\
  & \mc{6}{c|}{[$10^{-14}$~eV], NMO}
  \\
  \cline{2-7}
  & \mc{2}{c|}{DUNE} & \mc{2}{c|}{T2HK} & \mc{2}{c|}{DUNE+T2HK}
  \\
  & $3\sigma$ & $5\sigma$ & $3\sigma$ & $5\sigma$ & $3\sigma$ & $5\sigma$
  \\
  \hline
  $B-3L_e$ &  3.75 & 6.30 & 18.30 & 29.10 & 3.60 & 6.0 
  \\
  $L-3L_e$ & $\vert$ & $\vert$ & $\vert$ & $\vert$ & $\vert$ & $\vert$ 
  \\
  $B-\frac{3}{2}(L_{\mu}+L_{\tau})$ & $\vert$ & $\vert$ & $\vert$ & $\vert$ & $\vert$ & $\vert$ 
  \\ 
  $L_e-\frac{1}{2}(L_\mu+L_\tau)$ & $\vert$ & $\vert$ & $\vert$ & $\vert$ & $\vert$ & $\vert$ 
  \\
  $L_e+2L_\mu+2L_\tau$ & 22.40 & 29.0 & 50.0 & 62.80 & 4.0 & 7.0 
  \\
  $B_y+L_\mu+L_\tau$ & $\vert$ & $\vert$ & $\vert$ & $\vert$ & $\vert$ & $\vert$ 
  \\ 
  \hline
  $B-3L_\mu$ & 2.16 & 3.72 & 6.24 & 10.50 & 1.68 & 3.0  
  \\
  $L-3L_\mu$ & $\vert$ & $\vert$ & $\vert$ & $\vert$ & $\vert$ & $\vert$ 
  \\
  \hline
  $B-3L_\tau$ & 1.86 & 3.06 & 6.24 & 10.26 & 1.62 & 2.76 
  \\
  $L-3L_\tau$ & $\vert$ & $\vert$ & $\vert$ & $\vert$ & $\vert$ & $\vert$ 
  \\ 
  \hline
  $L_e-L_\mu$ & 1.90 & 3.20 & 5.90 & 9.90 & 1.50 & 2.65  
  \\
  \hline
  $L_e-L_\tau$ & 1.40 & 2.30 & 5.60 & 9.0  & 1.28 & 2.12  
  \\
  \hline
  $L_\mu-L_\tau$ & 1.03 & 1.70 & 3.20 & 5.40 & 0.91 & 1.50 
  \\
  $B-L_{e}-2L_{\tau}$ & $\vert$ & $\vert$ & $\vert$ & $\vert$ & $\vert$ & $\vert$ 
  \\  
  \hline
 \end{tabular}
 \caption{\textbf{\textit{Projected discovery prospects of the new matter potential induced by our candidate $U(1)^\prime$ symmetries.}}  As in table~\ref{tab:charges}, the symmetries are grouped according to the texture of the matter potential, $\mathbf{V}_{\rm LRI}$, that they induce.  Symmetries with equal or similar potential texture yield equal or similar discovery prospects.  See \figu{discovery_3s_5s_dune_t2hk} for a graphical representation of the values in this table, and sections~\ref{sec:stat_methods} and \ref{sec:discovery} for details.  The contents of this table are available in \Refe~\cite{GitHub_lrf-repo}.}
 \label{tab:discovery_strength}
\end{table}
%=================================================

\end{appendix}

%=================================================
\bibliographystyle{JHEP}
\bibliography{refer-lri.bib}
%=================================================

\end{document}